\documentclass[11pt,a4paper]{JHEP3}
\usepackage{amssymb}
\usepackage{amsthm}
\usepackage{amstext}
\usepackage{amsmath}
\usepackage{amsfonts}
\usepackage{verbatim}
\usepackage{latexsym}
\usepackage[dvips]{graphicx}
\usepackage{t1enc}
\usepackage{graphicx}
\usepackage[all]{xy}
\usepackage{makeidx}
\usepackage{slashed}
\usepackage{feynmf}

\DeclareMathOperator{\Tr}{Tr}

\numberwithin{equation}{section}

\title{Phases of Thermal $\mathcal{N}=2$ Quiver Gauge Theories}

\preprint{arXiv:0708.3199 [hep-th]}

\author{Kasper J. Larsen and Niels A. Obers\\
The Niels Bohr Institute, Blegdamsvej 17, 2100 Copenhagen \O, Denmark \\
E-mail: \email{kjlarsen@nbi.dk}, \email{obers@nbi.dk}}

\abstract{We consider large $N$ $U(N)^M$ thermal $\mathcal{N}=2$
quiver gauge theories on $S^1 \times S^3.$ We obtain a phase
diagram of the theory with $R$-symmetry chemical potentials,
separating a low-temperature/high-chemical potential region from a
high-temperature/low-chemical potential region. In close analogy
with the $\mathcal{N}=4$ SYM case, the free energy is of order
$\mathcal{O}(1)$ in the low-temperature region and of order
$\mathcal{O}(N^2M)$ in the high-temperature phase. We conclude
that the $\mathcal{N} = 2$ theory undergoes a first order Hagedorn
phase transition at the curve in the phase diagram separating
these two regions. We observe that in the region of zero
temperature and critical chemical potential the Hilbert space of
gauge invariant operators truncates to smaller subsectors. We
compute a 1-loop effective potential with non-zero VEV's for the
scalar fields in a sector where the VEV's are homogeneous and
mutually commuting. At low temperatures the eigenvalues of these
VEV's are distributed uniformly over an $S^5/\mathbb{Z}_M$ which
we interpret as the emergence of the $S^5/\mathbb{Z}_M$ factor of
the holographically dual geometry $AdS_5 \times S^5/\mathbb{Z}_M$.
Above the Hagedorn transition the eigenvalue distribution of the
Polyakov loop opens a gap, resulting in the collapse of the joint
eigenvalue distribution from $S^5/\mathbb{Z}_M \times S^1$ into
$S^6/\mathbb{Z}_M$.}

\begin{document}

\tableofcontents

\section{Introduction}

The phase structure of large $N$ $U(N)$ gauge theories at finite
temperature is in itself a very rich and interesting subject that
may provide qualitative insight into the phase structure of QCD.
Even more so, the AdS/CFT correspondence \cite{Maldacena:1997re,
Gubser:1998bc, Witten:1998qj} has provided a general framework for
translating results obtained in weakly coupled thermal gauge
theory into results about the finite temperature behavior of the
physics of black holes and stringy geometry at strong coupling.
One such connection was suggested by Witten \cite{Witten:1998zw}
who argued that the Hawking-Page phase transition
\cite{Hawking:1982dh} between thermal $AdS_5$ and the large
$AdS_5$ Schwarzschild black hole should have a holographic dual
description as a confinement/deconfinement transition in the dual
thermal field theory defined on the conformal boundary $S^1 \times
S^3$ of thermal $AdS_5$.

A general framework for studying large $N$ $U(N)$ gauge theories
on $S^3$ at finite temperature was given in \cite{Aharony:2003sx}.
In particular, this considered $\mathcal{N}=4$ $U(N)$ SYM theory
which was also independently studied in \cite{Sundborg:1999ue}.
Furthermore, for the $\mathcal{N}=4$ case the analysis was
extended in \cite{Yamada:2006rx, Harmark:2006di} to include
chemical potentials conjugate to the $R$-charges. In this way a
phase diagram of the theory as a function of both temperature and
chemical potentials was obtained. As one application of the phase
diagram, in \cite{Harmark:2006di} the observation was made that in
regions of small temperature and critical chemical potential
$\mathcal{N}=4$ SYM theory reduces to quantum mechanical
subsectors, including the $\mathrm{XXX}_{1/2}$ Heisenberg spin
chain.\footnote{Recently other decoupling limits have been found
in near-critical regions by extending this analysis to include the
chemical potentials conjugate to the angular momenta on $S^3$
\cite{Harmark:2007px}.}

Again for $\mathcal{N}=4$ SYM theory, the framework of
\cite{Aharony:2003sx, Sundborg:1999ue} was generalized in a
different direction in \cite{Hollowood:2006xb} by allowing
non-zero VEV's for the scalar fields. There a one-loop effective
potential for the theory at finite temperature on $S^3$ at weak 't
Hooft coupling was computed under the assumption that the VEV's of
the scalar fields are constant and diagonal
matrices.\footnote{This potential was computed earlier in
\cite{Yamada:2006rx} for the special case of zero Polyakov loop
eigenvalues.} The potential was used there to study the
manifestation of the Gregory-Laflamme instability\footnote{See
\cite{Harmark:2007md} for a recent review of the Gregory-Laflamme
instability.} of the small $AdS_5$ black hole from the weakly
coupled gauge theory point of view. The solutions to the equations
of motion obtained from the effective potential of
\cite{Hollowood:2006xb} were given in \cite{Gursoy:2007np} in
terms of a joint eigenvalue distribution of the Polyakov loop and
the scalar VEV's. Within the sector of constant and commuting scalar VEV's
it was found that the topology of the eigenvalue distribution of these
VEV's undergoes a phase transition $S^1 \times S^5 \to S^6$ at the
Hagedorn temperature. The authors interpreted the $S^5$ eigenvalue
distribution of the scalar VEV's as the emergence of
the $S^5$ factor of the holographically dual thermal $AdS_5 \times
S^5$ geometry. It should be noted that, while the truncation to
commuting matrices is consistent, this sector will not describe
the absolute minima of the action \cite{Aharony:2007rj}. For this reason
the observed phase transitions in the commuting saddles studied
in Ref.~\cite{Gursoy:2007np} are not transitions in the full gauge theory.

The discovery that the eigenvalues of scalar VEV's reconstruct the
dual spacetime geometry was originally made by Berenstein \emph{et
al.} in \cite{Berenstein:2004kk, Lin:2004nb, Berenstein:2005aa, Berenstein:2007wz} by
setting up matrix models for the various sectors of BPS operators
in the chiral ring. In particular the model for 1/8 BPS operators
was developed in \cite{Berenstein:2005aa} where the dynamics was
shown to reduce to that of the eigenvalues of three commuting
Hermitian matrices $Z,X,Y$ plus two fermionic
matrices\footnote{However, throughout the analysis of the dynamics
in \cite{Berenstein:2005aa}, the fermionic matrices $W_\alpha$ are
disregarded.} $W_\alpha$. The quantum mechanical Hamiltonian for
the eigenvalues involves an attractive harmonic oscillator part
and a repulsive Vandermonde type part. These forces are balanced
when the eigenvalues are localized to a hypersurface in
$\mathbb{C}^3$ which is taken to be an $S^5$ due to the $SO(6)$
invariance of the quantum Hamiltonian. This $S^5$ was identified
with the $S^5$ factor of the holographically dual geometry $AdS_5
\times S^5$.

The purpose of this paper is to investigate the phase structure of
$\mathcal{N}=2$ $U(N)^M$ quiver gauge theories at finite
temperature.\footnote{See also Refs. \cite{Schnitzer:2004qt,
Nakayama:2005mf, Hikida:2006qb, Schnitzer:2006xz, Grignani:2007xz}
for related work on other supersymmetric gauge theories.} Defined
at zero temperature and on a flat spacetime, these gauge theories
are $\mathcal{N}=2$ supersymmetric and conformally invariant
\cite{Douglas:1996sw, Kachru:1998ys}. We carry out the
investigation of the phase structure in two directions. First, we
consider the case of non-zero $R$-symmetry chemical potentials.
One interesting question here is whether the high-temperature
phase admits several solutions. A further point of interest is to
examine whether one can uncover information about closed
subsectors of the as yet not completely settled underlying spin
chain of $\mathcal{N}=2$ quiver gauge theory by studying the
near-critical chemical potential and low temperature regions of
the $(T,\mu)$ phase diagram of the theory as done for
$\mathcal{N}=4$ SYM theory \cite{Harmark:2006di, Harmark:2006ta}.

Another question of interest is to what extent the $S^5$
eigenvalue distribution of the $\mathcal{N}=4$ SYM scalar VEV's
found in \cite{Berenstein:2005aa} and \cite{Gursoy:2007np} can be
interpreted as the emergence of the $S^5$ factor of the dual
string theory geometry $AdS_5 \times S^5$. To examine this
question, we make use of the fact that $\mathcal{N}=2$ quiver
gauge theory can be realized as a $\mathbb{Z}_M$ projection of
$\mathcal{N}=4$ SYM theory. The holographically dual spacetime of
the $\mathcal{N}=2$ theory is thus $AdS_5 \times S^5/\mathbb{Z}_M$
where $\mathbb{Z}_M$ only acts on the $S^5$ factor. If the above
interpretation of emergent spacetime is correct, we should then
expect to find an $S^5/\mathbb{Z}_M$ eigenvalue distribution for
the VEV's of the scalar fields of $\mathcal{N}=2$ quiver gauge
theory. This has been studied via counting of BPS operators in
\cite{Berenstein:2005jq, Berenstein:2005ek, Berenstein:2006yy}.
Our approach to the problem is complementary in that it is valid
for weak 't Hooft coupling, and it is valid for all temperatures
in the range $0 \leq TR \ll \lambda^{-1/2}$ unlike
\cite{Berenstein:2005jq, Berenstein:2005ek, Berenstein:2006yy}
which is only valid for $T=0$. In parallel with Ref.~\cite{Gursoy:2007np},
we restrict to the sector of constant and commuting scalar VEV's.
Whereas this enables us to study phase transitions in the eigenvalue distributions,
revealing interesting dynamics, it does not necessarily reflect the
full phase structure. However, we find
it enlightening to see how the geometry of the dual AdS spacetime
is mirrored in the structure of the quantum effective action computed in this
sector.

The outline and summary of the results in this paper are as
follows. In Section 2 we give an introduction to $\mathcal{N}=2$
quiver gauge theory on $S^1 \times S^3$ with chemical potentials
conjugate to the $R$-charges. In Section 3 we evaluate the quantum
effective action of $\mathcal{N}=2$ quiver gauge theory with non-zero $R$-symmetry
chemical potentials and zero scalar VEV's in the $g_\mathrm{YM}
\to 0$ limit and express it in terms of single-particle partition
functions. We use the effective action to construct a matrix model
for $\mathcal{N}=2$ quiver gauge theory on $S^1 \times S^3$. The
model turns out to be an $M$-matrix model with adjoint and
bifundamental potentials.

In Section 4 we study the saddle points of the matrix model as
functions of temperature and chemical potential and thereby
examine the phase structure of the model. In the low-temperature
phase we find a saddle point corresponding to a uniform
distribution of the eigenvalues of the Polyakov loop\footnote{We
are using a somewhat sloppy terminology here: by `Polyakov loop'
we really mean the holonomy matrix of a closed curve winding about
the thermal circle and not just its trace. Throughout this paper
we will use the word to describe both and leave the precise
meaning to be determined from the context.}. In this phase the
free energy is $\mathcal{O}(1)$ with respect to $N$. This behavior
of the free energy suggests that the model in this phase describes
a non-interacting gas of color singlet states, and the phase is
therefore labelled ``confining''. This saddle point is observed to
become unstable when the temperature is raised above a certain
threshold temperature (which depends on the chemical potential).
The model then enters a new phase in which the free energy scales
as $N^2 M$ as $N \to \infty$. This phase is thus interpreted as
describing a non-interacting plasma of color non-singlet states
and is labelled ``deconfined''. The ``deconfinement'' transition
is of first order and is identified with a Hagedorn phase
transition. The condition of stability of the low-temperature
saddle point is translated into a phase diagram of the gauge
theory as a function of both temperature and chemical potentials.
We subsequently study the phase diagram in regions of small
temperature and critical chemical potential. We observe that the
Hilbert space of gauge invariant operators truncates to the
$SU(2)$ subsector when the chemical potential corresponding to the
$SU(2)_R$ factor of the $R$-symmetry group $SU(2)_R \times U(1)_R$
is turned on, whereas when both chemical potentials are turned on
and set equal, it truncates to a larger subsector that corresponds
to an orbifolded version of the $SU(2|3)$ sector found in
$\mathcal{N}=4$ SYM theory.

In Section 5 we develop a matrix model for $\mathcal{N}=2$ quiver
gauge theory on $S^1 \times S^3$ with non-zero VEV's for the
scalar fields and zero $R$-symmetry chemical potentials. We carry
out this computation in the special case where the background
fields are assumed to be ``commuting'' in a sense that conforms to
the quiver structure. Furthermore the background fields will be taken
to be static and spatially homogeneous in order to preserve the $SO(4)$
isometry of the spatial $S^3$ manifold.  The method employed for computing
the effective potential will be the standard background field
formalism. That is, we expand the quantum fields about classical
background fields and path integrate over the fluctuations,
discarding terms of cubic or higher order in the fluctuations. The
resulting fluctuation operators turn out to have a particular
tridiagonal structure in their quiver indices. By exploiting the
vacuum structure of the theory we find that the determinants
factorize, leading to an expression for the quantum effective
action of $\mathcal{N}=2$ $U(N)^M$ quiver gauge theory that
explicitly displays the $\mathbb{Z}_M$ structure of the theory.
Finally we generalize our results to a specific class of field
theories that can be obtained as $\mathbb{Z}_M$ projections of
$\mathcal{N}=4$ SYM theory, of which $\mathcal{N}=2$ quiver gauge
theory is a special case.

In Section 6 we find the minima of the matrix model of Section 5
in the large $N$ limit in a coarse grained approximation. We
consider the joint eigenvalue distribution of the scalar VEV's and
the Polyakov loop and find that the topology of the eigenvalue
distribution is tied to the Hagedorn phase transition. Below the
Hagedorn temperature the eigenvalues of the scalar VEV's are
distributed uniformly over an $S^5/\mathbb{Z}_M$ and the
eigenvalues of the Polyakov loop are distributed uniformly over an
$S^1$. Thus, the joint eigenvalue distribution is an
$S^5/\mathbb{Z}_M$ fibered trivially over $S^1$. We interpret this
$S^5/\mathbb{Z}_M$ as the emergence of the $S^5/\mathbb{Z}_M$
factor of the holographically dual $AdS_5 \times S^5/\mathbb{Z}_M$
geometry. Above the Hagedorn temperature the eigenvalue
distribution of the Polyakov loop becomes gapped and is thus an
interval. The scalar VEV's are now distributed uniformly over an
$S^5/\mathbb{Z}_M$ fibered over this interval, with the radius of
the $S^5/\mathbb{Z}_M$ at any point in the interval proportional
to the density of Polyakov loop eigenvalues at that point (for
fixed $TR$). The $S^5/\mathbb{Z}_M$ thus shrinks to zero radius at
the endpoints of the interval: the topology of the joint
eigenvalue distribution is an $S^6/\mathbb{Z}_M$ where the
$\mathbb{Z}_M$ is understood to act on the $S^5$ transverse to an
$S^1$ diameter. Finally we generalize our results to the
$\mathbb{Z}_M$ orbifold field theories discussed at the end of
Section 5. In particular we find that the geometry of the dual AdS
spacetime is mirrored in the structure of the quantum effective
action in a precise way within this class of orbifold field
theories.

In Section 7 we discuss the results we have obtained in this paper
and suggest directions for future study.
In Appendix A further details about $\mathcal{N}=2$ $U(N)^M$
quiver gauge theory are given, some of which the authors of this
paper have not found elsewhere in the literature. In particular,
we write the full Lagrangian density in terms of $SU(2)_R \times
U(1)_R$ invariants.
In Appendix B we give further technical details of the computation
of the quantum effective action obtained in Section 5.

\section{$\mathcal{N}=2$ quiver gauge theory with $R$-symmetry chemical potentials}

In this section we review $\mathcal{N}=2$ $U(N)^M$ quiver gauge
theories on $S^1 \times S^3$ with $R$-symmetry chemical
potentials. An introductory review of $\mathcal{N}=2$ quiver gauge
theories on $S^1 \times S^3$ is given in Section 2.1. Details,
some of which the authors have not found elsewhere in the
literature, are deferred to Appendix A. In Section 2.2 we then
write up the complete Lagrangian density including $R$-symmetry
chemical potentials.

\subsection{Review of $\mathcal{N}=2$ quiver gauge theory}

$\mathcal{N}=2$ quiver gauge theory with gauge group $U(N)^M$
arises as the world-volume theory of open strings ending on a
stack of $N$ D3-branes placed on the orbifold
$\mathbb{C}^3/\mathbb{Z}_M$. The gauge theory is thus
superconformal \cite{Kachru:1998ys} with 16 supercharges. It can
be obtained as a $\mathbb{Z}_M$ projection of $\mathcal{N}=4
\hspace{1.5mm} U(NM)$ SYM theory as explained in detail in
Appendix A. The resulting gauge group is $U(N)^M$ where all the
$U(N)$ factors of the gauge group have the same gauge coupling
constant $g_\mathrm{YM}$ associated with them. Letting $i = 1,
\ldots, M$ and identifying $i \simeq i + M$, the field content can be
summarized as follows. There are $M$ vector
multiplets\footnote{We will use an $\mathcal{N}=1$ notation
throughout since this proves convenient.} $(A_{\mu i}, \Phi_i,
\psi_{\Phi,i}, \psi_i)$ where $A_{\mu i}$ is the gauge field,
$\psi_i$ is the gaugino, $\Phi_i$ is a complex scalar field, and
$\psi_{\Phi,i}$ is the superpartner of $\Phi_i$. We take $\psi_i$
and $\psi_{\Phi,i}$ to be 2-component Weyl spinors. Furthermore
there are $M$ hypermultiplets $(A_{i,(i+1)}, B_{(i+1),i},
\chi_{A,i}, \chi_{B,i})$ where $A_{i,(i+1)}$ and $B_{(i+1),i}$ are
complex scalar fields and $\chi_{A,i}$ and $\chi_{B,i}$ are their
respective superpartners which we will take as 2-component Weyl
spinors. The fields in the $i$'th vector multiplet all transform
in the adjoint representation of the $i$'th $U(N)$ factor of the
gauge group. The fields in the $i$'th hypermultiplet transform in
a bifundamental representation of the $i$'th and $(i+1)$'th
factors. More specifically, letting $\mathbf{N}_i$ denote the
fundamental representation of the $i$'th $U(N)$ factor and
$\overline{\mathbf{N}}_i$ the corresponding antifundamental
representation, $A_{i,(i+1)}$ and its superpartner $\chi_{A,i}$
transform in the $\mathbf{N}_i \otimes
\overline{\mathbf{N}}_{i+1}$ representation, whereas $B_{(i+1),i}$
and its superpartner $\chi_{B,i}$ transform in the
$\overline{\mathbf{N}}_i \otimes \mathbf{N}_{i+1}$ representation.

The field content is conveniently summarized in the quiver diagram
in Figure 1. The diagram consists of $M$ nodes, labelled by
$i=1,\ldots,M$ with the identification $i \simeq i + M$. The
$i$'th node represents the $i$'th $U(N)$ gauge group factor.
Fields belonging to the $i$'th vector multiplet are drawn as
arrows that start and end on the $i$'th node. For the $i$'th
hypermultiplet, the fields transforming in the $\mathbf{N}_i
\otimes \overline{\mathbf{N}}_{i+1}$ representation are drawn as
arrows that start at the $i$'th node and end at the $(i+1)$'th
node; the fields transforming in the $\overline{\mathbf{N}}_i
\otimes \mathbf{N}_{i+1}$ are depicted as arrows going from the
$(i+1)$'th to the $i$'th node.


\begin{figure}[!h]
\begin{center}
\includegraphics[angle=0, width=0.6\textwidth]{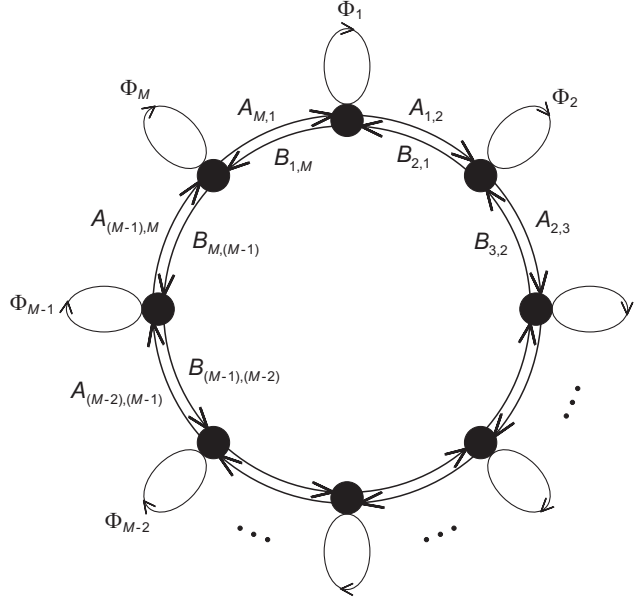}
 \textit{\caption{Quiver diagram summarizing the field content of
$\mathcal{N}=2 \: U(N)^M$ quiver gauge theory. Each of the black
dots (called nodes) represents a $U(N)$ gauge group factor. The
nodes are labelled by $i = 1, \ldots, M$ with the identification
$i \simeq i + M$. Arrows go from fundamental to antifundamental
representations of the corresponding gauge group factors. The
scalar fields $A_{i,(i+1)}, B_{(i+1),i}$ and $\Phi_i$ are shown in
the figure, whereas the gauge fields and all the superpartners
have been left implicit.}}
\end{center}
\end{figure}

The holographic dual of $\mathcal{N}=2$ quiver gauge theory was
found in \cite{Kachru:1998ys} to be Type IIB string theory on
$AdS_5 \times S^5/\mathbb{Z}_M$. The quotient $S^5/\mathbb{Z}_M$
is obtained by embedding $S^5$ in $\mathbb{C}^3$ where the action
of $\mathbb{Z}_M$ is as defined in (\ref{eq:orbifoldaction}). The
$AdS_5$ space has a radius given by $R_{AdS}^2 = \sqrt{4\pi g_s
(\alpha')^2 NM}$ where $g_s$ is the Type IIB string coupling.
There are also $NM$ units of 5-form RR-flux through the $AdS_5$.
Due to the orbifold action the volume of the quotient
$S^5/\mathbb{Z}_M$ equals the volume of the covering space $S^5$
divided by a factor $M$ where the $S^5$ has the same radius as
$AdS_5$. Similarly, there are $N$ units of 5-form RR-flux through
the $S^5/\mathbb{Z}_M$ factor which originate from $NM$ units of
flux in the covering space. Finally, we note that the Yang-Mills
coupling for each $U(N)$ gauge group factor $g_\mathrm{YM}$ is
related to the Type IIB coupling by $g_\mathrm{YM}^2 = 4\pi g_s
M$. This means that the 't Hooft coupling relevant for each factor
is $\lambda = g_\mathrm{YM}^2 N = 4\pi g_s NM$. This is the same
as the 't Hooft coupling on the original $NM$ D3-branes before
orbifolding, for which the Yang-Mills coupling was equal to $4\pi
g_s$. In the following we will often denote the Yang-Mills
coupling simply by $g$.

The action of $\mathcal{N}=2$ $U(N)^M$ quiver gauge theory defined
on $S^1 \times S^3$ is given as follows. To fix our conventions,
we set $F_{\mu \nu} = \partial_\mu A_\nu - \partial_\nu A_\mu +
ig[A_\mu, A_\nu]$ and $D_\mu = \partial_\mu + ig[A_\mu, \, \cdot
\, ].$ We will denote the circumference of the thermal circle
$S^1$ with $\beta$ and the radius of the spatial $S^3$ with $R$.
The Euclidean action of $\mathcal{N}=2$ quiver gauge theory on
$S^1 \times S^3$ is then
\begin{equation}
S = \int_{S^1 \times S^3} d^4 x \sqrt{|g|} \: \big(
\mathcal{L}_\mathrm{gauge} + \mathcal{L}_\mathrm{scalar} +
\mathcal{L}_\mathrm{ferm} \big)
\end{equation}
where the gauge boson, scalar field and spinor field Lagrangian
densities are given by, respectively\footnote{Note that for all
fields, including the Weyl spinors $\chi_A, \chi_B, \psi_\Phi,
\psi$, the bars denote the \emph{Hermitian} conjugate, not the
complex or Weyl conjugate. E.g., $(\overline{\chi_A})_{\alpha
\beta} = (\chi_A)^*_{\beta \alpha}$ where $\alpha, \beta$ are
gauge group indices and the $^*$ denotes complex conjugation.
Furthermore, in the third line of Eq.
(\ref{eq:quiveraction(scalar,nf)}), the notation means, e.g.,
$|[A,B]|^2 \equiv [A,B][\overline{A},\overline{B}]$.}
\begin{eqnarray}
\mathcal{L}_\mathrm{gauge} &=& \frac{1}{4} \Tr F_{\mu \nu} F_{\mu
\nu} \label{eq:quiveraction(gauge,nf)}\\
\mathcal{L}_\mathrm{scalar} &=& \Tr \Big[ \Big( D_\mu A \, D_\mu
\overline{A} + D_\mu B \, D_\mu \overline{B} + D_\mu \Phi \, D_\mu
\overline{\Phi} \Big) \nonumber \\
&\phantom{=}& \phantom{\Tr \Big[ \Big(} + R^{-2} \Big( A
\overline{A} + B \overline{B} + \Phi \overline{\Phi} \Big) + \,
\frac{1}{2} g^2 \Big( [A,\overline{A}] +
[B,\overline{B}] + [\Phi,\overline{\Phi}] \Big)^2\nonumber \\
&\phantom{=}& \phantom{\Tr \Big[ \Big(} - 2g^2 \Big( \big| [A,B]
\big|^2 + \big| [A,\Phi] \big|^2 + \big| [B,\Phi] \big|^2 \Big)
\Big] \label{eq:quiveraction(scalar,nf)} \\
\mathcal{L}_\mathrm{ferm} &=& i \Tr \Big( \overline{\chi_A} \,
\tau_\mu \hspace{-0.9mm} \stackrel{\leftrightarrow}{D}_\mu
\hspace{-0.9mm} \chi_A + \overline{\chi_B} \, \tau_\mu
\hspace{-0.9mm} \stackrel{\leftrightarrow}{D}_\mu \hspace{-0.9mm}
\chi_B + \overline{\psi} \, \tau_\mu \hspace{-0.9mm}
\stackrel{\leftrightarrow}{D}_\mu \hspace{-0.9mm} \psi + \psi_\Phi
\, \tau_\mu \hspace{-0.9mm} \stackrel{\leftrightarrow}{D}_\mu
\hspace{-0.9mm} \overline{\psi_\Phi} \Big) \phantom{aaaaaaaa} \nonumber \\
&\phantom{=}& + \hspace{0.5mm} \frac{g}{\sqrt{2}} \hspace{0.3mm}
\Tr \Big( \hspace{0.3mm} \overline{\chi_A} \hspace{0.3mm} \big(
[A,\psi_\Phi] - [\overline{B},\overline{\psi}] \big)
\hspace{0.5mm} + \hspace{0.5mm} \overline{\chi_B} \hspace{0.3mm}
\big( [\overline{A},\overline{\psi}] + [B,\psi_\Phi] \big)
\hspace{0.5mm} \nonumber \\
&\phantom{=}& \hspace{1.9cm} - \hspace{0.5mm} \overline{\psi}
\hspace{0.2mm} \big( [\overline{A},\overline{\chi_B}] -
[\overline{B},\overline{\chi_A}] \big) - \hspace{0.1cm} \psi_\Phi
\hspace{0.1mm} \big( [A,\overline{\chi_A}] + [B,\overline{\chi_B}]
\big) \nonumber \\
&\phantom{=}& \hspace{1.9cm} + \hspace{0.5mm} \chi_A \big(
[\overline{A},\overline{\psi_\Phi}] - [B,\psi] \big)
\hspace{0.5mm} + \hspace{0.5mm} \chi_B \big(
[A,\psi] + [\overline{B},\overline{\psi_\Phi}] \big) \nonumber \\
&\phantom{=}& \hspace{1.9cm} - \hspace{0.1cm} \psi \big(
[A,\chi_B] - [B,\chi_A] \big) \hspace{0.5mm} - \hspace{0.5mm}
\overline{\psi_\Phi} \hspace{0.1mm} \big( [\overline{A},\chi_A] +
[\overline{B},\chi_B] \big) \nonumber \phantom{\Big)} \\
&\phantom{=}& \hspace{1.9cm} + \hspace{0.1cm} \overline{\chi_A}
\hspace{0.4mm} [\overline{\Phi}, \overline{\chi_B}] \hspace{0.5mm}
- \hspace{0.5mm} \overline{\chi_B} \hspace{0.4mm}
[\overline{\Phi},\overline{\chi_A}] \hspace{0.5mm} +
\hspace{0.5mm} \overline{\psi} \hspace{0.4mm} [\Phi, \psi_\Phi]
\hspace{0.5mm} - \hspace{0.5mm} \psi_\Phi \hspace{0.25mm} [\Phi, \overline{\psi}] \nonumber \\
&\phantom{=}& \hspace{1.9cm} + \hspace{0.1cm} \chi_A
\hspace{0.3mm} [\Phi, \chi_B] \hspace{0.5mm} - \hspace{0.5mm}
\chi_B \hspace{0.3mm} [\Phi, \chi_A] \hspace{0.5mm} +
\hspace{0.5mm} \psi \hspace{0.2mm} [\overline{\Phi},
\overline{\psi_\Phi}] \hspace{0.5mm} - \hspace{0.5mm}
\overline{\psi_\Phi} \hspace{0.4mm} [\overline{\Phi}, \psi]
\hspace{1mm} \Big) \: . \label{eq:quiveraction(spinor,nf)}
\end{eqnarray}
The traces are taken over the $NM \times NM$ matrices. The spinor
fields $\chi_A, \chi_B, \psi_\Phi, \psi$ are undotted 2-component
Weyl spinors. We define $\tau_\mu = (1,i\boldsymbol{\sigma})$. The
operator $\stackrel{\leftrightarrow}{D}_\mu$ is defined by $\psi_1
\hspace{-0.9mm} \stackrel{\leftrightarrow}{D}_\mu \hspace{-0.9mm}
\psi_2 \equiv \frac{1}{2} \big( \psi_1 D_\mu \psi_2 - (D_\mu
\psi_1) \, \psi_2 \big)$. It is implied that the fields $A, B,
\Phi, A_\mu$ etc. take the orbifold projection invariant forms
given in Eqs.
(\ref{eq:projinvariantAmuA})-(\ref{eq:projinvariantBFi}) and
(\ref{eq:projinvariantpsichiA})-(\ref{eq:projinvariantchiBpsiPhi}).
Note that the scalar fields are conformally coupled to the
curvature of the spatial manifold $S^3$ through the term $R^{-2}
\Tr \big( A \overline{A} + B \overline{B} + \Phi \overline{\Phi}
\big)$ in (\ref{eq:quiveraction(scalar,nf)}). This effectively
induces a mass for the scalar fields.

The orbifolding breaks the $R$-symmetry group $SU(4)$ of
$\mathcal{N}=4$ SYM theory into $SU(2)_R \times U(1)_R$. As
described in Appendix A, $\Phi$ is associated with the $z_1$
direction of $\mathbb{C}^3$ which is inert under the action of the
orbifold group $\mathbb{Z}_M$, while $A$ and $B$ are associated
with $z_2$ and $z_3$ respectively. The $U(1)_R$ factor corresponds
to the transformation $z_1 \to e^{i\zeta} z_1$ and therefore acts
on the $\Phi$ fields by multiplying phase rotations. The $A$ and
$B$ fields have zero charge under $U(1)_R$. The $SU(2)_R$ symmetry
acts on the $A$ and $B$ fields and their Hermitian conjugates. In
fact, $(A, \overline{B})$ and $(-B,\overline{A})$ form $SU(2)_R$
doublets. Furthermore $(\overline{\psi}, \psi_\Phi)$ and
$(-\overline{\psi_\Phi}, \psi)$ are $SU(2)_R$ doublets whereas
$\chi_A$ and $\chi_B$ have zero charge under $SU(2)_R$. The gauge
field is not charged under $SU(2)_R \times U(1)_R$. We summarize
the $R$-charges in Table A.

\subsection{Lagrangian density with $R$-symmetry chemical potentials}

Given any non-Abelian symmetry group $G$, one can introduce
chemical potentials conjugate to the generators of a maximal torus
of $G$. In this section we will consider the case where $G$ is the
$R$-symmetry group $SU(2)_R \times U(1)_R$ of $\mathcal{N}=2$
quiver gauge theory. The maximal torus is $U(1) \times U(1)$. We
will denote the Cartan generators of $U(1)_R$ and $SU(2)_R$ by
$Q_1$ and $Q_2$, respectively, and the corresponding chemical
potentials by $\mu_1$ and $\mu_2$. For the $U(1)$ factor of the
maximal torus that corresponds to $U(1)_R$ the eigenvalues of the
Cartan generators can directly be read off from Table A. For the
$U(1) \subset SU(2)_R$ we choose as a basis for the Cartan
subalgebra the diagonal generator $\sigma_z$ so that the $SU(2)_R$
doublets will have well-defined charges under $U(1)$. (We choose
$\sigma_z$ rather than $\frac{1}{2} \sigma_z$ as the generator
$Q_2$ because we require $e^{i Q_2 \theta}$ to be invariant under
$\theta \to \theta + 2\pi$. Setting $Q_2 \equiv \sigma_z$ we have
$e^{i Q_2 \theta} = \mathrm{diag} (e^{i\theta}, e^{-i\theta})$
which is clearly invariant.) Therefore the charges under the
maximal torus $U(1)$ of $SU(2)_R$ will be 2 times the $SU(2)_R$
charges.

Thus for the bosonic fields,
\begin{eqnarray}
(\mu_a Q_a) \, A_{i,(i+1)} &=& \mu_2 \, A_{i,(i+1)} \label{eq:charges-under-Cartan-generators(A)}\\
(\mu_a Q_a) \, B_{(i+1),i} &=& \mu_2 \, B_{(i+1),i} \label{eq:charges-under-Cartan-generators(B)}\\
(\mu_a Q_a) \, \Phi_i &=& \mu_1 \Phi_i \label{eq:charges-under-Cartan-generators(Phi)}\\
(\mu_a Q_a) \, A_{\mu i} &=& 0 \: ,
\label{eq:charges-under-Cartan-generators(Amu)}
\end{eqnarray}
and for the fermionic fields,
\begin{eqnarray}
(\mu_a Q_a) \, \chi_{A,i} &=& -\textstyle{\frac{1}{2}} \mu_1 \, \chi_{A,i} \label{eq:charges-under-Cartan-generators(chiA)}\\
(\mu_a Q_a) \, \chi_{B,i} &=& -\textstyle{\frac{1}{2}} \mu_1 \, \chi_{B,i} \label{eq:charges-under-Cartan-generators(chiB)}\\
(\mu_a Q_a) \, \psi_i &=& \left(\textstyle{\frac{1}{2}}
\mu_1 -\mu_2\right) \psi_i \label{eq:charges-under-Cartan-generators(psiPhi)} \\
(\mu_a Q_a) \, \psi_{\Phi,i} &=& \left( -\textstyle{\frac{1}{2}}
\mu_1 - \mu_2 \right) \psi_{\Phi,i} \: .
\label{eq:charges-under-Cartan-generators(psi)}
\end{eqnarray}
The corresponding expressions for the Hermitian conjugate fields
are obtained by simply changing the signs of the chemical
potentials.

To obtain the Lagrangian density of $\mathcal{N}=2$ quiver gauge
theory with chemical potentials $\mu_a$ for the $SU(2)_R \times
U(1)_R$ Cartan generators, one makes the following substitution in
the Lagrangian density
\begin{equation}
D_\mu \longrightarrow D_\mu - \mu_a Q_a \delta_{\mu 0} \: .
\label{eq:introducechemicalpotential}
\end{equation}
Below we have written the Lagrangian densities for the fundamental
scalar and spinor fields of $\mathcal{N}=2$ quiver gauge theory.
This will be important for the analysis in the following sections
in order to distinguish the adjoint from the bifundamental
structures.

The Lagrangian density for the scalar fields with $R$-symmetry
chemical potentials is
\begin{eqnarray}
\mathcal{L}_\mathrm{scalar} &=&  \sum_{i=1}^M \Bigg\{
\hspace{0.4mm} \Tr \Big[ \Big( \partial_\mu A_{i,(i+1)} + ig
A_{\mu i} A_{i,(i+1)} - ig A_{i,(i+1)} A_{\mu (i+1)}
- \mu_2 \delta_{\mu 0} A_{i,(i+1)} \Big) \nonumber \\
&\phantom{=}& \hspace{1.8cm} \times \, \Big(
\partial_\mu \overline{ A_{i,(i+1)} } + ig A_{\mu (i+1)}
\overline{A_{i,(i+1)}} - ig \overline{ A_{i,(i+1)} } A_{\mu i}
+ \mu_2 \delta_{\mu 0} \overline{A_{i,(i+1)}} \Big) \Big] \nonumber \\
&\phantom{=}& + \, \Tr \Big[ \Big( \partial_\mu B_{(i+1),i} +
ig A_{\mu (i+1)} B_{(i+1),i} - ig B_{(i+1),i} A_{\mu i} - \mu_2 \delta_{\mu 0} B_{(i+1),i} \Big) \nonumber \\
&\phantom{=}& \hspace{1.6cm} \times \, \Big(
\partial_\mu \overline{B_{(i+1),i}} + ig A_{\mu i} \overline{B_{(i+1),i}} -
ig \overline{B_{(i+1),i}} A_{\mu (i+1)} + \mu_2 \delta_{\mu 0} \overline{B_{(i+1),i}} \Big) \Big] \nonumber \\
&\phantom{=}& + \, \Tr \Big[ \Big(
\partial_\mu \Phi_i + ig [A_{\mu i},\Phi_i] - \mu_1 \delta_{\mu 0} \Phi_i \Big) \Big(
\partial_\mu \overline{\Phi_i} + ig [A_{\mu i},\overline{\Phi_i}] + \mu_1 \delta_{\mu 0} \overline{\Phi_i} \Big) \Big] \nonumber \\
&\phantom{=}& + \: R^{-2} \Tr \Big( A_{i,(i+1)}
\overline{A_{i,(i+1)}} + \overline{B_{(i+1),i}} B_{(i+1),i} + \Phi_i \overline{\Phi_i} \Big) \nonumber \\
&\phantom{=}& + \: \frac{1}{2} g^2 \Tr \Big[ \Big( A_{i,(i+1)}
\overline{A_{i,(i+1)}} - \overline{A_{(i-1),i}} A_{(i-1),i} \nonumber \\
&\phantom{=}& \hspace{2.2cm} + \, B_{i,(i-1)}
\overline{B_{i,(i-1)}} - \overline{B_{(i+1),i}} B_{(i+1),i} +
[\Phi_i,\overline{\Phi_i}] \Big)^2 \Big] \nonumber \\
&\phantom{=}& - \: 2 g^2 \Tr \Big[ \Big( A_{i,(i+1)}
B_{(i+1),i} - B_{i,(i-1)} A_{(i-1),i} \Big) \nonumber \\
&\phantom{=}& \hspace{2.1cm} \times \, \Big(
\overline{A_{(i-1),i}} \: \: \overline{B_{i,(i-1)}} -
\overline{B_{(i+1),i}} \: \: \overline{A_{i,(i+1)}} \Big) \Big] \nonumber \\
&\phantom{=}& - \: 2 g^2 \Tr \Big[ \Big( A_{i,(i+1)} \Phi_{i+1} -
\Phi_i A_{i,(i+1)} \Big) \Big( \overline{A_{i,(i+1)}} \: \:
\overline{\Phi_i} - \overline{\Phi_{i+1}} \: \:
\overline{A_{i,(i+1)}} \Big) \Big] \nonumber \\
&\phantom{=}& - \: 2 g^2 \Tr \Big[ \Big( B_{(i+1),i} \Phi_i -
\Phi_{i+1} B_{(i+1),i} \Big) \Big(\overline{B_{(i+1),i}} \: \:
\overline{\Phi_{i+1}} - \overline{\Phi_i} \: \:
\overline{B_{(i+1),i}} \Big) \Big] \Bigg\} \: .
\label{eq:quiveraction(scalar,f,Rsymmchem)}
\end{eqnarray}
Here the traces are always taken over the gauge indices of the $N
\times N$ matrices. Observe that the chemical potentials $\mu_1$
and $\mu_2$ act like negative mass squares for $\Phi_i$ and
$A_{i,(i+1)}, B_{(i+1),i}$. On a compact spatial manifold such as
$S^3$, these terms are balanced by the positive mass square terms
induced by the conformal coupling to curvature. We immediately
observe from (\ref{eq:quiveraction(scalar,f,Rsymmchem)}) that
$\mathcal{N}=2$ quiver gauge theory on $S^1 \times S^3$ is
well-defined as long as $\mu_1, \mu_2 \leq R^{-1}$. If the
chemical potentials exceed this bound, the theory develops
tachyonic modes and there exists no stable ground state.

The Lagrangian density for the spinor fields with $R$-symmetry
chemical potentials is
\begin{eqnarray}
\mathcal{L}_\mathrm{ferm} &=& \sum_{i=1}^M \Bigg\{ \hspace{0.4mm}
\frac{i}{2} \Tr \Big( \overline{\chi_{A,i}} \: \tau_\mu \big
(\partial_\mu \chi_{A,i} + ig A_{\mu i} \chi_{A,i} - ig \chi_{A,i}
A_{\mu (i+1)}
+ \textstyle{\frac{1}{2}} \mu_1 \delta_{\mu 0} \chi_{A,i} \big) \Big) \nonumber \\
&\phantom{=}& \hspace{0.9cm} - \frac{i}{2} \Tr \Big( \big(
\partial_\mu \overline{\chi_{A,i}} + ig A_{\mu (i+1)}
\overline{\chi_{A,i}} - ig \overline{\chi_{A,i}} A_{\mu i} -
\textstyle{\frac{1}{2}} \mu_1 \delta_{\mu 0} \overline{\chi_{A,i}}
\big) \, \tau_\mu \, \chi_{A,i} \Big) \nonumber \\
&\phantom{=}& + \frac{i}{2} \Tr \Big( \overline{\chi_{B,i}} \:
\tau_\mu \big (\partial_\mu \chi_{B,i} + ig A_{\mu (i+1)}
\chi_{B,i} - ig \chi_{B,i} A_{\mu i} +
\textstyle{\frac{1}{2}} \mu_1 \delta_{\mu 0} \chi_{B,i} \big) \Big) \nonumber \\
&\phantom{=}& \hspace{0.5cm} - \frac{i}{2} \Tr \Big( \big(
\partial_\mu \overline{\chi_{B,i}} + ig A_{\mu i}
\overline{\chi_{B,i}} - ig \overline{\chi_{B,i}} A_{\mu (i+1)} -
\textstyle{\frac{1}{2}} \mu_1 \delta_{\mu 0} \overline{\chi_{B,i}}
\big) \, \tau_\mu \, \chi_{B,i} \Big) \nonumber
\end{eqnarray}
\begin{eqnarray}
&\phantom{=}& \hspace{-0.2cm} + \frac{i}{2} \Tr \Big(
\overline{\psi_i} \: \tau_\mu \big( \partial_\mu \psi_i +
ig[A_{\mu i}, \psi_i] - \left(\textstyle{\frac{1}{2}}\mu_1 - \mu_2
\right) \, \delta_{\mu 0} \psi_i \big) \Big) \nonumber \\
&\phantom{=}& \hspace{0.3cm} - \frac{i}{2} \Tr \Big(
\big(\partial_\mu \overline{\psi_i} + ig [A_{\mu i},
\overline{\psi_i}] + \left(\textstyle{\frac{1}{2}} \mu_1 - \mu_2
\right) \, \delta_{\mu 0} \overline{\psi_i} \big) \, \tau_\mu \,
\psi_i \Big) \nonumber \\
&\phantom{=}& \hspace{-0.2cm} + \frac{i}{2} \Tr \Big(
\psi_{\Phi,i} \: \tau_\mu \big( \partial_\mu
\overline{\psi_{\Phi,i}} + ig[A_{\mu i}, \overline{\psi_{\Phi,i}}]
- \left(\textstyle{\frac{1}{2}}\mu_1 + \mu_2 \right) \,
\delta_{\mu 0} \overline{\psi_{\Phi,i}} \big) \Big) \nonumber \\
&\phantom{=}& \hspace{0.3cm} - \frac{i}{2} \Tr \Big(
\big(\partial_\mu \psi_{\Phi,i} + ig [A_{\mu i}, \psi_{\Phi,i}] +
\left(\textstyle{\frac{1}{2}} \mu_1 + \mu_2 \right) \, \delta_{\mu
0} \psi_{\Phi,i} \big) \, \tau_\mu \, \overline{\psi_{\Phi,i}}
\Big) \nonumber \\
&\phantom{=}& + \frac{g}{\sqrt{2}} \Tr \Big( \epsilon^{cd} \big\{
\chi_{A,i} \hspace{0.2mm} (\overline{\lambda_i})_c ,
\hspace{0.5mm} (\overline{\chi_i})_d \big\} \hspace{0.8mm} +
\hspace{0.5mm} \epsilon^{cd} \big\{ \chi_{A,i}, \hspace{0.5mm}
(\overline{\chi_{i+1}})_c
\hspace{0.2mm} (\overline{\lambda_i})_d \big\} \nonumber \\
&\phantom{=}& \hspace{1.7cm} + \hspace{1mm} \epsilon^{cd} \big\{
\hspace{0.2mm} \overline{\chi_{A,i}} \hspace{0.5mm} (\lambda_i)_c
, \hspace{0.6mm} (\chi_{i+1})_d \big\} \hspace{0.6mm} +
\hspace{0.5mm} \epsilon^{cd}\big\{ \hspace{0.2mm}
\overline{\chi_{A,i}}, \hspace{0.6mm} (\chi_i)_c \hspace{0.2mm} (\lambda_i)_d \big\} \nonumber \\
&\phantom{=}& \hspace{1.7cm} + \hspace{1mm} \epsilon^{cd} \big\{
\chi_{B,i} \hspace{0.2mm} (\lambda_i)_c, \hspace{0.6mm}
(\overline{\chi_{i+1}})_d \big\} \hspace{0.8mm} + \hspace{0.6mm}
\epsilon^{cd} \big\{ \chi_{B,i}, \hspace{0.6mm}
(\overline{\chi_i})_c \hspace{0.2mm}
(\lambda_i)_d \big\} \nonumber \phantom{\Big)} \\
&\phantom{=}& \hspace{1.7cm} - \hspace{1mm} \epsilon^{cd} \big\{
\hspace{0.2mm} \overline{\chi_{B,i}} \hspace{0.6mm}
(\overline{\lambda_i})_c, \hspace{0.5mm} (\chi_i)_d \big\}
\hspace{0.6mm} - \hspace{0.5mm} \epsilon^{cd} \big\{
\hspace{0.2mm} \overline{\chi_{B,i}} , \hspace{0.5mm}
(\chi_{i+1})_c \hspace{0.2mm} (\overline{\lambda_i})_d \big\} \nonumber \phantom{\big)} \\
&\phantom{=}& \hspace{1.7cm} + \hspace{1mm} \epsilon^{cd} \big\{
(\chi_i)_c \hspace{0.2mm} \Phi_i, \hspace{0.5mm} (\chi_i)_d \big\}
\hspace{0.6mm} + \hspace{0.6mm} \epsilon^{cd} \big\{
(\overline{\chi_i})_c \hspace{0.4mm} \overline{\Phi_i},
\hspace{0.5mm} (\overline{\chi_i})_d \big\} \nonumber \\
&\phantom{=}& \hspace{1.7cm} + \hspace{1mm} \big\{ \chi_{A,i}
\hspace{0.2mm} \Phi_{i+1}, \hspace{0.6mm} \chi_{B,i} \big\}
\hspace{0.8mm} + \hspace{0.3mm} \big\{ \overline{\chi_{A,i}}
\hspace{0.5mm} \overline{\Phi_i}, \hspace{0.6mm}  \overline{\chi_{B,i}} \big\} \nonumber \\
&\phantom{=}& \hspace{1.7cm} - \hspace{1mm} \big\{ \chi_{B,i}
\hspace{0.2mm} \Phi_i , \hspace{0.6mm} \chi_{A,i} \big\}
\hspace{0.8mm} - \hspace{0.3mm} \big\{ \overline{\chi_{B,i}}
\hspace{0.6mm} \overline{\Phi_{i+1}} , \hspace{0.6mm}
\overline{\chi_{A,i}} \big\} \Big) \Bigg\} \: .
\label{eq:quiveraction(spinor,f,Rsymmchem)}
\end{eqnarray}
Here the traces are always taken over the gauge indices of the $N
\times N$ matrices. Note that the potential part of the Lagrangian
density has been written in terms of the $SU(2)_R$ doublets given
in Eqs.
(\ref{eq:SU(2)doublets(scalar)})-(\ref{eq:SU(2)doublets(spinor)})
for notational simplicity.

Finally, as the gauge fields have zero charge under $SU(2)_R
\times U(1)_R$, the gauge field part of the Lagrangian density is
unaffected by introducing the $R$-symmetry chemical potentials.
Nonetheless, we give the result here for convenience:
\begin{equation}
\mathcal{L}_\mathrm{gauge} = \frac{1}{4} \sum_{i=1}^M
\hspace{0.4mm} \Tr F_{\mu \nu}^i F_{\mu \nu}^i
\label{eq:quiveraction(gauge,f,Rsymmchem)}
\end{equation}
where of course $F_{\mu \nu}^i = \partial_\mu A_\nu^i -
\partial_\nu A_\mu^i + ig[A_\mu^i, A_\nu^i]$ and the trace is
taken over the gauge indices of the $N \times N$ matrices.

\section{Zero-coupling limit and the matrix model}

The matrix model we will consider is defined by integrating out
the fluctuations of the quantum fields. In Section 3.1 we
therefore first give a brief description of how to compute the
one-loop quantum effective action with non-zero chemical
potentials conjugate to the $R$-charges. The details of this
computation are well-described in the literature (see, e.g.,
Appendix A of \cite{Yamada:2006rx}). In Section 3.2 we then
proceed to construct the matrix model out of the 1-loop quantum
effective action.

\subsection{One-loop quantum effective action}

The partition function for the grand canonical ensemble has the
path integral representation
\begin{equation}
Z = \int \mathcal{D} A_\mu \: \mathcal{D} \phi \: \mathcal{D} \psi
\: e^{-\int_{S^1 \times S^3} d^4 x \: \sqrt{|g|} \:
(\mathcal{L}_\mathrm{gauge} + \mathcal{L}_\mathrm{scalar} +
\mathcal{L}_\mathrm{ferm})} \label{eq:formal-path-integral}
\end{equation}
with $\mathcal{L}_\mathrm{gauge}, \mathcal{L}_\mathrm{scalar}$ and
$\mathcal{L}_\mathrm{ferm}$ being the Lagrangian densities with
$R$-symmetry chemical potentials given by Eqs.
(\ref{eq:quiveraction(gauge,f,Rsymmchem)}),
(\ref{eq:quiveraction(scalar,f,Rsymmchem)}) and
(\ref{eq:quiveraction(spinor,f,Rsymmchem)}), respectively, and
where the measures $\mathcal{D} A_\mu$, $\mathcal{D} \phi$ and
$\mathcal{D} \psi$ are the products of the measures over all the
gauge fields, scalar fields and spinor fields, respectively. We
will obtain an effective action from this expression by taking the
free limit $g \rightarrow 0$ of the tree-level action. However,
since the theory is defined on a compact spatial $S^3$ one must
impose the Gauss law constraint that all states be gauge
invariant. We perform the projection onto gauge invariant states
by using $A_{0i}$ as a Lagrange multiplier,
\begin{equation}\label{eq:zero-more-decomposition}
A_{\mu i}(x) \longrightarrow \widetilde{A}_{\mu i}(x) +
\delta_{\mu 0} a_i/g
\end{equation}
where $\widetilde{A}_{0 i}$ integrates to zero over $S^1 \times
S^3$ and $a_i$ are constant Hermitian matrices which by gauge
invariance can be assumed diagonal, $a_i = \mathrm{diag}(q_i^1,
\ldots, q_i^N)$. To obtain the correct zero coupling limit one
inserts the decomposition (\ref{eq:zero-more-decomposition}) into
the action given through
(\ref{eq:quiveraction(scalar,f,Rsymmchem)})-(\ref{eq:quiveraction(gauge,f,Rsymmchem)})
and then takes the $g\to 0$ limit.

As the quantum fields are defined on $S^1 \times S^3$ one
decomposes them into Fourier modes on $S^1$ and $S^3$ spherical
harmonics. More specifically, let $\tau$ denote the direction
along the $S^1$. We will use the convention that any field $\phi$
defined on $S^1 \times S^3$ has the Fourier mode decomposition
\begin{equation}
\phi(\tau,\boldsymbol{x}) = \sum_{k=-\infty}^\infty e^{i \omega_k
\tau} \phi^{[k]}(\boldsymbol{x})
\end{equation}
where the quantized Matsubara frequencies are $\omega_k =
\frac{2\pi k}{\beta}$ for bosons and $\omega_k = \frac{(2k +
1)\pi}{\beta}$ for fermions giving, respectively, periodic and
antiperiodic boundary conditions around the thermal
circle.\footnote{However, for the Fadeev-Popov ghosts the boundary
conditions are taken periodic.} One then decomposes the spatial
components of the gauge field into spherical harmonics on $S^3$ by
writing them as a sum of a transverse (i.e. divergenceless) vector
field $\mathbf{A}_i^\bot$ and a longitudinal vector field $\nabla
F_i$ where $F_i$ is a scalar function. That is, for $k = 1, 2, 3$
we decompose
\begin{equation}
\widetilde{A}_i^k = (A_i^\bot)^k + (\nabla F_i)^k
\label{eq:decomposition-of-spatial-gauge-field}
\end{equation}
and insert the expression on the right hand side into the action
given through
(\ref{eq:quiveraction(scalar,f,Rsymmchem)})-(\ref{eq:quiveraction(gauge,f,Rsymmchem)}).

\begin{center}
\begin{tabular}{|c c|c|c|c|} \hline \hspace{0.8cm} \phantom{\Big(}
quantum field & & eigenvalue & \hspace{0.1mm} notation in text
\hspace{0.1mm} & \hspace{0.3mm} degeneracy ($D_h$) \\
\hline \phantom{\Big(} transverse vector & $\mathbf{A}^\bot$ &
$-(h+1)^2 \, R^{-2}$ & $-\Delta_g^2$ &
$2h(h+2)$ \\
\hline \phantom{\Big(} \hspace{-1mm} longitudinal vector & $\nabla
F$ & $-h(h+2) \, R^{-2}$ & $-\Delta_s^2$ &
$(h+1)^2$ \\
\hline \phantom{\Big(} \hspace{-1mm} real scalar & $A^0, \phi$ &
$-h(h+2) \, R^{-2}$ & $-\Delta_s^2$ & $(h+1)^2$ \\
\hline \phantom{\Big(} \hspace{-1mm} Weyl spinor & $\psi$ &
$-\left(h+\frac{1}{2} \right)^2 R^{-2}$ & $-\Delta_f^2$ & $h(h+1)$ \\
\hline
\end{tabular}
\end{center}
\vspace{0.2cm} {\small{\textbf{Table 1.} \textit{Eigenvalues and
corresponding degeneracies of the $S^3$ spatial Laplacian
$\nabla^2 \equiv
\partial_1^2 + \partial_2^2 + \partial_3^2$ for various quantum
fields defined on $S^3$. Here $R$ denotes the radius of $S^3$. The
irreducible representations of the $SO(4)$ isometry group are
labelled by the angular momentum $h$ which has the range $h = 0,
1, 2, \ldots$ for all the fields except for the longitudinal
vector field $\nabla F$ where $h$ starts from 1.}}}
\\
\\
The quantum effective action $\Gamma \equiv - \ln Z$ is defined by
integrating out all fluctuating fields (cf.
(\ref{eq:formal-path-integral})), leaving an expression that only
depends on the zero mode $a_i$. It is convenient to express
$\Gamma$ as a functional of the holonomy matrix of a closed curve
wound around the thermal circle, i.e. $U_i \equiv e^{i\beta a_i}$
after decomposing the gauge field according to
(\ref{eq:zero-more-decomposition}) and taking $g \to 0$. By
performing the traces over the Matsubara frequencies and over the
angular momenta $h$, with appropriate eigenvalues of the Laplacian
$\nabla^2$ on $S^3$ and the associated degeneracies (cf. Table 1)
one finds the following expression for the quantum effective
action in terms of the variables $x \equiv e^{-\beta}$ and $y_j
\equiv e^{\beta \mu_j}$
\begin{eqnarray}
\Gamma[U_i] &=& -\sum_{i=1}^M \sum_{l=1}^\infty \Bigg[ \,
\frac{1}{l} \left( \frac{6x^{2l} - 2x^{3l}}{(1-x^l)^3} \right) +
\frac{1}{l} \left( \frac{x^l + x^{2l}}{(1-x^l)^3} \right) \big(
y_1^l + y_1^{-l} \big) \nonumber \\
&\phantom{=}& \hspace{2cm} + \, \frac{(-1)^{l+1}}{l} \left(
\frac{2x^{3l/2}}{(1-x^l)^3} \right) \hspace{-0.7mm} \big(
y_1^{l/2} + y_1^{-l/2} \big) \big( y_2^l + y_2^{-l} \big) \,
\Bigg] \Big( \Tr U_i^l \, \Tr U_i^{-l} \Big) \nonumber \\
&\phantom{=}& - \, \sum_{i=1}^M \sum_{l=1}^\infty \Bigg[ \,
\frac{1}{l} \left( \frac{x^l + x^{2l}}{(1-x^l)^3} \right) \big(
y_2^l + y_2^{-l} \big) + \frac{(-1)^{l+1}}{l} \left(
\frac{2x^{3l/2}}{(1-x^l)^3} \right) \big( y_1^{l/2} + y_1^{-l/2}
\big) \, \Bigg] \nonumber \\
&\phantom{=}& \hspace{2cm} \times \: \Big( \Tr U_i^l
\hspace{0.3mm} \Tr U_{i+1}^{-l} \hspace{0.3mm} + \hspace{0.3mm}
\Tr U_i^{-l} \hspace{0.3mm} \Tr U_{i+1}^l \Big) \: .
\label{eq:quantum-effective-action}
\end{eqnarray}
Note that the adjoint holonomy factors come from the vector
multiplets $\big( A_{\mu i}, \Phi_i, \psi_{\Phi,i}, \psi_i \big)$,
and the bifundamental factors come from the hypermultiplets $\big(
A_{i,(i+1)}, B_{(i+1),i}, \chi_{A,i}, \chi_{B,i} \big)$. For later
convenience we define here the total single-particle partition
functions for the bosonic and fermionic sectors of the vector and
hypermultiplets:
\begin{eqnarray}
z_\mathrm{ad}^B (x;y_1,y_2) &\equiv& \frac{6x^2 - 2x^3}{(1-x)^3} +
\frac{x+x^2}{(1-x)^3} \hspace{0.3mm} \big(y_1 + y_1^{-1} \big) \label{eq:spPF(ad,bos)}\\
z_\mathrm{ad}^F (x;y_1,y_2) &\equiv& \frac{2x^{3/2}}{(1-x)^3}
\hspace{0.3mm} \big( y_1^{1/2} + y_1^{-1/2} \big) \big( y_2 + y_2^{-1} \big) \label{eq:spPF(ad,ferm)} \\
z_\mathrm{bi}^B (x;y_1,y_2) &\equiv& \frac{x + x^2}{(1-x)^3}
\hspace{0.3mm} \big( y_2 + y_2^{-1} \big) \label{eq:spPF(bi,bos)} \\
z_\mathrm{bi}^F (x;y_1,y_2) &\equiv& \frac{2x^{3/2}}{(1-x)^3}
\hspace{0.3mm} \big( y_1^{1/2} + y_1^{-1/2} \big) \: .
\label{eq:spPF(bi,ferm)}
\end{eqnarray}
These results are consistent with Ref. \cite{Aharony:2003sx}, Eqs.
(3.17)-(3.18), where the summation over representations is taken
to run over the adjoint and the bifundamental representations, and
the charges $Q$ are taken as $\beta$ times the Cartan charges
$Q_1, Q_2$ given implicitly through
(\ref{eq:charges-under-Cartan-generators(A)})-(\ref{eq:charges-under-Cartan-generators(psi)}).

\subsection{The matrix model}

The matrix model we will consider is defined by the partition
function
\begin{equation}
Z_{\mathrm{MM}} = \int \prod_{i=1}^M \big[ \mathcal{D} U_i \big]
\exp \big( \hspace{-0.5mm} - \hspace{-0.3mm} \Gamma[U_i]
\hspace{0.3mm} \big) \label{eq:matrix-model(definition)}
\end{equation}
where $\Gamma[U_i]$ is given in
(\ref{eq:quantum-effective-action}). It is convenient for taking
the continuum limit to rewrite $\Gamma[U_i]$ directly in terms of
the zero modes $a_i$. To simplify the notation, define the
rescaled zero mode $\alpha_i \equiv \beta a_i$ so that $U_i =
e^{i\alpha_i}$. Hence
\begin{equation}
Z_{\mathrm{MM}} = \int \prod_{i=1}^M \big[\mathcal{D} \alpha_i
\big] \, \exp \Bigg( - \sum_{m\neq n} \Big(
V_\mathrm{\hspace{0.3mm} ad}(\alpha_i^m - \alpha_i^n) +
V_\mathrm{\hspace{0.3mm} bi} (\alpha_i^m - \alpha_{i+1}^n ) \Big)
\Bigg)
\end{equation}
where the adjoint and bifundamental potentials are, respectively
\begin{eqnarray}
V_\mathrm{\hspace{0.3mm} ad}(\theta) &\equiv& - \ln \left| \,
\sin\left( \frac{\theta}{2} \right) \right| - \sum_{l=1}^\infty
\frac{1}{l} \hspace{0.3mm} \left( z_\mathrm{ad}^B
(x^l;y_1^l,y_2^l) \hspace{0.3mm} + \hspace{0.3mm} (-1)^{l+1}
z_\mathrm{ad}^F (x^l;y_1^l,y_2^l) \right) \cos (l\theta) \nonumber \\
&=& \ln 2 + \sum_{l=1}^\infty \frac{1}{l} \hspace{0.3mm} \left( 1
\hspace{0.3mm} - \hspace{0.3mm} z_\mathrm{ad}^B(x^l;y_1^l,y_2^l)
\hspace{0.3mm} - \hspace{0.3mm} (-1)^{l+1} z_\mathrm{ad}^F(x^l;y_1^l,y_2^l)
\right) \cos (l\theta) \label{eq:adjointV} \\
V_\mathrm{\hspace{0.3mm} bi}(\theta) &\equiv& - \sum_{l=1}^\infty
\frac{2}{l} \hspace{0.3mm} \left( z_\mathrm{bi}^B
(x^l;y_1^l,y_2^l) \hspace{0.3mm} + \hspace{0.3mm} (-1)^{l+1}
z_\mathrm{bi}^F(x^l;y_1^l,y_2^l) \right) \cos(l\theta) \: .
\label{eq:bifundamentalV}
\end{eqnarray}

We will now take the continuum limit $N \to \infty.$ It is
convenient to introduce eigenvalue distributions
$\rho_i(\theta_i)$ proportional to the density of the eigenvalues
$e^{i\theta_i}$ of $U_i$ at the angle $\theta_i \in [-\pi,\pi]$.
Here $\rho_i$ must be everywhere non-negative, and we choose its
normalization so that for any fixed $i$
\begin{equation}
\int_{-\pi}^\pi d\theta_i \, \rho_i(\theta_i) = 1 \: .
\label{eq:normalization-of-rho}
\end{equation}
Furthermore we define the Fourier modes of $\rho_i$ and
$V_\mathrm{ad}$ and $V_\mathrm{bi}$:
\begin{equation}
\rho_i^l \equiv \int_{-\pi}^\pi d\theta_i \, \rho_i(\theta_i)
\hspace{0.4mm} \cos(l\theta_i) \: , \hspace{0.4cm} V_\mathrm{ad}^l
\equiv \int_{-\pi}^\pi d\theta \, V_\mathrm{ad} (\theta)
\hspace{0.4mm} \cos(l \theta) \: , \hspace{0.4cm} V_\mathrm{bi}^l
\equiv \int_{-\pi}^\pi d\theta \, V_\mathrm{bi} (\theta)
\hspace{0.4mm} \cos(l \theta) \label{eq:Fourier-modes}
\end{equation}
so that, assuming $\rho_i$, $V_\mathrm{ad}$, $V_\mathrm{bi}$ to be
even functions, we have the Fourier expansions
\begin{equation}
\rho_i(\zeta) = \frac{1}{\pi} \sum_{l=0}^\infty \hspace{0.2mm}
\rho_i^l \hspace{0.2mm} \cos(l \zeta) \: ,
\label{eq:Fourier-expansion-of-rho} \hspace{0.4cm}
V_\mathrm{ad}(\zeta) = \frac{1}{\pi} \sum_{l=0}^\infty
\hspace{0.2mm} V_\mathrm{ad}^l \hspace{0.2mm} \cos(l \zeta) \: ,
\hspace{0.4cm} V_\mathrm{bi}(\zeta) = \frac{1}{\pi}
\sum_{l=0}^\infty \hspace{0.2mm} V_\mathrm{bi}^l \hspace{0.2mm}
\cos(l \zeta) \: .
\end{equation}

The continuum limit is obtained by making the
substitution\footnote{Here it is implied that the content of the
brackets $\big[ \cdots \big]$ carries an $i$ label.}
\begin{equation}
\frac{1}{N} \sum_{n=1}^N \big[ \cdots \big]  \: \longrightarrow \:
\int_{-\pi}^\pi d\theta_i \hspace{0.7mm} \rho_i (\theta_i)
\hspace{0.5mm} \big[ \cdots \big]
\label{eq:cont.limit(Polyakov-loop)}
\end{equation}
Furthermore we replace the path integral measure
$\big[\mathcal{D}\alpha_i \big] \longrightarrow \big[\mathcal{D}
\lambda_i \big]$. Thus, in the continuum limit the path integral
of the matrix model takes the form
\begin{equation}
Z_{\mathrm{MM}} = \int \prod_{i=1}^M \big[ \mathcal{D} \lambda_i
\big] \hspace{0.3mm} \exp\big( \hspace{-0.7mm} - \hspace{-0.5mm}
S_{\mathrm{MM}} [\boldsymbol{\rho}] \hspace{0.3mm} \big)
\label{eq:MM-PF(cont.limit)}
\end{equation}
where the action for the eigenvalue distribution functions
$\boldsymbol{\rho}$ is
\begin{equation}
S_{\mathrm{MM}} [\boldsymbol{\rho}] = \frac{N^2}{\pi} \sum_{i=1}^M
\sum_{l=1}^\infty \Big( (\rho_i^l)^2
V_\mathrm{ad}^l(T;\mu_1,\mu_2) \, + \, \rho_i^l \hspace{0.2mm}
\rho_{i+1}^l V_\mathrm{bi}^l(T;\mu_1,\mu_2) \Big) \: .
\label{eq:MM-action(cont.limit)}
\end{equation}
To summarize, the matrix model under study is defined by Eqs.
(\ref{eq:MM-PF(cont.limit)})-(\ref{eq:MM-action(cont.limit)}).

\section{Phase structure}

The term $-\ln \left| \sin \left( \frac{\theta}{2} \right)
\right|$ in the adjoint potential (\ref{eq:adjointV}) originating
from the change of measure is a temperature-independent repulsive
potential. On the other hand, the remaining parts of the adjoint
and bifundamental potentials
(\ref{eq:adjointV})-(\ref{eq:bifundamentalV}) provide an
attractive force\footnote{The fact that the remaining parts of
(\ref{eq:adjointV})-(\ref{eq:bifundamentalV}) are attractive
potentials can be shown following the argument in
\cite{Aharony:2003sx}, footnote 32.} which grows from zero to
infinite strength as the temperature is raised from zero to
infinity. One would therefore expect that at low temperatures, the
stable saddle points of the matrix model are characterized by the
eigenvalues of the holonomy matrices $U_i$ spreading out uniformly
over the unit circle, whereas at high temperatures the attractive
potential causes them to localize \cite{Aharony:2003sx}.

\subsection{Low-temperature solution and phase transition}

We now consider the saddle points of the matrix model action
(\ref{eq:MM-action(cont.limit)}),
\begin{equation}
0 = \frac{\partial S_{\mathrm{MM}}}{\partial \rho_i^l} =
\frac{N^2}{\pi} \Big( 2 \rho_i^l V_\mathrm{ad}^l \, + \,
\big(\rho_{i-1}^l + \rho_{i+1}^l \big) V_\mathrm{bi}^l \Big) \: .
\label{eq:saddlepointsofSMM}
\end{equation}
For $M\geq 2$, this condition translates into $M$ linear equations
in $M$ unknowns:
\begin{equation}
2 \rho_i^l V_\mathrm{ad}^l \, + \, \big(\rho_{i-1}^l +
\rho_{i+1}^l \big) V_\mathrm{bi}^l = 0 \: .
\end{equation}
The determinant of this system of equations is generically
non-zero, so we find the unique solution $\rho_i^l = 0,$
corresponding to the flat distribution $\rho_i = \frac{1}{2\pi}$.
Thus we conclude that the eigenvalues of the holonomy matrices
$U_i$ are distributed uniformly on each of the $M$ unit circles.
This defines the low-temperature solution of the matrix model.

The leading $\mathcal{O}(N^2)$ contribution to the free energy
computed from the path integral (\ref{eq:MM-PF(cont.limit)}) comes
from the action $S_\mathrm{MM}[\boldsymbol{\rho}]$. However, as
$\rho_i^l = 0$, the first non-zero contribution to the free energy
in this phase comes from a Gaussian integral over the fluctuations
about the solution $\rho_i = \frac{1}{2\pi}$. The free energy is
therefore of $\mathcal{O}(1)$ with respect to $N$, suggesting that
the theory in this phase describes a non-interacting gas of color
singlet states. Furthermore, we note that the Polyakov loop $W(C)
\equiv \Tr \mathcal{P} \exp \big(ig \int_0^\beta dx^0 A_i^0 \big)$
has zero expectation value since the trace averages to zero in the
uniform eigenvalue distribution. In particular, this implies that
the $\mathbb{Z}_N$ center symmetry is left unbroken in this phase.
Accordingly, we label this phase ``confining''.

For $M \geq 2$ the solution $\rho_i = \frac{1}{2\pi}$ will be a
minimum of the action until we reach values of $(T;\mu_1,\mu_2)$
for which
\begin{equation}
0 = \det H_{ij} = \left| \hspace{0.3mm} \frac{\partial^2
S_{\mathrm{MM}}}{\partial \rho_i^l \, \partial \rho_j^l} \,
\right| \label{eq:phase-transition-condition}
\end{equation}
for any fixed $l$. When the temperature or the chemical potentials
are raised above these critical values, the flat distribution
becomes an unstable saddle point of the matrix model,
and the model thus enters a new phase which we will discuss in the
next section. For now we note that
(\ref{eq:phase-transition-condition}) defines a phase transition
condition of the matrix model.

It will be convenient to express the Hessian matrix in terms of
the variables $\xi_l \equiv 2V_\mathrm{ad}^l$ and $\eta_l \equiv
V_\mathrm{bi}^l$. Note first that in the special case $M=2$, due
to the identification $i \simeq i + M = i + 2,$ the Hessian matrix
obtained from (\ref{eq:MM-action(cont.limit)}) takes the
form\footnote{We omit here, and in the following, the overall
factor of $\frac{N^2}{\pi}$ in Eq.
(\ref{eq:MM-action(cont.limit)}) for notational simplicity.}
\begin{equation}
H = \left( \begin{array}{cc} \xi_l & 2 \eta_l \\ 2 \eta_l & \xi_l
\end{array} \right) \: . \label{eq:Hessian(caseM=2)}
\end{equation}
The determinant factorizes as $\det H = -4 \big( \eta_l -
\frac{1}{2} \xi_l \big) \big( \eta_l + \frac{1}{2} \xi_l \big)$.
For $M\geq 3$ the Hessian matrix is a tridiagonal, periodically
continued matrix:
\begin{equation}
H_{ij} = \left\{ \begin{array}{cl} \xi_l &
\hspace{0.3cm}\mathrm{for} \: \, j=i
\\ \eta_l & \hspace{0.3cm} \mathrm{for} \: \, j = i \pm 1
\end{array} \right. \label{eq:Hessian(tridiagform)}
\end{equation}
where, as usual, we make the identifications $i \simeq i+M$ and $j
\simeq j+M$. The determinant of $H$ factorizes as
follows\footnote{This formula is a special case of
(\ref{eq:determinant_circulant_matrix}).}
\begin{equation}
\det \left( \begin{array}{cccc} \xi & \eta & & \hspace{0.2cm}\eta \\
\eta & \xi & \ddots & \\
& \ddots & \ddots & \hspace{0.2cm} \eta \\
\eta & & \eta & \hspace{0.2cm} \xi
\end{array}\right) \hspace{1.5mm} = \hspace{1.5mm} \prod_{j=1}^M \hspace{0.2mm}
\left( \xi + 2\cos\left( \frac{2\pi j}{M} \right) \eta \right) \:
. \label{eq:Hessian-determinant-factorizes}
\end{equation}
Thus, the determinant of $H$ vanishes on any of the lines $\xi_l +
2\cos\left( \frac{2\pi j}{M} \right) \eta_l = 0$ for
$j=1,\ldots,M$. To single out the physically relevant condition
for the vanishing of $\det H$ we will first consider the case
$M=12$ to gain intuition. For $M=12$ the determinant in particular
factorizes as
\begin{equation}
\det H = -36 \, \xi_l^2 \hspace{0.3mm} \big(\eta_l^2 - \xi_l^2
\big)^2 \left( \eta_l^2 - \frac{\xi_l^2}{4}\right) \left( \eta_l^2
- \frac{\xi_l^2}{3}\right)^2
\end{equation}
where $l$ is fixed. In Figure 2 we have divided the
$(\xi_l,\eta_l)$ plane into regions where $H$ is positive-definite
(denoted by $+$) and where $H$ is indefinite (denoted by $-$).

Thus regions marked by $+$ correspond to a local extremum
(minimum) of $S_{\mathrm{MM}}$, and regions marked by $-$
correspond to unstable saddle points. In Figure 2 we have
furthermore marked the region occupied by the $\mathcal{N}=2$
quiver gauge theory matrix model in the low temperature phase by
plotting $(\xi_1,\eta_1)$ for $(T;\mu_1,\mu_2) = (0.1;0.8,0.8)$.
For fixed chemical potentials, $z_\mathrm{ad}^B, z_\mathrm{ad}^F,
z_\mathrm{bi}^B, z_\mathrm{bi}^F$ all increase monotonically with
the temperature. Therefore, as the temperature increases, the dot
in Figure 2 will move as indicated and hit the instability line
$\eta_l = - \frac{1}{2} \xi_l$ at the phase transition
temperature.

By the same analysis, for any $M \geq 2$ the phase transition
occurs at the instability line $\eta_l = \alpha(M) \hspace{0.4mm}
\xi_l$ where $\alpha(M)$ is the numerically smallest negative
slope of the zero lines of the Hessian determinant. For all $M
\geq 3$ we find from (\ref{eq:Hessian-determinant-factorizes})
that $\alpha(M) = -\frac{1}{2}$ (corresponding to $j=M$). For
$M=2$ we also find $\alpha(M) = -\frac{1}{2}.$ Indeed, note that
for $M \geq 2$ the matrix obtained by substituting $\eta_l =
-\frac{1}{2} \hspace{0.4mm} \xi_l$ in Eqs.
(\ref{eq:Hessian(caseM=2)}) and (\ref{eq:Hessian(tridiagform)})
will have a zero eigenvalue (with $(1, 1, \ldots, 1)$ as an
eigenvector) and hence zero determinant.

\begin{figure}[!h]
\begin{center}
\includegraphics[angle=0, width=0.5\textwidth]{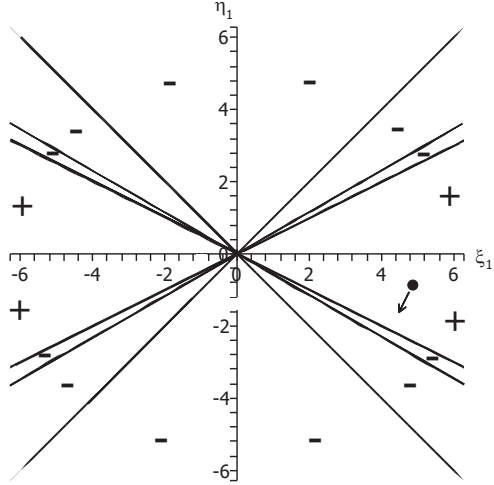}
\textit{\caption{Regions of positive-definiteness and
indefiniteness of $H$ for the case $M=12$. Regions where $H$ is
positive-definite (corresponding to a local minimum of
$S_{\mathrm{MM}}$) are marked by $+$; regions where $H$ is
indefinite (corresponding to an unstable saddle point of
$S_{\mathrm{MM}}$) are marked by $-$. The lines represent the
locus of $\det H = 0.$ The physically accessible region of the
$(\xi_l,\eta_l)$ plane is bounded from above by the $\xi_l$ axis
and from below by the line of the numerically smallest negative
slope. This is illustrated by the dot which corresponds to
$(T;\mu_1,\mu_2)$ $= (0.1;0.8,0.8)$ and $l=1$. The arrow shows how
the dot will move as the temperature is increased, keeping $\mu_1$
and $\mu_2$ fixed.}}
\end{center}
\end{figure}

\subsubsection*{The large $M$ limit}

As a consistency check, we can derive that $\lim_{M \to \infty}
\alpha(M) = -\frac{1}{2}$ by a different route. We take the
continuum limit $M \to \infty$ in the quiver direction. The quiver
label $i$ thus becomes a continuous angular parameter $\vartheta$
which we take to be $2\pi$-periodic; i.e., we identify $\vartheta
\simeq \vartheta + 2\pi$. Accordingly we make the substitutions
\begin{eqnarray}
(\rho_i^l)^2 \: &\longrightarrow& \: (\rho^l(\vartheta))^2 \\
(\rho_i^l \rho_{i+1}^l) \: &\longrightarrow& \: -\frac{1}{2}
\big(\dot{\rho}^l(\vartheta)\big)^2 + (\rho^l(\vartheta))^2
\end{eqnarray}
where $\dot{\phantom{\rho}}$ denotes $\frac{d}{d\vartheta}$. The
matrix model action (\ref{eq:MM-action(cont.limit)}) thus
becomes\footnote{The extra prefactor $\frac{M}{2\pi}$ comes from
changing the counting measure over $i$ to the measure $d
\vartheta$.}
\begin{equation}
S_{\mathrm{MM}}[\boldsymbol{\rho}] \: = \: \frac{N^2 M}{(2\pi)^2}
\sum_{l=1}^\infty \int_0^{2\pi} \hspace{-0.5mm} d\vartheta \left[
\hspace{0.3mm} (\xi_l + 2\eta_l) \hspace{0.3mm}
(\rho^l(\vartheta))^2 - \eta_l \big( \dot{\rho}^l (\vartheta)
\big)^2 \hspace{0.3mm} \right] \: .
\label{eq:MM-action(quiver-cont.limit)}
\end{equation}
The Euler-Lagrange equations obtained from this action are those
of a harmonic oscillator,
\begin{equation}
\eta_l \ddot{\rho}^{\, l}(\vartheta) \, + \, (\xi_l + 2\eta_l) \,
\rho^l (\vartheta) = 0
\end{equation}
where $l = 1, 2, \ldots$. Note here that it is the bifundamental
contribution in (\ref{eq:MM-action(cont.limit)}) that gives rise
to the derivative term in (\ref{eq:MM-action(quiver-cont.limit)})
and in turn to the mass term for the harmonic oscillator. Thus,
the harmonic oscillator EOM's in the large $M$ limit is a pure
`quiver phenomenon'. Solutions to these equations will become
unstable when the tension $\tau \equiv (\xi_l + 2\eta_l)$ goes
from $\tau > 0$ to $\tau < 0$. Thus, for large $M$, the phase
transition will occur when $\eta_l = -\frac{1}{2} \xi_l$,
consistent with what we found above.

We now return to the phase transition condition $\eta_l =
\alpha(M) \hspace{0.5mm} \xi_l$. Since $z_\mathrm{ad}^B,
z_\mathrm{ad}^F, z_\mathrm{bi}^B, z_\mathrm{bi}^F$ are all
monotonically increasing as functions of $x$ and $0 \leq x < 1$,
the $l=1$ condition is the strongest. Therefore, the phase
transition condition for $M \geq 2$ is
\begin{equation}\label{eq:phasetransitionconditiongeq2}
\mathrm{for} \:\: M \geq 2 : \big( z_\mathrm{ad}^B (x;y_1,y_2) +
z_\mathrm{ad}^F (x;y_1,y_2) \big) + 2\big( z_\mathrm{bi}^B
(x;y_1,y_2) + z_\mathrm{bi}^F (x;y_1,y_2) \big) = 1 \: .
\end{equation}
Finally, in the special case $M = 1$ we immediately obtain
$V_\mathrm{ad}^l + V_\mathrm{bi}^l = 0$ from
(\ref{eq:saddlepointsofSMM}) due to the identification $i \simeq i
+ M = i + 1$. Putting $l = 1,$ this is precisely the phase
transition condition (\ref{eq:phasetransitionconditiongeq2}). We
thus conclude that for any $M$ the phase transition condition is
\begin{equation}
\big( z_\mathrm{ad}^B (x;y_1,y_2) + z_\mathrm{ad}^F (x;y_1,y_2)
\big) + 2\big( z_\mathrm{bi}^B (x;y_1,y_2) + z_\mathrm{bi}^F
(x;y_1,y_2) \big) = 1 \: .
\label{eq:phase-transition-condition(final)}
\end{equation}
In Figure 3 below we have plotted the curves in the $(T,\mu)$
plane obtained from this condition for the cases $(\mu_1,\mu_2) =
(\mu,0)$ ; $(\mu_1,\mu_2) = (0,\mu)$ and $(\mu_1,\mu_2) =
(\mu,\mu)$. For each of these cases, the relevant curve defines
the phase diagram of $\mathcal{N}=2$ quiver gauge theory as a
function of both temperature and chemical potential. Note that, as
discussed in Section 2.2, if one or both of the chemical
potentials are larger than the inverse radius of the spatial
manifold $S^3$, the theory develops tachyonic modes and becomes
ill-defined. Therefore the line $\mu = 1/R$ defines a boundary of
the phase diagram.

The phase transition condition
(\ref{eq:phase-transition-condition(final)}) defines a phase
transition temperature $T_H(\mu_1, \mu_2)$ as a function of the
chemical potentials. We will refer to $T_H(\mu_1, \mu_2)$ as the
Hagedorn temperature of $\mathcal{N}=2$ quiver gauge theory. This
terminology will be justified in Section 4.2. We remark that the
Hagedorn temperature at zero chemical potential is
\begin{equation}
T_H = -\frac{1}{\ln(7 - 4\sqrt{3})} \approx 0.37966
\end{equation}
in units of $R^{-1}$, the inverse radius of the $S^3$. This is
exactly the Hagedorn temperature for $\mathcal{N}=4$ SYM theory
(cf. \cite{Aharony:2003sx, Sundborg:1999ue}). The origin of this
fact can be traced to the observation in \cite{Bershadsky:1998cb}
that in the large $N$ limit the correlation functions of
$\mathcal{N}=4 \:\: U(N)$ SYM theory equal the corresponding
correlation functions of the $\mathcal{N}=2$ quiver gauge theories
obtained from orbifold projections. Since our computations rely on
perturbation theory (namely, taking the $g \to 0$ limit of the
action and then performing Gaussian path integrations), and we are
furthermore taking the $N \to \infty$ limit, we should expect that
the matrix model defined out of the quantum effective action will
have the same behavior for the $\mathcal{N}=2$ quiver gauge theory
as for the $\mathcal{N}=4$ SYM theory.

Furthermore, for small chemical potentials the Hagedorn
temperature is given by
\begin{equation}
T_H(\mu_1, \mu_2) = \frac{1}{\beta_0} + c \big(\mu_1^2 + 2\mu_2^2
\big) + c_{11} \mu_1^4 + c_{12}\mu_1^2 \mu_2^2 + c_{22}\mu_2^4 +
\mathcal{O}(\mu_i^6) \label{eq:TH-small-mu(1)}
\end{equation}
where the coefficients are
\begin{eqnarray}
\beta_0 &=&  -\ln(7-4\sqrt{3}) \: , \quad c = -
\frac{\sqrt{3}}{18} \: , \quad c_{11} =
-\textstyle{\frac{\beta_0}{864}} \left( \textstyle{\frac{362
\beta_0 - 209 \sqrt{3} \beta_0 + 2896\sqrt{3}
- 5016}{-627 + 362\sqrt{3}}} \right) \phantom{aaaaaaaa} \label{eq:TH-small-mu(2)} \\
c_{12} &=&  \textstyle{\frac{\beta_0}{216}} \left(
\textstyle{\frac{1810\beta_0 - 1045\sqrt{3} \beta_0 - 2896\sqrt{3}
+ 5016}{-627 + 362\sqrt{3}}} \right) \: , \quad c_{22} =
\textstyle{\frac{\beta_0}{108}} \left( \textstyle{\frac{362
\beta_0 - 209\sqrt{3} \beta_0 - 1448\sqrt{3} + 2508}{-627 +
362\sqrt{3}}} \right) \label{eq:TH-small-mu(3)}
\end{eqnarray}

\begin{figure}[!h]
\begin{center}
\includegraphics[angle=0, width=0.6\textwidth]{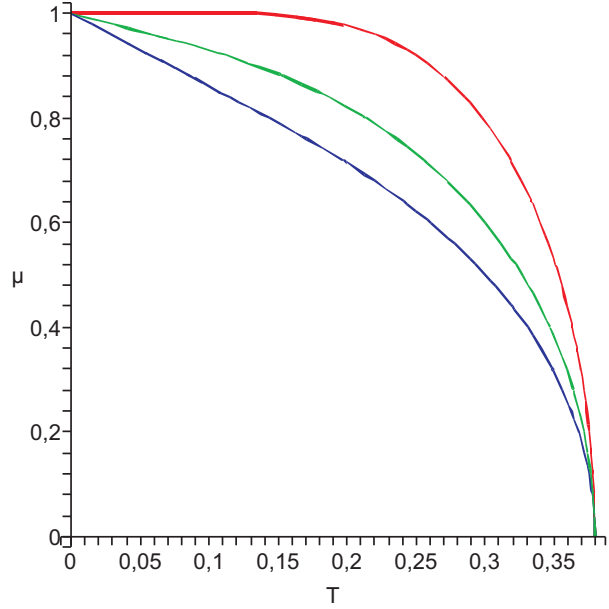}
\textit{\caption{Phase diagram of $\mathcal{N}=2$ quiver gauge
theory. The outermost curve is the transition line corresponding
to $(\mu_1,\mu_2) = (\mu,0)$. It has slope $0$ in the neighborhood
of the point $(T,\mu) = (0,1)$. The inbetween curve corresponds to
$(\mu_1,\mu_2) = (0,\mu)$, with slope $-\ln 2$ near $(0,1)$. The
innermost curve corresponds to $(\mu_1,\mu_2) = (\mu,\mu)$, with
slope $-\ln 4$ near $(0,1)$. The phase transition temperature at
zero chemical potential is common for the three curves and equals
$T = -\frac{1}{\ln(7 - 4\sqrt{3})} \approx 0.37966$ as in the
$\mathcal{N}=4$ SYM case.}}
\end{center}
\end{figure}

\subsection{Solution above the Hagedorn temperature}

As the temperature is increased beyond $T > T_H$, the attractive
terms in the pairwise potential continue to increase in strength,
and so the eigenvalues will become increasingly localized. The
precise distribution can be determined, following
\cite{Aharony:2003sx}, by the condition that a single additional
eigenvalue $\alpha_i$ added on the $i$'th circle experiences no
net force from the other eigenvalues on the circles $i-1,$ $i$ and
$i+1$:
\begin{equation}
0 \hspace{1.2mm} = \hspace{1.2mm} \int_{-\pi}^\pi d\zeta
\hspace{1.0mm} 2 V_\mathrm{ad}'(\alpha_i - \zeta) \hspace{0.4mm}
\rho_i(\zeta) \hspace{0.8mm} + \hspace{0.3mm}\int_{-\pi}^\pi
d\zeta \hspace{1.0mm} V_\mathrm{bi}'(\alpha_i - \zeta)
\hspace{0.2mm} \big( \rho_{i-1}(\zeta) + \rho_{i+1}(\zeta) \big)
\label{eq:no-force-condition}
\end{equation}
where $V_\mathrm{ad}$ and $V_\mathrm{bi}$ are given in
(\ref{eq:adjointV}) and (\ref{eq:bifundamentalV}), respectively.
This provides $M$ equilibrium conditions for the lattice
action
\begin{equation} S_\mathrm{latt} \hspace{1.5mm} = \hspace{1.5mm} N
\hspace{0.2mm} \sum_{i=1}^M \sum_{l=1}^\infty \hspace{0.4mm}
\frac{a_l \rho^l_i \hspace{0.4mm} + \hspace{0.4mm} b_l
\rho_{i-1}^l \hspace{0.3mm} + \hspace{0.4mm} b_l \rho_{i+1}^l}{l}
\hspace{0.8mm} \Big( \hspace{-0.4mm} \Tr U_i^l + \Tr U_i^{-l}
\Big) \label{eq:lattice-action}
\end{equation}
where
\begin{eqnarray}
a_l &=& z_\mathrm{ad}^B (x^l;
y_1^l, y_2^l) \hspace{1.0mm} + \hspace{1.0mm} (-1)^{l+1} z_\mathrm{ad}^F (x^l; y_1^l, y_2^l) \\
b_l &=& z_\mathrm{bi}^B (x^l; y_1^l, y_2^l) \hspace{1.0mm} +
\hspace{0.8mm} (-1)^{l+1} z_\mathrm{bi}^F (x^l; y_1^l, y_2^l) \: .
\end{eqnarray}

The exact solution of (\ref{eq:lattice-action}) was found in
\cite{Jurkiewicz:1982iz}. It takes the form
\begin{equation}
\rho_i(\theta) = \frac{1}{\pi} \left( \sin^2 \left(
\textstyle{\frac{\theta_0^i}{2}} \right) - \sin^2 \left(
\textstyle{\frac{\theta - \alpha_i }{2}} \right) \right)^{1/2}
\sum_{k=1}^\infty Q_k^i \cos \big( (k-1/2) (\theta -\alpha_i)
\big) \: , \hspace{0.2cm} i=1 \ldots M
 \label{eq:exact-sol-of-equilib-cond1}
\end{equation}
where
\begin{equation}
Q_k^i = 2\sum_{j=0}^\infty \big( a_{j+k} \rho_i^{j+k} + b_{j+k}
\rho_{i-1}^{j+k} + b_{j+k} \rho_{i+1}^{j+k} \big) P_j (\cos
\theta_0^i) \: . \label{eq:exact-sol-of-equilib-cond2}
\end{equation}
The support of $\rho_i$ is $[\alpha_i-\theta_0^i,\alpha_i +
\theta_0^i]$. Here one must impose the consistency requirement
\begin{equation}
\rho_i^n = \int_{-\pi}^\pi d\theta \hspace{0.4mm} \rho_i(\theta)
\cos(n\theta) \: . \label{eq:consistency-requirement}
\end{equation}
For simplicity the following analysis will be carried out only in
the truncated case $a_{m > 1} = b_{m > 1} = 0$ which shares the
same qualitative behavior with the general case. For $n=1$, the
consistency condition (\ref{eq:consistency-requirement}) then
becomes
\begin{eqnarray}
\rho_i^1 &=& \frac{2}{\pi} \big( a_1 \rho_i^1 + b_1 \rho_{i-1}^1 +
b_1 \rho_{i+1}^1 \big) \int_{-\theta_0^i}^{\theta_0^i} d\xi\left(
\sin^2 \left( \textstyle{\frac{\theta_0^i}{2}} \right) - \sin^2
\left( \textstyle{\frac{\xi}{2}} \right) \right)^{1/2} \cos \left(
\textstyle{\frac{\xi}{2}} \right) \cos(\xi + \alpha_i) \nonumber \\
&=& \cos \alpha_i \big( a_1 \rho_i^1 + b_1 \rho_{i-1}^1 + b_1
\rho_{i+1}^1 \big) (2s_i^2 - s_i^4)  \label{eq-for-rho_i(1)}
\end{eqnarray}
where $s_i^2 \equiv \sin^2 \left( \frac{\theta_0^i}{2} \right)$.
In analogy with \cite{Aharony:2003sx}, $\theta_0^i$ is determined
from $Q_1^i = Q_0^i + 2$, leading to the $M$ equations
\begin{equation}
1 = 2s_i^2 \big( a_1 \rho_i^1 + b_1 \rho_{i-1}^1 + b_1
\rho_{i+1}^1 \big) \: . \label{eq:eq-for-s_i}
\end{equation}
By means of (\ref{eq:eq-for-s_i}), one can rewrite
(\ref{eq-for-rho_i(1)}) as the set of $M$ coupled equations
\begin{equation}
a_1 (\rho_i^1)^2 + \big( b_1 \rho_{i-1}^1 + b_1 \rho_{i+1}^1 - a_1
\cos \alpha_i \big) \rho_i^1 + \cos\alpha_i \big(
\textstyle{\frac{1}{4}} - b_1 \rho_{i-1}^1 - b_1 \rho_{i+1}^1
\big) = 0 \: . \label{eq-for-rho_i(2)}
\end{equation}
If one allows some of the $\alpha_i$ to be nonzero, one finds for
any fixed $\mu_1,\mu_2$ a range of temperatures above
$T_H(\mu_1,\mu_2)$ where (\ref{eq-for-rho_i(2)}) has no solution
such that all $s_i$, given by (\ref{eq:eq-for-s_i}), satisfy $0
\leq s_i^2 \leq 1$. Thus one must have $\alpha_1 = \cdots =
\alpha_M = 0$.

With these centers of masses of the eigenvalue
distributions, requiring that $0 \leq s_i^2 \leq 1$ leads to a
unique solution of (\ref{eq-for-rho_i(2)}). This solution has all
$\rho_i^1$  equal, as well as all $s_i$ equal and given by
\begin{equation}
s_i^2 \hspace{0.7mm} = \hspace{0.8mm} \sin^2 \left(
\frac{\theta_0^i}{2} \right) \hspace{0.7mm} = \hspace{0.7mm} 1 -
\sqrt{1-\frac{1}{a_1 + 2b_1}} \: .
\label{eq:support-of-above-T_H-solution}
\end{equation}
With the assumption $a_{m > 1} = b_{m > 1} = 0$ the exact solution
(\ref{eq:exact-sol-of-equilib-cond1})-(\ref{eq:exact-sol-of-equilib-cond2})
thus truncates to
\begin{equation}
\rho_i (\theta) \hspace{1.0mm} = \hspace{1.0mm} \frac{1}{\pi
s_i^2} \sqrt{s_i^2 - \sin^2\left( \frac{\theta}{2}\right)}
\hspace{0.7mm} \cos \left( \frac{\theta}{2} \right) \: .
\label{eq:approximate-above-T_H-solution}
\end{equation}
It is immediately clear from
(\ref{eq:support-of-above-T_H-solution}) that for temperatures
above the Hagedorn temperature one has $\theta_0^i < \pi$; i.e.,
the eigenvalue distribution becomes gapped.
In particular we note that the phase above the Hagedorn
temperature has unbroken quiver translational invariance; i.e.
$\rho_i = \rho_{i+1}$.
The unbroken quiver translational invariance is expected on more general
grounds due to the perturbative equivalence between $\mathcal{N}=4$ SYM theory
and $\mathcal{N}=2$ quiver gauge theory \cite{Bershadsky:1998cb}, although it
should be noted that in Ref.~\cite{Bershadsky:1998cb}
the gauge theories are studied on $\mathbb{R}^4$ rather than $S^1 \times S^3$.%
\footnote{Note that non-perturbative effects could potentially destroy
the quiver translational invariance. However, the computation carried
out in this section is only valid perturbatively, so we would not
expect to see such effects. See
Refs.~\cite{Erlich:1998gb, Kovtun:2003hr, Kovtun:2004bz, Kovtun:2005kh, Armoni:2007jt}
for work on non-perturbative equivalence between parent/daughter gauge
theories related by orbifold and orientifold projections.
}

\subsubsection*{Free energy slightly above the Hagedorn temperature}

Using the Hagedorn temperature for small chemical potentials given
in (\ref{eq:TH-small-mu(1)})-(\ref{eq:TH-small-mu(3)}) we can
compute the free energy slightly above the Hagedorn temperature in
analogy with \cite{Harmark:2006di}. Defining $\Delta T \equiv T -
T_H(\mu_1, \mu_2)$, we find for $0 < \Delta T \ll 1$ the
perturbative expansion
\begin{eqnarray}
\frac{F}{N^2 M} &=& -\beta_0 \frac{3}{8} \left( 1 - \beta_0
\frac{2\sqrt{3} + \beta_0}{36} \big( \mu_1^2 + 2\mu_2^2 \big) +
\mathcal{O}(\mu_i^4) \right)
\Delta T \phantom{aaaaaaaaaaaaaaaaaaaaa}\nonumber \\
&\phantom{=}& - \beta_0^2 \sqrt{\frac{3}{8}} \left( 1 - \beta_0
\frac{4 + \sqrt{3} \hspace{0.5mm} \beta_0}{24\sqrt{3}} \big(
\mu_1^2 + 2 \mu_2^2 \big) + \mathcal{O}(\mu_i^4) \right) \Delta
T^{3/2} + \mathcal{O}(\Delta T^2) \: .
\label{eq:free-energy(slightly-above-T-H)}
\end{eqnarray}

\subsubsection*{High-temperature behavior of free energy}

In the $T \to \infty$ limit the pairwise attractive potentials
grow to infinite strength, so the eigenvalues of the holonomy
matrices $U_i$ localize to extremely small intervals; i.e. the
eigenvalue distribution functions will become delta functions,
$\rho_i(\theta_i) \to \delta(\theta_i)$. (This is also clear from
(\ref{eq:support-of-above-T_H-solution}) since for $T \to \infty$
one has $a_1, b_1 \to \infty$ and thus $\theta_0 \to 0$. The
normalization condition (\ref{eq:normalization-of-rho}) then
implies $\rho_i(\theta_i) \to \delta(\theta_i)$.) Therefore
$\rho_i^l \to 1$, and so from (\ref{eq:MM-action(cont.limit)}) and
(\ref{eq:spPF(ad,bos)})-(\ref{eq:spPF(bi,ferm)}) we find that the
free energy in the $T \to \infty$ limit is
\begin{equation}
F \hspace{1mm} = \hspace{1mm} - N^2 M \left( \frac{\pi^2 T^4}{6} +
\frac{T^2}{4} \big( \mu_1^2 + 2 \mu_2^2 \big) - \frac{1}{32 \pi^2}
\big( \mu_1^4 - 4\mu_1^2 \mu_2^2 \big) \right) \hspace{-0.3mm}
\mathrm{Vol}(S^3) \: . \label{eq:free-energy(large-T-with-mu)}
\end{equation}
Here we have applied the polylogarithm regularization procedure
described in Appendix E of Ref. \cite{Harmark:1999xt} in order to
obtain (\ref{eq:free-energy(large-T-with-mu)}).\footnote{Note that
there is a minus sign missing on the right hand side of (E.4) for
the $n \neq 1$ case.} We note that the free energy scales as $N^2
M$ as $N \to \infty$. This is to be expected from the orbifold
projection invariant form of the fields
(\ref{eq:projinvariantAmuA})-(\ref{eq:projinvariantBFi}) and
(\ref{eq:projinvariantpsichiA})-(\ref{eq:projinvariantchiBpsiPhi}),
given that the free energy scales as $N^2$ for $\mathcal{N}=4$
$U(N)$ SYM theory for high temperatures in the $N \to \infty$
limit (cf. Eq. (5.62) of \cite{Aharony:2003sx}).

The fact that the free energies
(\ref{eq:free-energy(slightly-above-T-H)}) and
(\ref{eq:free-energy(large-T-with-mu)}) are both of
$\mathcal{O}(N^2 M)$ with respect to $N$ suggests that the gauge
theory in the phase above the Hagedorn temperature describes a
non-interacting plasma of color non-singlet states. Furthermore,
from the fact that the eigenvalue distribution
(\ref{eq:support-of-above-T_H-solution})-(\ref{eq:approximate-above-T_H-solution})
is gapped we can immediately conclude that the Polyakov loop
$W(C)$ has non-zero expectation value as the trace does not
average to zero in this case. In particular, this implies that the
$\mathbb{Z}_N$ center symmetry is spontaneously broken in this
phase. Accordingly, we label this phase ``deconfined''. Thus, we
conclude that the phase transition defined by Eq.
(\ref{eq:phase-transition-condition(final)}) is a
confinement/deconfinement phase transition. Since furthermore the
derivative of the free energy with respect to the temperature is
discontinuous at the phase transition temperature $T_H(\mu_1,
\mu_2)$, we conclude that the transition is of first order.
Furthermore, cf. \cite{Sundborg:1999ue, Aharony:2003sx}, we
identify it with a Hagedorn phase transition, and $T_H(\mu_1,
\mu_2)$ is thus the Hagedorn temperature of $\mathcal{N}=2$ quiver
gauge theory.

\subsubsection*{Twisted partition function}

In analogy with \cite{Unsal:2007fb}, one may study the
twisted partition function for the quiver gauge theory, taking
the boundary conditions for the spinor fields on the $S^1$
to be periodic rather than antiperiodic.  In this case the
Matsubara frequencies for the spinor fields will be the same as
for the bosonic fields, and the twisted partition function
$\widetilde{Z} = \Tr (-1)^F e^{-\beta H} =
e^{-\widetilde{\Gamma}[U_i]}$ may be obtained directly from
(\ref{eq:quantum-effective-action}) by replacing $(-1)^{l+1}
\longrightarrow (-1)$. Following \cite{Unsal:2007fb} we
choose to exhibit the $\mathbb{Z}_M$ symmetry of the (twisted) partition function
by rewriting the adjoint and bifundamental holonomy factors in terms of
eigenvectors under quiver node displacements $i \to i+1$. Indeed,
define for $\omega \equiv e^{2\pi i/M}$,
\begin{equation}
\Omega_k^l \hspace{0.5mm} \equiv \hspace{0.5mm} \sum_{j=1}^M
\omega^{-kj} \hspace{0.5mm} U_j^l \: .
\end{equation}
Under the quiver node displacement $U_i^l \hspace{1mm}
\longrightarrow \hspace{1mm} U_{i+1}^l$ we find $\Omega_k^l
\hspace{1mm} \longrightarrow \hspace{1mm} \omega^k \Omega_k^l$ so
that $\Omega_k^l$ is an eigenvector under the displacement with
the eigenvalue $\omega^k$. Writing the holonomy factors in terms
of $\Omega_k^l$, the twisted quantum effective action takes the
form
\begin{eqnarray} \widetilde{\Gamma}[U_i] &=& - \frac{1}{M}
\sum_{k=1}^M \sum_{l=1}^\infty \Bigg[ \, \frac{1}{l} \left(
\frac{6x^{2l} - 2x^{3l}}{(1-x^l)^3} \right) + \frac{1}{l} \left(
\frac{x^l + x^{2l}}{(1-x^l)^3} \right) \big(
y_1^l + y_1^{-l} \big) \nonumber \\
&\phantom{=}& \hspace{2.5cm} - \, \frac{1}{l} \left(
\frac{2x^{3l/2}}{(1-x^l)^3} \right) \hspace{-0.7mm} \big(
y_1^{l/2} + y_1^{-l/2} \big) \big( y_2^l + y_2^{-l} \big) \,
\Bigg] \Big( \Tr \Omega_k^l \Tr \Omega_{-k}^{-l} \Big) \nonumber \\
&\phantom{=}& - \, \frac{1}{M} \sum_{k=1}^M \sum_{l=1}^\infty
\Bigg[ \, \frac{1}{l} \left( \frac{x^l + x^{2l}}{(1-x^l)^3}
\right) \big( y_2^l + y_2^{-l} \big) - \frac{1}{l} \left(
\frac{2x^{3l/2}}{(1-x^l)^3} \right) \big( y_1^{l/2} + y_1^{-l/2}
\big) \, \Bigg] \nonumber \\
&\phantom{=}& \hspace{2.5cm} \times \: \omega^{-k} \Big(  \Tr
\Omega_k^l \Tr \Omega_{-k}^{-l} + \hspace{0.3mm} \Tr \Omega_k^{-l}
\Tr \Omega_{-k}^l \Big) \: .
\end{eqnarray}
It would be interesting to study the phase structure for the twisted
partition function.

\subsection{Quantum mechanical sectors}

Since $\mathcal{N}=2$ quiver gauge theory is a conformal field
theory, we can exploit the state/opera\-tor correspondence and map
the Hamiltonian $H$ to the dilatation operator $D$. As a
consequence, the partition function of thermal $\mathcal{N}=2$
quiver gauge theory in the grand canonical ensemble takes the form
\begin{equation}
Z(T; \mu_1, \mu_2) = \Tr_\mathcal{H} \big( e^{-\beta D + \beta
\mu_i Q_i} \big) \: .
\label{eq:partition-function-grand-canonical-ensemble}
\end{equation}
Here the trace is taken over the entire Hilbert space
$\mathcal{H}$ of gauge invariant operators. For weak 't Hooft
coupling $\lambda \ll 1$, the dilatation operator $D$ can be
expanded perturbatively\footnote{This was shown for
$\mathcal{N}=4$ $U(N)$ SYM theory in \cite{Beisert:2003tq,
Beisert:2004ry}.}
\begin{equation}
D = D_0 + \sum_{n=2}^\infty \lambda^{n/2} D_n \: .
\label{eq:dilat.op(pert.expansion)}
\end{equation}
We let $Q$ denote the total charge with respect to the Cartan
generators of $SU(2)_R \times U(1)_R$, $Q = Q_1 + Q_2$, with $\mu$
as the associated chemical potential.\footnote{Recall that in
Section 2.2 we defined the generator of the Cartan subalgebra of
$SU(2)_R$ to be $\sigma_z$ rather than $\frac{1}{2} \sigma_z$ so
that we have the associated charges $Q_1, Q_2$ implicitly given
through Eqs.
(\ref{eq:charges-under-Cartan-generators(A)})-(\ref{eq:charges-under-Cartan-generators(psi)}).
It is these charges we are referring to here, rather than the
$R$-charges given in Table A.} Taking $\lambda = 0$, the partition
function (\ref{eq:partition-function-grand-canonical-ensemble})
can be rewritten as
\begin{equation}
Z(T; \mu) = \Tr_\mathcal{H} \exp \Big( -\beta (D_0 - Q) - \beta (1
- \mu) Q \Big) \: . \label{eq:partion-function(GCE,rewritten)}
\end{equation}
Following \cite{Harmark:2006di}, we now consider the region of
small temperature and near-critical chemical potential
\begin{equation}
T \ll 1 \: , \hspace{1.3cm} 1 - \mu \ll 1 \: .
\label{eq:near-critical-region}
\end{equation}
In this region, the Hilbert space of gauge invariant operators of
$\mathcal{N}=2$ quiver gauge theory truncates to certain
subsectors. To show this, first observe that in the region
(\ref{eq:near-critical-region}), operators with $D_0 > Q$ appear
with an extremely small weight factor in the partition function
(\ref{eq:partion-function(GCE,rewritten)}) since $\beta \gg 1$. On
the other hand, for operators with $D_0 = Q$, the weight factor is
non-negligible precisely because $1 - \mu \ll 1$. Therefore, the
partition function (\ref{eq:partion-function(GCE,rewritten)}) is
dominated by contributions from operators belonging to the
subsector
\begin{equation}
\mathcal{H}_0 \equiv \big\{ \hspace{0.3mm} \mathcal{O} \in
\mathcal{H} \hspace{1.2mm} \big| \hspace{1.2mm} (D_0 - Q)
\mathcal{O} \hspace{0.5mm} = \hspace{0.5mm} 0 \big\} \: .
\end{equation}
We thus conclude that by taking the near-critical limit
\begin{equation}
x \hspace{0.6mm} \longrightarrow \hspace{0.9mm} 0 \: ,
\hspace{1.2cm} xy \hspace{2mm} \mathrm{fixed}
\label{eq:near-critical-limit} \: ,
\end{equation}
the full Hilbert space $\mathcal{H}$ of gauge-invariant operators
effectively truncates to the subsector $\mathcal{H}_0$. We will
consider three concrete examples of this truncation below,
obtained by either turning off one of the $R$-symmetry chemical
potentials, or by putting them equal. As we remark below, the
resulting subsectors are in a certain sense quantum mechanical.

\subsubsection*{Case 1: The 1/2 BPS sector}

We take $(\mu_1 , \mu_2) = (\mu, 0)$, and thus the total Cartan
charge is $Q = Q_1$. Taking the near-critical limit
(\ref{eq:near-critical-limit}) of the partition function
(\ref{eq:matrix-model(definition)}) then yields
\begin{equation}
Z(x;y) \hspace{1.5mm} \longrightarrow \hspace{1.5mm} \int
\prod_{i=1}^M \big[ \mathcal{D} U_i \big] \exp \left( \sum_{i=1}^M
\sum_{l=1}^\infty \frac{(xy)^l}{l} \Tr U_i^l \Tr U_i^{-l} \right)
\: .
\end{equation}
Since the scalar field $\Phi_i$ has $D_0 = Q = 1$, we therefore
conclude that the Hilbert space of gauge invariant operators
truncates to the 1/2 BPS sector spanned by multi-trace operators
of the form
\begin{equation}
\Tr \big( \Phi_{i_1}^{J_1} \big) \Tr \big( \Phi_{i_2}^{J_2} \big)
\cdots \Tr \big( \Phi_{i_k}^{J_k} \big) \: .
\end{equation}

It is clear that in the near-critical limit
(\ref{eq:near-critical-limit}) all operators with covariant
derivatives decouple. Thus all modes originating from defining a
field theory on the spatial manifold $S^3$ are removed, and the
locality of the field theory is lost. In this sense the resulting
subsector of the field theory is quantum mechanical.

\subsubsection*{Case 2: The $SU(2)$ sector}

We take $(\mu_1, \mu_2) = (0, \mu)$, and thus the total Cartan
charge is $Q = Q_2$. Taking the near-critical limit
(\ref{eq:near-critical-limit}) of the partition function
(\ref{eq:matrix-model(definition)}) then yields
\begin{equation}
Z(x;y) \hspace{1.5mm} \longrightarrow \hspace{1.5mm} \int
\prod_{i=1}^M \big[ \mathcal{D} U_i \big] \exp \left( \sum_{i=1}^M
\sum_{l=1}^\infty \frac{2(xy)^l}{l} \Tr U_i^l \Tr U_{i+1}^{-l}
\right) \: .
\end{equation}
Since the scalar fields $A_{i,(i+1)}$ and $B_{(i+1),i}$ both have
$D_0 = Q = 1$, we therefore conclude that the Hilbert space of
gauge invariant operators truncates to the $SU(2)$ sector spanned
by multi-trace operators of the form
\begin{equation}
\prod_{j=1}^k \hspace{0.3mm} \Tr \big( Z_{1
\boldsymbol{\to}}^{(j)} Z_{2 \boldsymbol{\to} }^{(j)} \cdots
Z_{J_j \boldsymbol{\to}}^{(j)} \big) \label{eq:SU(2)-sector}
\end{equation}
where any letter $Z^{(j)}_{i_j \boldsymbol{\to}}$ is one of the
scalars $A_{i,(i+1)}$ or $B_{(i+1),i}$. The subscripts
`${\boldsymbol{\to}}$' denote that the quiver labels on the fields
in question must trace out a closed loop on the quiver diagram in
Figure 1 so as to ensure gauge invariance. I.e., an example of a
gauge invariant single-trace operator is $\Tr \big( A_{i,(i+1)}
A_{(i+1),(i+2)} B_{(i+2),(i+1)} B_{(i+1),i} \big)$.

\subsubsection*{Case 3: The $SU(2|3)/\mathbb{Z}_M$ sector}

We take $(\mu_1, \mu_2) = (\mu,\mu)$ and thus the total Cartan
charge is $Q = Q_1 + Q_2$. Taking the near-critical limit
(\ref{eq:near-critical-limit}) of the partition function
(\ref{eq:matrix-model(definition)}) then yields
\begin{eqnarray}
Z(x;y) &\longrightarrow& \int \prod_{i=1}^M \big[ \mathcal{D} U_i
\big] \exp \left[ \sum_{i=1}^M \sum_{l=1}^\infty \left(
\frac{(xy)^l + 2(-1)^{l+1} (xy)^{3l/2}}{l}\right) \Tr U_i^l \Tr
U_i^{-l} \right. \nonumber \\
&\phantom{=}& \hspace{3.0cm} \left. + \hspace{0.4mm} \sum_{i=1}^M
\sum_{l=1}^\infty \frac{2(xy)^l}{l} \Tr U_i^l \Tr U_{i+1}^{-l}
\right] \: .
\end{eqnarray}
Since the scalar fields $A_{i,(i+1)}, B_{(i+1),i}, \Phi_i$ all
have $D_0 = Q = 1$, and the Weyl spinor field
$\overline{\psi_{\Phi,i}}$ has $D_0 = Q = \frac{3}{2}$, we
therefore conclude that the Hilbert space of gauge invariant
operators truncates to a subsector spanned by multi-trace
operators of the form
\begin{equation}
\prod_{j=1}^k \hspace{0.3mm} \Tr \big( W_{1
\boldsymbol{\to}}^{(j)} W_{2 \boldsymbol{\to} }^{(j)} \cdots
W_{J_j \boldsymbol{\to}}^{(j)} \big)
\end{equation}
where any letter $W^{(j)}_{i_j \boldsymbol{\to}}$ is either one of
the scalars $A_{i,(i+1)}, B_{(i+1),i}, \Phi_i$, or the Weyl spinor
$\overline{\psi_{\Phi,i}}$. Otherwise, the notation is as
explained below (\ref{eq:SU(2)-sector}).

It would be interesting to study this subsector further and
determine its symmetry group. This group is presumably a subgroup
of the $SU(2|3)$ symmetry observed in the $\mathcal{N}=4$ SYM case
\cite{Harmark:2006di}, and determined by the way the
$\mathbb{Z}_M$ orbifolding breaks the embedding of $SU(2|3)$ into
the full $\mathcal{N}=4$ superconformal group $PSU(2,2|4)$.

In \cite{Harmark:2006di} the authors considered weakly coupled
$\mathcal{N}=4$ $U(N)$ SYM theory on $S^1 \times S^3$ with
$R$-symmetry chemical potentials in similar near-critical regions
of the phase diagram as studied here. It was found that the
Hilbert space of gauge invariant operators truncates to similar
subsectors as identified here, namely the 1/2 BPS sector, the
$SU(2)$ subsector or the $SU(2|3)$ subsector, depending on which
chemical potentials are turned on. Furthermore, the analysis in
\cite{Harmark:2006di} was generalized to small, but non-zero 't
Hooft coupling $\lambda$ by utilizing the 1-loop correction $D_2$
to the dilatation operator (cf. the perturbative expansion
(\ref{eq:dilat.op(pert.expansion)})). In the large $N$ limit,
$D_2$ restricted to the $SU(2)$ subsector becomes the Hamiltonian
of an $SU(2)$ spin chain; and restricted to the $SU(2|3)$
subsector it becomes the Hamiltonian of an $SU(2|3)$ spin chain.
What is remarkable is that in both these cases, the spin chains
are integrable \cite{Minahan:2002ve, Beisert:2003tq,
Beisert:2004ry}, and that the truncated Hilbert spaces can be
identified with subsectors of the complete dilatation operator of
$\mathcal{N}=4$ $U(N)$ SYM theory that are expected to be closed
to any order in perturbation theory.

For $\mathcal{N}=2$ quiver gauge theory, the full dilatation
operator along with possible integrable subsectors is not yet
completely settled, so we are not able to immediately generalize
our results to small, but non-zero 't Hooft coupling $\lambda$.
However, we note that much progress has been made in this area. In
particular, anomalous dimensions of various operators, the
anomalous dimension matrix restricted to various subsectors, Bethe
ans\"{a}tze and integrability have been investigated in
\cite{Itzhaki:2002kh, Alishahiha:2002ev, Kim:2002fp, Mukhi:2002ck,
Bertolini:2002nr, Wang:2003cu, DiVecchia:2004jw, Ideguchi:2004wm,
DeRisi:2004bc, Sadri:2005gi, Beisert:2005he, Astolfi:2006is}.

\section{One-loop quantum effective action with scalar VEV's}

In this section we will extend the matrix model for
$\mathcal{N}=2$ quiver gauge theory on $S^1 \times S^3$ in Section
3 to include non-zero VEV's for the scalar fields. To this end we
calculate the quantum effective action at weak 't Hooft coupling
to 1 loop in a slice of the configuration space of the background
fields. To simplify the calculation we restrict to the case of
zero $R$-symmetry chemical potentials. The potential we compute
will be valid within the temperature range $0 \leq TR \ll
\lambda^{-1/2}$. The origin of the bound $TR \ll \lambda^{-1/2}$
comes from the fact that $R^{-1}$ provides a cutoff on the
momentum integrals that appear in the loop diagrams that contribute to the
effective action. Provided that $R^{-1}$ is much larger than the inverse Debye
length, one avoids infrared divergences which would require a resummation of the
thermal mass of the fields.

The method employed for computing the effective potential will be
the standard background field formalism. That is, we expand the
quantum fields about classical background fields and path
integrate over the fluctuations, discarding terms of cubic or
higher order in the fluctuations. The background fields will be
taken to be static and spatially homogeneous; thus, the potential
obtained from the computation will be a static effective
potential. Furthermore, we carry out the computation only in a
slice of the configuration space in which the background fields
are mutually ``commuting'' in a sense that conforms to the quiver
structure.

We now proceed with a more detailed description of the
calculation. For convenience we first rescale all the fields in
the $\mathcal{N}=2$ quiver gauge theory Lagrangian density (as
given in Eqs. (\ref{eq:quiverLagrangian}),
(\ref{eq:quiveraction(scalar,f)}),
(\ref{eq:quiveraction(gauge,f)}) and
(\ref{eq:quiveraction(spinor,f)})) with a factor of
$g_\mathrm{YM}^{\phantom{1}}$ as follows
\begin{equation}
\phi \hspace{2mm} \longrightarrow \hspace{2mm}
\frac{1}{g_\mathrm{YM}^{\phantom{1}}} \hspace{0.5mm} \phi \: .
\label{eq:redef-of-fields-with-gYM}
\end{equation}
We then expand the quantum fields about classical background
fields by applying the following transformations to the Lagrangian
density
\begin{eqnarray}
A_{i,(i+1)} \hspace{1mm} &\longrightarrow& \hspace{1mm}
A_{i,(i+1)} \hspace{0.4mm} + \hspace{0.4mm}  a_{i,(i+1)} \label{eq:adding_background_field(a)}\\
B_{(i+1),i} \hspace{1mm} &\longrightarrow& \hspace{1mm}
B_{(i+1),i} \hspace{0.4mm} + \hspace{0.4mm}  b_{(i+1),i} \label{eq:adding_background_field(b}\\
\Phi_i \hspace{1mm} &\longrightarrow& \hspace{1mm}
\Phi_i \hspace{0.4mm} + \hspace{0.4mm}  \phi_i \label{eq:adding_background_field(phi)}\\
A_{\mu \hspace{0.4mm} i} \hspace{1mm} &\longrightarrow&
\hspace{1mm} A_{\mu \hspace{0.4mm} i} \hspace{0.4mm} +
\hspace{0.4mm} \delta_{\mu 0} \hspace{0.4mm} \alpha_i \: .
\label{eq:adding_background_field(alpha)}
\end{eqnarray}
The background fields $a_{i,(i+1)}, b_{(i+1),i}, \phi_i$ and
$\alpha_i$ are assumed to solve the Euler-Lagrange EOM's so that
they are the VEV's of the corresponding fluctuating fields. We
take the background fields to be static and spatially homogeneous,
i.e. constant on $S^1 \times S^3$. This is to preserve the $SO(4)$
isometry of $S^3$ as we will not examine the more exotic phases in
which the vacuum spontaneously breaks rotational invariance.

The terms of the Lagrangian density arising after the
transformations
(\ref{eq:adding_background_field(a)})-(\ref{eq:adding_background_field(alpha)})
are grouped by their order in the fluctuating fields. The terms of
zeroth order are grouped into a tree-level Lagrangian density. The
terms linear in the fluctuating fields combine to vanish as the
background fields are solutions to the Euler-Lagrange EOM's. We
discard terms containing fluctuating fields to cubic or higher
order.\footnote{Note that with the redefinition of fields in
(\ref{eq:redef-of-fields-with-gYM}), discarding terms of cubic or
higher order in the fluctuations is the analog of taking the
$g_\mathrm{YM}^{\phantom{1}} \longrightarrow 0$ limit in Section
3.1.} The quantum corrections to the tree-level Lagrangian density
thus arise from path integrations over the terms quadratic in the
fluctuations. The result will thus be valid to 1-loop order in the
loop expansion.

It is technically difficult to compute the quantum corrections to
the effective potential for arbitrary background fields. We will
therefore only carry out the computation assuming that the
background fields satisfy the constraints given below. These
constraints are analogous to requiring that the background fields
commute, while at the same time they respect the quiver structure
of the theory.

First, the Polyakov loops must ``commute'' with the scalar VEV's:
\begin{eqnarray}
\alpha_i \hspace{0.5mm} a_{i,(i+1)} - a_{i,(i+1)} \hspace{0.5mm}
\alpha_{i+1} &=&  0 \: , \hspace{2cm} \alpha_{i+1} \hspace{0.5mm}
\overline{a_{i,(i+1)}}
- \overline{a_{i,(i+1)}} \hspace{0.5mm} \alpha_i \hspace{0.5mm} = \hspace{0.5mm} 0 \nonumber \\
\alpha_{i+1} \hspace{0.5mm} b_{(i+1),i} - b_{(i+1),i}
\hspace{0.5mm} \alpha_i &=&  0 \: , \hspace{2.08cm} \alpha_i
\hspace{0.5mm} \overline{b_{(i+1),i}} -
\overline{b_{(i+1),i}} \hspace{0.5mm} \alpha_{i+1} \hspace{0.5mm} = \hspace{0.5mm} 0 \label{eq:constraints1} \\
\big[ \alpha_i, \phi_i \big] &=& 0 \: , \hspace{4.73cm} \big[
\alpha_i, \overline{\phi_i} \big] \hspace{0.5mm} = \hspace{0.5mm}
0 \: . \nonumber
\end{eqnarray}
Second, the scalar VEV's must ``commute'' among themselves:
\begin{eqnarray}
a_{i,(i+1)} \hspace{0.5mm} \overline{a_{i,(i+1)}} -
\overline{a_{(i-1),i}} \hspace{0.5mm} a_{(i-1),i} &=& 0 \: ,
\hspace{1.1cm} b_{i,(i-1)} \hspace{0.5mm} \overline{b_{i,(i-1)}} -
\overline{b_{(i+1),i}} \hspace{0.5mm}
b_{(i+1),i} \hspace{0.5mm} = \hspace{0.5mm} 0  \phantom{aaaaa} \nonumber \\
\big[ \phi_i, \overline{\phi_i} \big] &=& 0 \: , \hspace{1.03cm}
a_{(i-1),i} \hspace{0.5mm} \overline{b_{(i+1),i}} -
\overline{b_{i,(i-1)}}
\hspace{0.5mm} a_{i,(i+1)} \hspace{0.5mm} = \hspace{0.5mm} 0  \nonumber \\
\overline{a_{i,(i+1)}} \hspace{0.5mm} b_{i,(i-1)} - b_{(i+1),i}
\hspace{0.5mm} \overline{a_{(i-1),i}} &=& 0 \: , \hspace{2.12cm}
a_{i,(i+1)} \hspace{0.5mm} \overline{\phi_{i+1}} -
\overline{\phi_i} \hspace{0.5mm}
a_{i,(i+1)} \hspace{0.5mm} = \hspace{0.5mm} 0  \nonumber \\
\overline{a_{i,(i+1)}} \hspace{0.5mm} \phi_i - \phi_{i+1}
\hspace{0.5mm} \overline{a_{i,(i+1)}} &=& 0 \: , \hspace{2.19cm}
b_{(i+1),i} \hspace{0.5mm} \overline{\phi_i} -
\overline{\phi_{i+1}} \hspace{0.5mm}
b_{(i+1),i} \hspace{0.5mm} = \hspace{0.5mm} 0  \label{eq:constraints2} \\
\overline{b_{(i+1),i}} \hspace{0.5mm} \phi_{i+1} - \phi_i
\hspace{0.5mm} \overline{b_{(i+1),i}} &=& 0 \: , \hspace{1.03cm}
a_{i,(i+1)} \hspace{0.5mm}
b_{(i+1),i} - b_{i,(i-1)} \hspace{0.5mm} a_{(i-1),i} \hspace{0.5mm} = \hspace{0.5mm} 0 \nonumber \\
\overline{a_{(i-1),i}} \hspace{0.8mm} \overline{b_{i,(i-1)}} -
\overline{b_{(i+1),i}} \hspace{0.8mm} \overline{a_{i,(i+1)}} &=& 0
\: , \hspace{2.12cm} a_{i,(i+1)}
\hspace{0.5mm} \phi_{i+1} - \phi_i \hspace{0.5mm} a_{i,(i+1)} \hspace{0.5mm} = \hspace{0.5mm} 0 \nonumber \\
\overline{a_{i,(i+1)}} \hspace{0.5mm} \overline{\phi_i} -
\overline{\phi_{i+1}} \hspace{0.5mm} \overline{a_{i,(i+1)}} &=& 0
\: , \hspace{2.2cm} b_{(i+1),i}
\hspace{0.5mm} \phi_i - \phi_{i+1} \hspace{0.5mm} b_{(i+1),i} \hspace{0.5mm} = \hspace{0.5mm}  0 \nonumber \\
\overline{b_{(i+1),i}} \hspace{0.8mm} \overline{\phi_{i+1}} -
\overline{\phi_i} \hspace{0.8mm} \overline{b_{(i+1),i}} &=& 0 \: .
\nonumber
\end{eqnarray}
\\
Since the zero modes $a_{i,(i+1)}, b_{(i+1),i}, \phi_i$ and
$\alpha_i$ are constant over $S^1 \times S^3$, the tree-level
action is obtained from the tree-level Lagrangian density by
simply multiplying the volume of $S^1 \times S^3$. After imposing
the constraints (\ref{eq:constraints1})-(\ref{eq:constraints2})
the tree-level action takes the form
\begin{equation}
S^{(0)} \hspace{0.7mm} = \hspace{1.2mm} \frac{2\pi^2 \beta
R}{g_\mathrm{YM}^2} \hspace{0.8mm} \sum_{i=1}^M \hspace{0.5mm} \Tr
\Big( a_{i,(i+1)} \hspace{0.5mm} \overline{a_{i,(i+1)}} +
b_{(i+1),i} \hspace{0.5mm} \overline{b_{(i+1),i}} + \phi_i
\hspace{0.5mm} \overline{\phi_i} \Big) \: .
\label{eq:tree-level_action}
\end{equation}

We choose an $R_\xi$ gauge defined by adding the gauge fixing
action
\begin{eqnarray}
S_{\mathrm{g.f.}} &=& \frac{1}{g_\mathrm{YM}^2} \frac{1}{2\xi}
\sum_{i=1}^M \hspace{-0.5mm} \int \hspace{-0.5mm} d^4 x \sqrt{|g|}
\hspace{0.4mm} \Tr \hspace{-0.3mm} \Bigg[
\partial_\mu A_{\mu i} \hspace{-0.2mm} + \hspace{-0.2mm} i
[\alpha_i,A_{0i}] \hspace{-0.2mm} + \hspace{-0.2mm} i\xi \Bigg(
\hspace{-1mm} \Big( \overline{a_{(i-1),i}} A_{(i-1),i}
\hspace{-0.2mm} -
\hspace{-0.2mm} A_{i,(i+1)} \overline{a_{i,(i+1)}} \Big) \nonumber \\
&\phantom{=}& \hspace{0.4cm} + \, \Big( a_{i,(i+1)} \hspace{0.5mm}
\overline{A_{i,(i+1)}} - \overline{A_{(i-1),i}} \hspace{0.5mm}
a_{(i-1),i} \Big) + \Big( \overline{b_{(i+1),i}} \hspace{0.5mm}
B_{(i+1),i} - B_{i,(i-1)} \hspace{0.5mm} \overline{b_{i,(i-1)}} \Big) \nonumber \\
&\phantom{=}& \hspace{0.4cm} + \, \Big( b_{i,(i-1)} \hspace{0.5mm}
\overline{B_{i,(i-1)}} - \overline{B_{(i+1),i}} \hspace{0.5mm}
b_{(i+1),i} \Big) + \big[ \overline{\phi_i}, \Phi_i \big] + \big[
\phi_i, \overline{\Phi_i} \big] \hspace{0.3mm} \Bigg) \Bigg]^2 \:
. \label{eq:gauge_fixing_action}
\end{eqnarray}
We will furthermore choose the Feynman gauge $\xi = 1$ for
convenience. The virtue of this gauge fixing action is that, using
(\ref{eq:constraints1})-(\ref{eq:constraints2}), it cancels terms
appearing in the Lagrangian density after the transformations
(\ref{eq:adding_background_field(a)})-(\ref{eq:adding_background_field(alpha)})
that contain both gauge field and scalar field fluctuations. Thus,
one can do the path integrations over the gauge field fluctuations
and over the scalar field fluctuations separately.

\subsubsection*{Specification of the vacuum}

We will restrict to the case where all the zero modes
$a_{i,(i+1)}, b_{(i+1),i}, \phi_i$ and $\alpha_i$ are taken to be
diagonal $N\times N$ matrices.\footnote{When the VEV's are allowed
to be off-diagonal, satisfying the constraints
(\ref{eq:constraints1})-(\ref{eq:constraints2}) along with the
quiver $M$-periodicity (i.e., $a_{(i+M),(i+M+1)} = a_{i,(i+1)}$
etc.) ultimately leads to relations between $N$ and $M$, such as
$N \hspace{0.5mm} | \hspace{0.5mm} M$.} The most general ansatz
satisfying all the constraints
(\ref{eq:constraints1})-(\ref{eq:constraints2}) is given by
\begin{eqnarray}
a_{i,(i+1)} &=& \mathrm{diag} \big( e^{i \theta_1^i} ,
\hspace{0.4mm} \ldots, \hspace{0.4mm} e^{i \theta_N^i} \big) \hspace{0.5mm} a_{(i-1),i} \label{eq:quivervacuum(a,apriori)}\\
b_{(i+1),i} &=& \mathrm{diag} \big( e^{-i \theta_1^i} ,
\hspace{0.4mm} \ldots, \hspace{0.4mm} e^{-i \theta_N^i} \big) \hspace{0.5mm} b_{i,(i-1)} \label{eq:quivervacuum(b,apriori)} \\
\phi_i &=& \phi_{i+1} \label{eq:quivervacuum(phi,apriori)} \\
\alpha_i &=& \alpha_{i+1} \: .
\label{eq:quivervacuum(alpha,apriori)}
\end{eqnarray}
If we furthermore require the vacuum to respect the gauge
invariance and quiver translational invariance of the action along
with the quiver $M$-periodicity, the most general form is
\begin{eqnarray}
a_{i,(i+1)} &=& \omega^k \hspace{0.4mm} a_{(i-1),i} \label{eq:quivervacuum(a)}\\
b_{(i+1),i} &=& \omega^{-k} \hspace{0.4mm} b_{i,(i-1)} \label{eq:quivervacuum(b)} \\
\phi_i &=& \phi_{i+1} \label{eq:quivervacuum(phi)} \\
\alpha_i &=& \alpha_{i+1} \label{eq:quivervacuum(alpha)}
\end{eqnarray}
where $\omega = e^{2\pi i /M}$ and $k \in \mathbb{Z}$. This is the
vacuum we will adhere to in the computations throughout this
section. We will find that the expression for the quantum
effective action is independent of the value of $k$ in
(\ref{eq:quivervacuum(a)})-(\ref{eq:quivervacuum(b)}).

\subsection{Quantum corrections from bosonic fluctuations}

There are radiative corrections to the tree-level potential coming
from path integrations over the part of the action that is
quadratic in the bosonic fluctuations. Below we present in a
bilinear form the part of the action that is quadratic in the
bosonic fluctuations, as it appears after being added to the gauge
fixing action (\ref{eq:gauge_fixing_action}) and the Fadeev-Popov
ghost action, and the constraints
(\ref{eq:constraints1})-(\ref{eq:constraints2}) have been imposed.
The path integrals will then be Gaussian and can be evaluated
easily.

First we introduce some notation. Define
\begin{eqnarray}
\mathbf{A}_{\mu \hspace{0.4mm} mn} &\equiv& \left( \begin{array}{c} (A_{\mu \hspace{0.4mm} 1})_{mn} \\
\vdots \\ (A_{\mu \hspace{0.4mm} M})_{mn} \end{array} \right) \: ,
\hspace{1.77cm} \mathbf{A}_{mn} \hspace{0.05cm} \equiv
\hspace{0.05cm} \left( \begin{array}{c} (A_{1,2})_{mn} \\
\vdots \\ (A_{M,1})_{mn} \end{array} \right) \: , \\
\mathbf{B}_{mn} &\equiv& \left( \begin{array}{c} (B_{1,M})_{mn} \\
\vdots \\ (B_{M,(M-1)})_{mn} \end{array} \right) \: ,
\hspace{1cm} \boldsymbol{\Phi}_{mn} \hspace{0.05cm} \equiv
\hspace{0.05cm} \left( \begin{array}{c} (\Phi_1)_{mn} \\
\vdots \\ (\Phi_M)_{mn} \end{array} \right)
\end{eqnarray}
so that, e.g.,
\begin{equation}
(\mathbf{A}^T)_{mn} = \Big( (A_{1,2})_{mn}, \, \ldots, \,
(A_{M,1})_{mn} \Big) \hspace{0.7cm} \mathrm{and} \hspace{0.7cm}
\mathbf{A}^*_{mn}
= \left( \begin{array}{c} (\overline{A_{1,2}})_{nm} \\
\vdots \\ (\overline{A_{M,1}})_{nm} \end{array} \right) \: .
\end{equation}
Furthermore, we define for fixed $m,n$ the fluctuation operators
$\Box_g^{mn}, \Box_\mathbf{A}^{mn}, \Box_\mathbf{B}^{mn}$ and
$\Box_\mathbf{\Phi}^{mn}$ as certain $M \times M$ matrices
(labelled by $i,j=1,\ldots,M$) whose detailed form is given in
(\ref{eq:fluctuation-operator(Box-g)})-(\ref{eq:fluctuation-operator(Box-Phi)}).
Then the part of the action that is quadratic in the bosonic
fluctuations (including the Fadeev-Popov ghosts
$\overline{c_i},c_i$) can be written in the form ($k = 1,2,3$)
\begin{eqnarray}
S_b &=& \frac{1}{g_\mathrm{YM}^2} \sum_{m,n=1}^N \int d^4 x
\sqrt{|g|} \hspace{0.9mm} \Big( \textstyle{\frac{1}{2}}
(\mathbf{A}_k^{\bot \, T})_{mn} \hspace{0.4mm} \Box_g^{mn}
\hspace{0.4mm} (\mathbf{A}_k^\bot)_{nm} + \textstyle{\frac{1}{2}}
(\partial_k \mathbf{F}^T)_{mn} \hspace{0.4mm} \Box_g^{mn}
\hspace{0.4mm} (\partial_k \mathbf{F})_{nm} \phantom{aaaaa} \nonumber \\
&\phantom{=}& \hspace{4.0cm} + \textstyle{\frac{1}{2}}
\hspace{0.5mm} (\mathbf{A}_0^T)_{mn} \, \Box_g^{mn} \,
\mathbf{A}_{0 \hspace{0.4mm} nm} + \,
(\overline{\mathbf{c}}^T)_{mn} \, \Box_g^{mn} \, \mathbf{c}^*_{mn}
\nonumber \\
&\phantom{=}& \hspace{1.8cm} + \hspace{1.0mm} (\mathbf{A}^T)_{mn}
\, \Box_\mathbf{A}^{mn} \, \mathbf{A}^*_{mn} \, + \,
(\mathbf{B}^T)_{mn} \, \Box_\mathbf{B}^{mn} \, \mathbf{B}^*_{mn} +
\hspace{1.0mm} (\boldsymbol{\Phi}^T)_{mn} \,
\Box_{\boldsymbol{\Phi}}^{mn} \, {\boldsymbol{\Phi}}^*_{mn} \Big)
\label{eq:quadratic-part-of-bos.-action}
\end{eqnarray}
where, as in Section 3.1, the spatial components of the gauge
field have been decomposed into a transversal (i.e.,
divergenceless) part $(A_i^\bot)^k$ and a longitudinal part
$(\nabla F_i)^k$. Thereby all the fields have been written in
terms of $S^3$ spherical harmonics. The path integrations over the
bosonic fluctuations $A_{i,(i+1)}, B_{(i+1),i}, \Phi_i$ and
$A_{\mu \hspace{0.4mm} i}$ can now readily be done and yield the
formal expression\footnote{We are using a rather sloppy notation
here as the term involving $\Box_g^{mn}$ is to be interpreted as
the total contribution from the path integrations over the
transversal and longitudinal parts of the spatial components of
the gauge field, the time component of the gauge field and the
Fadeev-Popov ghosts. The individual contributions are explicitly
written out in (\ref{eq:quantum_effective_action(bos,apriori)})
below.}
\begin{eqnarray}
\Gamma_\mathrm{bos}\big[ \hspace{0.2mm} \alpha_i , \hspace{0.2mm}
a_{i,(i+1)}, \hspace{0.2mm} b_{(i+1),i}, \hspace{0.2mm} \phi_i
\hspace{0.2mm} \big] &=& \frac{1}{2} \hspace{0.5mm} \sum_{m,n=1}^N
\hspace{0.2mm} \Tr \hspace{0.4mm} \ln \hspace{0.4mm} \det
\Box_g^{mn} \, + \,
\sum_{m,n=1}^N \hspace{0.2mm} \Tr \hspace{0.4mm} \ln \hspace{0.4mm}
\det \Box_\mathbf{A}^{mn} \phantom{aaaaaa}\nonumber \\
&\phantom{=}& + \hspace{1mm} \sum_{m,n=1}^N \hspace{0.2mm} \Tr
\hspace{0.4mm} \ln \hspace{0.4mm} \det \Box_\mathbf{B}^{mn} \, +
\, \sum_{m,n=1}^N \hspace{0.2mm} \Tr \hspace{0.4mm} \ln
\hspace{0.4mm} \det \Box_{\boldsymbol{\Phi}}^{mn} \: .
\label{eq:formal_bosonic_QEA}
\end{eqnarray}
Here the traces are taken over the Matsubara frequencies and over
the $S^3$ spherical harmonics, and the determinants are taken over
the $i,j$ indices of the operators $\Box_g^{mn},
\Box_\mathbf{A}^{mn}, \Box_\mathbf{B}^{mn}$ and
$\Box_{\boldsymbol{\Phi}}^{mn}$. Let us define here for
convenience
\begin{eqnarray}
v_{i,j; \hspace{0.7mm} n,m} &\equiv& 2 \hspace{0.3mm} \Big( \big(
(a_{i,(i+1)})_{nn} - \omega^{-j} \hspace{0.3mm}
(a_{i,(i+1)})_{mm}\big) \big( (\overline{a_{i,(i+1)}})_{nn} -
\omega^j \hspace{0.3mm} (\overline{a_{i,(i+1)}})_{mm}\big) \nonumber \\
&\phantom{=}& \hspace{0.5cm} + \hspace{0.8mm} \big(
(b_{(i+1),i})_{nn} - \omega^j \hspace{0.3mm} (b_{(i+1),i})_{mm}
\big) \big( (\overline{b_{(i+1),i}})_{nn} - \omega^{-j}
\hspace{0.3mm} (\overline{b_{(i+1),i}})_{mm} \big) \nonumber \\
&\phantom{=}& \hspace{0.5cm} + \hspace{0.8mm} \big( (\phi_i)_{nn}
- (\phi_i)_{mm} \big) \big( (\overline{\phi_i})_{nn} -
(\overline{\phi_i})_{mm} \big) \Big) \: .
\label{eq:VEV_structure_of_QEA}
\end{eqnarray}
Now we apply the determinant formula
(\ref{eq:determinant_circulant_matrix}) to the formal expression
(\ref{eq:formal_bosonic_QEA}) for $\Gamma_\mathrm{bos}$. Then we
take the traces over the Matsubara frequencies and over the $S^3$
spherical harmonics, labelled by the angular momentum $h$ (see
Table 1). This yields the following result
\begin{eqnarray}
\Gamma_\mathrm{bos} &=& \frac{1}{2M} \hspace{0.5mm} \sum_{i,j=1}^M
\sum_{m,n=1}^N \sum_{k=-\infty}^\infty \Tr_{h \geq 0}
\hspace{0.5mm} \ln \Big[
\big( \omega_k + (\alpha_i^{nn} - \alpha_i^{mm}) \big)^2 + \Delta_g^2 + v_{i,j; \hspace{0.7mm} n,m} \Big] \nonumber \\
&\phantom{=}& \hspace{-1.5mm} + \hspace{0.6mm} \frac{1}{2M}
\hspace{0.5mm} \sum_{i,j=1}^M \sum_{m,n=1}^N
\sum_{k=-\infty}^\infty \Tr_{h > 0} \hspace{0.5mm} \ln \Big[ \big(
\omega_k + (\alpha_i^{nn} - \alpha_i^{mm}) \big)^2 + \Delta_s^2 +
v_{i,j; \hspace{0.7mm} n,m} \Big] \nonumber \\
&\phantom{=}& \hspace{-1.5mm} + \left( \frac{1}{2} - 1 \right)
\frac{1}{M} \hspace{0.5mm} \sum_{i,j=1}^M \sum_{m,n=1}^N
\sum_{k=-\infty}^\infty \Tr_{h \geq 0} \hspace{0.5mm} \ln \Big[
\big( \omega_k + (\alpha_i^{nn} - \alpha_i^{mm}) \big)^2 + \Delta_s^2 + v_{i,j; \hspace{0.7mm} n,m} \Big] \nonumber \\
&\phantom{=}& + \hspace{0.5mm} \frac{3}{M} \hspace{0.5mm}
\sum_{i,j=1}^M \sum_{m,n=1}^N \sum_{k=-\infty}^\infty \Tr_{h \geq
0} \hspace{0.5mm} \ln \Big[ \big( \omega_k + (\alpha_i^{nn} -
\alpha_i^{mm}) \big)^2 + \Delta_s^2 + R^{-2} + v_{i,j;
\hspace{0.7mm} n,m} \Big] \: . \nonumber \\
\label{eq:quantum_effective_action(bos,apriori)}
\end{eqnarray}
Here the first line comes from the path integrations over the
transverse part of the spatial gauge field, and the second line
from the integrations over the longitudinal part. The third line
comes from integrating over the temporal component of the gauge
field and the Fadeev-Popov ghosts, contributing with the weights
$\frac{1}{2}$ and $-1$, respectively. Finally, the fourth line
comes from path integrating over the conformally coupled scalar
fluctuations. Note that there is an exact cancellation between the
contributions of all $h > 0$ spherical harmonics in the second and
third line. As we will see in Section 6, the surviving
contribution from the $h=0$ scalar spherical harmonic will be the
dominating radiative correction in the low-temperature regime.

After performing the summations over the Matsubara frequencies and
writing out the traces over the $S^3$ spherical harmonics with the
appropriate eigenvalues of $\nabla^2$ and their degeneracies (cf.
Table 1) we find
\begin{eqnarray}
\Gamma_\mathrm{bos} &=& \frac{1}{2M} \sum_{i,j=1}^M \sum_{m,n=1}^N
\Bigg[ - \beta \big( v_{i,j; \hspace{0.7mm} n,m} \big)^{1/2} + 2
\hspace{0.2mm} \sum_{l=1}^\infty \frac{1}{l} e^{-\beta l (v_{i,j;
\hspace{0.5mm}
n,m})^{1/2}} \cos \big( \beta l (\alpha_i^{nn}- \alpha_i^{mm}) \big) \nonumber \\
&\phantom{=}& \hspace{2.8cm} + \sum_{h=0}^\infty \hspace{0.4mm}
2h(h+2) \hspace{0.3mm} \Bigg( \beta \big( (h+1)^2 R^{-2} + v_{i,j;
\hspace{0.7mm} n,m} \big)^{1/2} \nonumber \\
&\phantom{=}& \hspace{3.5cm} - \hspace{0.8mm} 2 \hspace{0.3mm}
\sum_{l=1}^\infty \frac{1}{l} e^{-\beta l ( (h+1)^2 R^{-2} +
v_{i,j; \hspace{0.5mm} n,m})^{1/2}}
\cos \big( \beta l (\alpha_i^{nn}- \alpha_i^{mm}) \big) \Bigg) \nonumber \\
&\phantom{=}& \hspace{2.8cm} + \hspace{0.9mm} 6 \hspace{0.3mm}
\sum_{h=0}^\infty \hspace{0.4mm}(h+1)^2 \hspace{0.3mm} \Bigg(
\beta \big( (h+1)^2 R^{-2} + v_{i,j; \hspace{0.7mm} n,m} \big)^{1/2} \nonumber \\
&\phantom{=}& \hspace{3.5cm} - \hspace{0.8mm} 2 \hspace{0.3mm}
\sum_{l=1}^\infty \frac{1}{l} e^{-\beta l ( (h+1)^2 R^{-2} +
v_{i,j; \hspace{0.4mm} n,m})^{1/2}} \cos \big( \beta l
(\alpha_i^{nn}- \alpha_i^{mm}) \big) \Bigg) \Bigg] \nonumber
\\ \label{eq:quantum_effective_action(bos)}
\end{eqnarray}
where $v_{i,j; \hspace{0.7mm} n,m}$ is defined in
(\ref{eq:VEV_structure_of_QEA}). This is the complete result for
the contribution to the quantum effective action coming from
bosonic fluctuations.

\subsection{Quantum corrections from fermionic fluctuations}

The fluctuating fermionic fields will also give rise to radiative
corrections that can be computed much along the lines of the
bosonic corrections. It is convenient to carry out the calculation
using $\mathcal{N}=4$ SYM notation for the Weyl spinor fields. The
quiver structure of the action is taken into account by including
appropriate factors $\Omega_c$ in the fluctuation operator as
explained in Appendix B.2.

The fermionic part of the Lagrangian density can be written in
$\mathcal{N}=4$ SYM notation (cf. (\ref{eq:defquiverspinors(nf)}))
in the following bilinear form
\begin{equation}
\mathcal{L}_\mathrm{ferm} = \frac{1}{g_\mathrm{YM}^2}
\sum_{i,j=1}^M \sum_{m,n=1}^N \big( \hspace{0.1mm}
(\overline{\lambda_p})_{i;\hspace{0.3mm} mn}, (\lambda_p)_{i;
\hspace{0.3mm} mn} \big) \hspace{0.5mm}
\mathbf{D}^{mn}_{ij} \hspace{0.3mm} \left( \hspace{-0.6mm}
\begin{array}{c} (\lambda_q)_{j; \hspace{0.3mm} nm} \\
(\overline{\lambda_q})_{j; \hspace{0.3mm} nm} \end{array}
\hspace{-0.5mm} \right) \label{eq:quadratic-part-of-ferm.-action}
\end{equation}
where the fluctuation operator $\mathbf{D}^{mn}_{ij}$ is given in
(\ref{eq:ferm.-fluctuation-operator}). Taking the determinant of
$\mathbf{D}^{mn}_{ij}$ as explained in Appendix B.2, and taking
the traces over the fermionic Matsubara frequencies $\omega_k
\equiv \frac{(2k+1)\pi}{\beta}$ and over the $S^3$ spherical
harmonics, one finds the following result
\begin{eqnarray}
\Gamma_\mathrm{ferm} &=&  -\frac{4}{M} \sum_{i,j=1}^M
\sum_{m,n=1}^N \sum_{h=0}^\infty \hspace{0.5mm} h(h+1)
\hspace{0.3mm} \Bigg( \hspace{-0.3mm} \beta \left( \hspace{-0.5mm}
\left(h + \textstyle{\frac{1}{2}} \right)^2 R^{-2} + v_{i,j;
\hspace{0.7mm} n,m}\right)^{1/2} \nonumber
\\
&\phantom{=}& + \hspace{0.8mm} 2 \hspace{0.3mm}  \sum_{l=1}^\infty
\frac{(-1)^{l+1}}{l} \hspace{0.5mm} e^{-\beta l \left(
\left(h+\frac{1}{2} \right)^2 R^{-2} \hspace{0.7mm} +
\hspace{0.7mm} v_{i,j; \hspace{0.5mm} n,m} \right)^{1/2}} \cos
\big( \beta l (\alpha_i^{nn} - \alpha_i^{mm}) \big)
\hspace{-0.3mm} \Bigg) \phantom{aaaaa}
\label{eq:quantum_effective_action(ferm)}
\end{eqnarray}
where $v_{i,j; \hspace{0.7mm} n,m}$ is defined in
(\ref{eq:VEV_structure_of_QEA}). The factor 4 comes from
performing 4 path integrations. This is the complete result for
the contribution to the quantum effective action coming from
fermionic fluctuations.

We conclude that the quantum effective action of $\mathcal{N}=2$
quiver gauge theory with constant scalar field VEV's satisfying
(\ref{eq:constraints1})-(\ref{eq:constraints2}) is given by
\begin{equation}
\Gamma_\mathrm{eff} \hspace{1.0mm} = \hspace{1.0mm}  S^{(0)} +
\Gamma_\mathrm{bos} + \Gamma_\mathrm{ferm}
\label{eq:quantum-effective-action(total)}
\end{equation}
where $S^{(0)}$ is the tree-level action
\begin{equation}
S^{(0)} = \frac{2\pi^2 \beta R}{g_\mathrm{YM}^2} \hspace{0.8mm}
\sum_{i=1}^M \hspace{0.5mm} \sum_{n=1}^N \Big( (a_{i,(i+1)})_{nn}
\hspace{0.2mm} (\overline{a_{i,(i+1)}})_{nn} + (b_{(i+1),i})_{nn}
\hspace{0.2mm} (\overline{b_{(i+1),i}})_{nn} + (\phi_i)_{nn}
\hspace{0.2mm} (\overline{\phi_i})_{nn} \Big) \: .
\label{eq:tree-level_action(complete)}
\end{equation}
and $\Gamma_\mathrm{bos}$ and $\Gamma_\mathrm{ferm}$ are given in
(\ref{eq:quantum_effective_action(bos)}) and
(\ref{eq:quantum_effective_action(ferm)}), respectively, with
$v_{i,j; \hspace{0.7mm} n,m}$ given in
(\ref{eq:VEV_structure_of_QEA}).

Note that the tree-level potential
(\ref{eq:tree-level_action(complete)}) is attractive, whereas the
1-loop quantum corrections in
(\ref{eq:quantum_effective_action(bos)}) and
(\ref{eq:quantum_effective_action(ferm)}) are repulsive. As we
will see in Section 6, the competition between an attractive and a
repulsive part of the potential will cause the equilibrium
configurations of the eigenvalues of the scalar VEV's to be
hypersurfaces.

\subsection{Generalization to other $\mathbb{Z}_M$ orbifold field theories}

The computations in this section and in Appendix B can immediately
be generalized to field theories obtained as $\mathbb{Z}_M$
projections of $\mathcal{N}=4$ $U(NM)$ SYM theory where the action
of $\mathbb{Z}_M$ is that in (\ref{eq:orbifoldaction}) with
$\omega$ replaced by $\omega^p$ for $p \in \mathbb{Z}$. For these
theories\footnote{These theories have also been considered in,
e.g., Refs. \cite{Kachru:1998ys, Bershadsky:1998mb,
Bershadsky:1998cb, Ideguchi:2004wm}.}, the quantum fields must
satisfy the $\mathbb{Z}_M$ invariance conditions obtained from
(\ref{eq:Z_M-invariance-condition-for-bos-fields}) and
(\ref{eq:Z_M-invariance-condition-for-ferm-fields}) by replacing
$\omega \to \omega^p$. In turn, the fields will take
$\mathbb{Z}_M$ projection invariant forms analogous to
(\ref{eq:projinvariantAmuA})-(\ref{eq:projinvariantBFi}) and
(\ref{eq:projinvariantpsichiA})-(\ref{eq:projinvariantchiBpsiPhi}),
except that the bifundamental fields will have non-zero entries on
the $p$'th super- or sub-diagonal. That is, $A$ and $B$ will have
the non-zero entries $A_{i,(i+p)}$ and $B_{(i+p),i}$,
respectively, and analogously for the respective superpartners
$\chi_A$ and $\chi_B$. As a result, the fluctuation operators
$\Box_g^{mn}, \Box_\mathbf{A}^{mn}, \Box_\mathbf{B}^{mn}$ and
$\Box_\mathbf{\Phi}^{mn}$ in
(\ref{eq:fluctuation-operator(Box-g)})-(\ref{eq:fluctuation-operator(Box-Phi)})
and $\Delta_{ij}$ in (\ref{eq:fluctuation-operator(Delta)}) will
have non-zero entries on the $p$'th super- and sub-diagonals.
Therefore, using the generalized determinant formula\footnote{We
emphasize that the entries of the $M \times M$ matrix in
(\ref{eq:generalized-determinant-formula}) are allowed to be
complex numbers.} (where $\omega \equiv e^{2\pi i/M}$)
\begin{equation}
\det \left( \begin{array}{ccccc} z_1 & z_2 & z_3 & \cdots & z_M \\
z_M & z_1 & z_2 & \cdots & z_{M-1} \\
z_{M-1} & z_M & z_1 & \cdots & z_{M-2} \\
\vdots & \vdots & \vdots & \ddots & \vdots \\
z_2 & z_3 & z_4 & \cdots & z_1
\end{array} \right) = \prod_{j=1}^M \big( z_1 + \omega^j z_2 +
\omega^{2j} z_3 + \cdots + \omega^{(M-1)j} z_M \big)
\label{eq:generalized-determinant-formula}
\end{equation}
we see that the fluctuation determinants factorize as in
(\ref{eq:determinant_circulant_matrix}), with $\omega^{pj}$
replacing $\omega^j$. We conclude that the quantum effective
action of these more general $\mathbb{Z}_M$ orbifold field
theories is given by the expression
(\ref{eq:quantum-effective-action(total)}) where $S^{(0)}$ is
given in (\ref{eq:tree-level_action(complete)}) and
$\Gamma_\mathrm{bos}$ and $\Gamma_\mathrm{ferm}$ are given in
(\ref{eq:quantum_effective_action(bos)}) and
(\ref{eq:quantum_effective_action(ferm)}), respectively. The only
change is that $v_{i,j; \hspace{0.7mm} n,m}$ now takes the form
\begin{eqnarray}
v_{i,j; \hspace{0.7mm} n,m} &\equiv& 2 \hspace{0.3mm} \Big( \big(
(a_{i,(i+p)})_{nn} - \omega^{-pj} \hspace{0.3mm}
(a_{i,(i+p)})_{mm}\big) \big( (\overline{a_{i,(i+p)}})_{nn} -
\omega^{pj} \hspace{0.3mm} (\overline{a_{i,(i+p)}})_{mm}\big) \nonumber \\
&\phantom{=}& \hspace{0.5cm} + \hspace{0.8mm} \big(
(b_{(i+p),i})_{nn} - \omega^{pj} \hspace{0.3mm} (b_{(i+p),i})_{mm}
\big) \big( (\overline{b_{(i+p),i}})_{nn} - \omega^{-pj}
\hspace{0.3mm} (\overline{b_{(i+p),i}})_{mm} \big) \nonumber \\
&\phantom{=}& \hspace{0.5cm} + \hspace{0.8mm} \big( (\phi_i)_{nn}
- (\phi_i)_{mm} \big) \big( (\overline{\phi_i})_{nn} -
(\overline{\phi_i})_{mm} \big) \Big) \: .
\label{eq:VEV-structure-of-generalized-QEA}
\end{eqnarray}

\section{Topology transition and emergent spacetime}

In this section we will find the solutions minimizing the
effective potential computed in Section 5 (given in
(\ref{eq:quantum-effective-action(total)}),
(\ref{eq:tree-level_action(complete)}),
(\ref{eq:quantum_effective_action(bos)}),
(\ref{eq:quantum_effective_action(ferm)}) and
(\ref{eq:VEV_structure_of_QEA})) within the temperature range $0
\leq TR \ll \lambda^{-1/2}$. We stress that, since the effective
action of Section 5 is only valid within a sector of constant
background fields satisfying
(\ref{eq:constraints1})-(\ref{eq:constraints2}), the minima we
find in this section are not the absolute minima of the gauge
theory, and the phase transitions within this sector of background
fields do not necessarily extend to phase transitions in the full
gauge theory (cf. \cite{Aharony:2007rj}). Nonetheless, we will see
that the matrix model of Section 5 exhibits some interesting
dynamics.

The resulting distributions of eigenvalues will preserve the
$SU(2) \times U(1)$ $R$-symmetry of $\mathcal{N}=2$ quiver gauge
theory. As in Ref. \cite{Gursoy:2007np} we believe that due to the
preserved $R$-symmetry, the minima found here are indeed the
global minima of the effective action (within the sector of
constant ``commuting'' VEV's). The key observation needed for
obtaining the solutions is that both in the low-tempera\-ture
regime and above the Hagedorn temperature $T_H$, the eigenvalue
distributions for the scalar VEV's and the Polyakov loop can be
solved for separately. As we will see, the Hagedorn transition
causes a change in the topology of the joint eigenvalue
distribution when the temperature is raised above $T_H$.

\subsection{Low-temperature eigenvalue distribution}

For temperatures low compared to the inverse radius of the $S^3$
(i.e., $TR \ll 1$), one can consistently discard terms in the
quantum effective potential that are suppressed by Boltzmann
factors\footnote{We will verify a posteriori that this procedure
is valid for all temperatures below the Hagedorn temperature.},
and so one obtains the following low-temperature limit of the
effective potential
\begin{eqnarray}
\Gamma_{TR \ll 1} &=& \frac{2\pi^2 \beta R}{g_\mathrm{YM}^2}
\sum_{i=1}^M \sum_{n=1}^N \Big( (a_{i,(i+1)})_{nn}
(\overline{a_{i,(i+1)}})_{nn} + (b_{(i+1),i})_{nn}
(\overline{b_{(i+1),i}})_{nn} + (\phi_i)_{nn}
(\overline{\phi_i})_{nn} \Big) \nonumber \\
&\phantom{=}& \hspace{-0.8cm} - \frac{\beta}{2M} \sum_{i,j=1}^M
\sum_{m,n=1}^N \Big[ \hspace{0.2mm} 2 \hspace{0.3mm} \Big( \big(
(a_{i,(i+1)})_{nn} - \omega^{-j} \hspace{0.3mm}
(a_{i,(i+1)})_{mm}\big) \big( (\overline{a_{i,(i+1)}})_{nn} -
\omega^j \hspace{0.3mm} (\overline{a_{i,(i+1)}})_{mm}\big) \nonumber \\
&\phantom{=}& \hspace{2.2cm} + \hspace{0.8mm} \big(
(b_{(i+1),i})_{nn} - \omega^j \hspace{0.3mm} (b_{(i+1),i})_{mm}
\big) \big( (\overline{b_{(i+1),i}})_{nn} - \omega^{-j}
\hspace{0.3mm} (\overline{b_{(i+1),i}})_{mm} \big) \nonumber \\
&\phantom{=}& \hspace{2.2cm} + \hspace{0.8mm} \big( (\phi_i)_{nn}
- (\phi_i)_{mm} \big) \big( (\overline{\phi_i})_{nn} -
(\overline{\phi_i})_{mm} \big) \Big) \Big]^{1/2} \: .
\label{eq:low-T-QEA}
\end{eqnarray}
We observe that the eigenvalues of the Polyakov loop are not
coupled to the eigenvalues of the scalar VEV's. Therefore, for low
temperatures, the distribution of the Polyakov loop eigenvalues
will be the same as in the case with zero scalar VEV's treated in
Section 4. Thus, we immediately conclude from Section 4.1 that the
eigenvalues $e^{i\alpha_i^{nn}}$ of the Polyakov loop (for $i$
fixed) are uniformly distributed over $S^1$ for any temperature
below the Hagedorn temperature. Note that for a uniform
distribution of the angles $\alpha_i^{nn}$, the terms multiplied
by Boltzmann factors in (\ref{eq:quantum_effective_action(bos)})
and (\ref{eq:quantum_effective_action(ferm)}) vanish exactly.
Therefore we can consistently discard these terms as long as the
temperature is below $T_H$.

In order to find the minima of (\ref{eq:low-T-QEA}) we make the
observation that by making the identifications
\begin{eqnarray}
a_{i,(i+1)} &\cong& \omega^{-1} a_{i,(i+1)} \label{eq:orbifold_identification(a)}\\
b_{(i+1),i} &\cong& \omega \hspace{0.5mm} b_{(i+1),i} \label{eq:orbifold_identification(b)} \\
\phi_i &\cong& \phi_i \label{eq:orbifold_identification(phi)}
\end{eqnarray}
and applying them recursively to (\ref{eq:low-T-QEA}), the
low-temperature effective potential reduces to
\begin{eqnarray}
\Gamma_{TR \ll 1} &=& \frac{2\pi^2 \beta R}{g_\mathrm{YM}^2}
\sum_{i=1}^M \sum_{n=1}^N \Big( (a_{i,(i+1)})_{nn}
(\overline{a_{i,(i+1)}})_{nn} + (b_{(i+1),i})_{nn}
(\overline{b_{(i+1),i}})_{nn} + (\phi_i)_{nn}
(\overline{\phi_i})_{nn} \Big) \nonumber \\
&\phantom{=}& \hspace{-0.8cm} - \frac{\beta}{2} \sum_{i=1}^M
\sum_{m,n=1}^N \Big[ \hspace{0.2mm} 2 \hspace{0.3mm} \Big( \big(
(a_{i,(i+1)})_{nn} - (a_{i,(i+1)})_{mm}\big) \big(
(\overline{a_{i,(i+1)}})_{nn} -
(\overline{a_{i,(i+1)}})_{mm}\big) \nonumber \\
&\phantom{=}& \hspace{2.2cm} + \hspace{0.8mm} \big(
(b_{(i+1),i})_{nn} - (b_{(i+1),i})_{mm} \big) \big(
(\overline{b_{(i+1),i}})_{nn} - (\overline{b_{(i+1),i}})_{mm} \big) \nonumber \\
&\phantom{=}& \hspace{2.2cm} + \hspace{0.8mm} \big( (\phi_i)_{nn}
- (\phi_i)_{mm} \big) \big( (\overline{\phi_i})_{nn} -
(\overline{\phi_i})_{mm} \big) \Big) \Big]^{1/2} \: .
\label{eq:low-T-QEA-after-orbifolding}
\end{eqnarray}
It is important to note that the identifications
(\ref{eq:orbifold_identification(a)})-(\ref{eq:orbifold_identification(phi)})
correspond uniquely to the effective potential. That is, if one
replaces $\omega$ by $\omega^q$ in
(\ref{eq:orbifold_identification(a)})-(\ref{eq:orbifold_identification(b)}),
the potential (\ref{eq:low-T-QEA}) will not reduce to
(\ref{eq:low-T-QEA-after-orbifolding}) for general $M$. To see
this, note that, since all $M$ powers of $\omega$ appear in
(\ref{eq:low-T-QEA}), the order of $\omega^q$ must be $M$. Thus we
must have $\gcd(q,M)=1$ for all $M$ which implies $q=1$.

We now proceed with finding the minima of
(\ref{eq:low-T-QEA-after-orbifolding}). These will be minima of
(\ref{eq:low-T-QEA}) where the identifications
(\ref{eq:orbifold_identification(a)})-(\ref{eq:orbifold_identification(phi)})
have been made. It is convenient to introduce the dimensionless
variables
\begin{equation}
(\theta_i)_n \hspace{0.7mm} \equiv \hspace{0.7mm} \beta
(\alpha_i)_{nn} \: , \label{eq:defoftheta}
\end{equation}
\begin{equation}
(z_i)_{n,1} \hspace{0.7mm} \equiv \hspace{0.7mm} \beta
(\phi_i)_{nn} \: , \hspace{1.0cm} (z_i)_{n,2} \hspace{0.7mm}
\equiv \hspace{0.7mm} \beta (a_{i,(i+1)})_{nn} \: , \hspace{1.0cm}
(z_i)_{n,3} \hspace{0.7mm} \equiv \hspace{0.7mm} \beta
\hspace{0.1mm} (b_{(i+1),i})_{nn} \label{eq:defofz's}
\end{equation}
and
\begin{equation}
(\boldsymbol{z}_i)_n \equiv \big( (z_i)_{n,1}, (z_i)_{n,2},
(z_i)_{n,3} \big)
\end{equation}
so that $(\boldsymbol{z}_i)_n \in \mathbb{C}^3$ for fixed $i$ and
$n$. Furthermore we introduce a norm on $\mathbb{C}^3$ defined by
\begin{equation}
\| \boldsymbol{w} - \boldsymbol{z} \| \hspace{1.5mm} \equiv
\hspace{1.5mm} \left( \sum_{c=1}^3 \big| (w_c) - (z_c) \big|^2
\right)^{1/2}
\end{equation}
where $| \cdot |$ denotes the modulus. Written in this notation,
(\ref{eq:low-T-QEA-after-orbifolding}) takes the form
\begin{equation}
\Gamma_{TR \ll 1} = \frac{2\pi^2 R}{g_\mathrm{YM}^2 \beta}
\sum_{i=1}^M \sum_{n=1}^N \big\| (\boldsymbol{z}_i)_n \big\|^2 -
\frac{1}{\sqrt{2}} \sum_{i=1}^M \sum_{m,n=1}^N \big\|
(\boldsymbol{z}_i)_n - (\boldsymbol{z}_i)_m \big\| \: .
\label{eq:low-T-QEA(norm-notation)}
\end{equation}

We will now take the continuum limit $N \to \infty$ and describe
the eigenvalues of the Polyakov loop and the scalar VEV's by a
joint eigenvalue distribution $\rho_i(\theta_i,\boldsymbol{z}_i)$
proportional to the density of eigenvalues at the point
$(\theta_i, \boldsymbol{z}_i)$ (for some fixed $i$) and normalized
as $\int d\theta_i \hspace{0.4mm} d^3 \boldsymbol{z}_i
\hspace{0.6mm} \rho_i(\theta_i,\boldsymbol{z}_i) = 1$. The
continuum limit is obtained by applying the substitution
\begin{equation}
\frac{1}{N} \sum_{n=1}^N \big[ \cdots \big] \longrightarrow \int
d\theta_i \hspace{0.5mm} d^3 \boldsymbol{z}_i \hspace{0.6mm}
\rho_i (\theta_i, \boldsymbol{z}_i) \hspace{0.4mm} \big[ \cdots
\big] \label{eq:cont.limit(all_VEV's)}
\end{equation}
in analogy with (\ref{eq:cont.limit(Polyakov-loop)}). Here it is
implied that the content of the brackets $\big[ \cdots \big]$
carries an $i$ label. In the continuum limit, the equation of
motion for $\boldsymbol{z}_i$ obtained from
(\ref{eq:low-T-QEA(norm-notation)}) reads
\begin{equation}
\frac{\sqrt{2} \hspace{0.2mm} \pi R}{\lambda \beta} \hspace{0.5mm}
\boldsymbol{z}_i = \int_{D_i} d^3 \boldsymbol{z}_i' \hspace{0.9mm}
\rho_i (\boldsymbol{z}_i') \frac{\boldsymbol{z}_i -
\boldsymbol{z}_i'}{\| \boldsymbol{z}_i - \boldsymbol{z}_i' \|} \:
. \label{eq:low-T-EOM}
\end{equation}
Here $\rho_i (\cdot)$ is defined as the average
$\rho_i(\boldsymbol{z}_i) \equiv \int_{-\pi}^\pi d\theta_i
\hspace{0.5mm} \rho_i(\theta_i,\boldsymbol{z}_i)$, and $D_i
\subseteq \mathbb{C}^3$ denotes the support for $\rho_i$. The
solution to (\ref{eq:low-T-EOM}) is given by the eigenvalue
distribution
\begin{equation}
\rho_i (\boldsymbol{z}_i) = \frac{\delta(\| \boldsymbol{z}_i \| -
r_i)}{2\pi^4 r_i^5} \label{eq:low-T-saddle-point}
\end{equation}
where the radius $r_i$ is given by
\begin{equation}
r_i = \frac{\lambda \beta}{\sqrt{2} \hspace{0.2mm} \pi^3 R}
\frac{1024}{945} \label{eq:radius_of_low-T_S5}
\end{equation}
as can be checked straightforwardly. That is, (\ref{eq:low-T-EOM})
is satisfied for any $\boldsymbol{z}_i$ when the eigenvalues are
distributed uniformly over an $S^5$ with the radius
(\ref{eq:radius_of_low-T_S5}). Since
(\ref{eq:low-T-QEA(norm-notation)}) was obtained from the
low-temperature effective potential (\ref{eq:low-T-QEA}) by making
the orbifold identifications
(\ref{eq:orbifold_identification(a)})-(\ref{eq:orbifold_identification(phi)}),
we thus conclude that the minimum of (\ref{eq:low-T-QEA}) is a
uniform distribution of the eigenvalues of the scalar VEV's over
$S^5/\mathbb{Z}_M$ where the action of $\mathbb{Z}_M$ is precisely
as in (\ref{eq:orbifoldaction}). This is consistent with
\cite{Berenstein:2006yy}, as one should expect in the low
temperature limit where thermal effects are small. Since we found
that the eigenvalues of the Polyakov loop are distributed
uniformly over an $S^1$ for temperatures below the Hagedorn
temperature, we conclude furthermore that the joint eigenvalue
distribution of the scalar VEV's and the Polyakov loop is
$S^5/\mathbb{Z}_M \times S^1$ in this temperature range.

It is remarkable that the eigenvalues of the scalar VEV's localize
to a hypersurface in $\mathbb{C}^3$ rather than spreading out over
the configuration space. The physical origin of the localization
is essentially common for the matrix model developed here and the
matrix model of \cite{Berenstein:2006yy}, namely the competition
between an attractive part of the quantum effective potential, and
a repulsive part where the latter is generated by the path
integrations. We interpret the eigenvalue distribution of the
scalar VEV's as the emergence of the $S^5/\mathbb{Z}_M$ factor of
the holographically dual $AdS_5 \times S^5/\mathbb{Z}_M$ string
theory geometry. Finally we note that the hypersurface
$S^5/\mathbb{Z}_M$ has the isometry group $SU(2) \times U(1)$,
resulting from breaking the $SU(4)$ isometry via the orbifold
identifications
(\ref{eq:orbifold_identification(a)})-(\ref{eq:orbifold_identification(phi)}).
Since this is the full $R$-symmetry group $SU(2)_R \times U(1)_R$
of $\mathcal{N}=2$ quiver gauge theory we believe (cf.
\cite{Gursoy:2007np}) that the minimum found here is indeed the
global minimum of the effective action of Section 5.

\subsection{Eigenvalue distribution above the Hagedorn temperature}

In the matrix model treated in Sections 3 and 4 where the VEV's of
the scalar fields were zero we observed that as the temperature is
increased above $T_H \approx 0.38 \hspace{0.6mm} R^{-1}$, the
Polyakov loop eigenvalue distributions open a gap. In this section
we will examine how this phase transition manifests itself in the
general case with non-zero scalar VEV's.

From the radius (\ref{eq:radius_of_low-T_S5}) one in particular
finds that for low temperatures $\| \boldsymbol{z}_i \| \gg
\lambda$, so that the tree-level term dominates over the quantum
correction by a factor $\sim \frac{\| \boldsymbol{z}_i
\|}{\lambda} \gg 1$. On the other hand, around the Hagedorn
temperature $T_H$ one finds $\| \boldsymbol{z}_i \| \sim \lambda$,
and the tree-level term and the quantum corrections come within
the same order of magnitude. It is therefore natural to re-express
the effective potential in terms of the new variables
\begin{equation}
(\zeta_i)_{n,1} \hspace{0.5mm} \equiv \hspace{0.5mm} \lambda^{-1}
(z_i)_{n,1} \: , \hspace{1.0cm} (\zeta_i)_{n,2} \hspace{0.5mm}
\equiv \hspace{0.5mm} \lambda^{-1} (z_i)_{n,2} \: , \hspace{1.0cm}
(\zeta_i)_{n,3} \hspace{0.5mm} \equiv \hspace{0.5mm} \lambda^{-1}
(z_i)_{n,3} \: . \label{eq:defofzeta's}
\end{equation}
The computations in this section will be valid for temperatures in
the range $0 \leq TR \ll \lambda^{-1/2}$. Since we can no longer
neglect the terms multiplied by Boltzmann factors, we have to
consider the full quantum effective action as computed in Section
5 (given in (\ref{eq:quantum-effective-action(total)}),
(\ref{eq:tree-level_action(complete)}),
(\ref{eq:quantum_effective_action(bos)}),
(\ref{eq:quantum_effective_action(ferm)}) and
(\ref{eq:VEV_structure_of_QEA})). Once again, we apply the
orbifold identifications
(\ref{eq:orbifold_identification(a)})-(\ref{eq:orbifold_identification(phi)}),
and express the result in terms of the variables $\theta_i,
\boldsymbol{\zeta}_i$. However, the rescaling with the 't Hooft
coupling $\lambda$ in (\ref{eq:defofzeta's}) will reorganize the
perturbative expansion of the effective potential into
\begin{equation}
\Gamma_\mathrm{eff} \hspace{1mm} = \hspace{1mm}
\Gamma^{(0)}[\theta_i] + \lambda \hspace{0.4mm}
\Gamma^{(1)}[\theta_i,\boldsymbol{\zeta}_i] +
\mathcal{O}(\lambda^2) \: . \label{eq:QEA(total,reorganized)}
\end{equation}
Here the 0-loop term is
\begin{eqnarray}
\Gamma^{(0)}[\theta_i] &=& \sum_{i=1}^M \sum_{m,n=1}^N
\sum_{l=1}^\infty \frac{1}{l} \Big[ \hspace{0.2mm} 1 - \left(
z_\mathrm{ad}^B(e^{-\beta l R^{-1}};1,1) \hspace{0.5mm} +
\hspace{0.5mm} 2 \hspace{0.2mm} z_\mathrm{bi}^B(e^{-\beta l
R^{-1}};1,1) \right) \phantom{aaaaaaaaaaaaa} \nonumber \\
&\phantom{=}& \hspace{-0.3cm} - \hspace{0.7mm} (-1)^{l+1} \left(
z_\mathrm{ad}^F(e^{-\beta l R^{-1}};1,1) \hspace{0.5mm} +
\hspace{0.5mm} 2 \hspace{0.3mm} z_\mathrm{bi}^F(e^{-\beta l
R^{-1}};1,1) \right) \hspace{-0.7mm} \Big] \cos \hspace{-0.4mm}
\big( l (\theta_i)_n - l(\theta_i)_m \big)
\end{eqnarray}
where $z_\mathrm{ad}^B, z_\mathrm{ad}^F, z_\mathrm{bi}^B,
z_\mathrm{bi}^F$ are given in Eqs. (\ref{eq:spPF(ad,bos)}),
(\ref{eq:spPF(ad,ferm)}), (\ref{eq:spPF(bi,bos)}),
(\ref{eq:spPF(bi,ferm)}), respectively, and $y_1 = y_2 = 1$ in
this case since we are taking $\mu_1 = \mu_2 = 0$ here.

The 1-loop term in (\ref{eq:QEA(total,reorganized)}) is given by
\begin{eqnarray}
\Gamma^{(1)}[\theta_i,\boldsymbol{\zeta}_i] &=& \frac{2\pi^2
RN}{\beta} \sum_{i=1}^M \sum_{n=1}^N \big\|
(\boldsymbol{\zeta}_i)_n \big\|^2 \nonumber
\phantom{aaaaaaaaaaaaaaaaaaaaaaaaaaaaaaaaaaaaaaaa}\\
&\phantom{=}& \hspace{0.1cm} - \frac{1}{\sqrt{2}} \sum_{i=1}^M
\sum_{m,n=1}^N \big\| (\boldsymbol{\zeta}_i)_n -
(\boldsymbol{\zeta}_i)_m \big\| \left( 1 + 2\sum_{l=1}^\infty \cos
\big( l(\theta_i)_n - l(\theta_i)_m \big) \right) .
\label{eq:1-loop_term_of_QEA(total,reorganized)}
\end{eqnarray}
From the expansion (\ref{eq:QEA(total,reorganized)}) it is
immediately obvious that to leading order in $\lambda$ the
$\theta_i$ are unaffected by the $\boldsymbol{\zeta}_i$.
Therefore, to leading order, the eigenvalue distributions of the
$\theta_i$ are the same as they were in the case with zero scalar
VEV's treated in Section 4. The eigenvalue distributions of the
scalar VEV's can therefore be found by minimizing
$\Gamma^{(1)}[\theta_i,\boldsymbol{\zeta}_i]$. Taking the large
$N$ limit of (\ref{eq:1-loop_term_of_QEA(total,reorganized)})
according to (\ref{eq:cont.limit(all_VEV's)}) one finds
\begin{eqnarray}
\frac{1}{N^2}\Gamma^{(1)} &=&  \frac{2\pi^2 R}{\beta}
\hspace{-0.5mm} \sum_{i=1}^M \int d\theta_i \hspace{0.5mm} d^3
\boldsymbol{\zeta}_i \hspace{0.8mm}
\rho_i(\theta_i,\boldsymbol{\zeta}_i) \hspace{0.5mm} \|
\boldsymbol{\zeta}_i \|^2 \nonumber
\phantom{aaaaaaaaaaaaaaaaaaaaaaaaaaaaaaaa}\\
&\phantom{=}& \hspace{0.5cm} - \sqrt{2} \hspace{0.6mm} \pi
\sum_{i=1}^M \int d\theta_i \hspace{0.7mm} d^3\boldsymbol{\zeta}_i
\hspace{0.7mm} d^3\boldsymbol{\zeta}_{i}' \hspace{0.9mm}
\rho_i(\theta_i,\boldsymbol{\zeta}_i) \hspace{0.3mm}
\rho_i(\theta_i,\boldsymbol{\zeta}'_{i}) \hspace{0.7mm} \|
\boldsymbol{\zeta}_i - \boldsymbol{\zeta}'_{i} \| \: .
\label{eq:large_N_limit_of_1-loop_term}
\end{eqnarray}
Here we have used the identity $1 + 2\sum_{l=1}^\infty \cos \big(
l(\theta_i)_n - l(\theta_i)_m \big) = 2\pi \delta\big(
(\theta_i)_n - (\theta_i)_m \big)$ which is simply the Fourier
expansion of the delta function.

Now we proceed to minimize the action
(\ref{eq:large_N_limit_of_1-loop_term}). Since the eigenvalue
distributions for the Polyakov loop and the scalar VEV's can be
solved for separately, the joint eigenvalue distribution
factorizes:
\begin{equation}
\rho_i (\theta_i, \boldsymbol{\zeta}_i) = \frac{\rho_i (\theta_i)
\hspace{0.5mm} \delta\big(\| \boldsymbol{\zeta}_i \| -
r_i(\theta_i) \big)}{\| \boldsymbol{\zeta}_i\|^5 \hspace{0.2mm} (1
+ (dr_i/d\theta_i)^2)^{1/2} \hspace{0.2mm} \mathrm{Vol}(S^5)} \: .
\label{eq:JED_factorizes}
\end{equation}
Inserting (\ref{eq:JED_factorizes}) into the 1-loop term
(\ref{eq:large_N_limit_of_1-loop_term}) one finds
\begin{eqnarray}
\frac{1}{N^2} \Gamma^{(1)} &=& \frac{2\pi^2 R}{\beta} \sum_{i=1}^M
\int d\theta_i \hspace{0.8mm} \rho_i(\theta_i) \hspace{0.5mm} r_i
(\theta_i)^2 - 2\pi C \sum_{i=1}^M \int d\theta_i \hspace{0.7mm}
\rho_i(\theta_i)^2 \hspace{0.2mm} r_i (\theta_i)
\phantom{aaaaaaaaaaaaa} \nonumber \\
&=& \frac{2\pi^2 R}{\beta} \sum_{i=1}^M \int d\theta_i \left[
\hspace{0.2mm} \rho_i (\theta_i) \left( r_i (\theta_i) -
\frac{C\beta}{2\pi R} \hspace{0.3mm} \rho_i (\theta_i) \right)^2 -
\frac{C^2 \beta^2}{4\pi^2 R^2} \hspace{0.3mm} \rho_i (\theta_i)^3
\right]
\end{eqnarray}
where $C = \frac{2048 \sqrt{2}}{945 \pi}$. The final term only
contributes to the 2-loop order distribution of the Polyakov loop
eigenvalues and can therefore be ignored. Hence for a minimum we
have
\begin{equation}
r_i (\theta_i) = \frac{C \beta}{2\pi R} \hspace{0.5mm} \rho_i
(\theta_i) \: . \label{eq:radius-of-S5-over-rho-support}
\end{equation}
As we know from Section 4.2, when the temperature is raised above
the Hagedorn temperature $T_H$, the Polyakov loop eigenvalue
distribution becomes gapped and is thus an interval
$[-\theta_0,\theta_0]$. The scalar VEV eigenvalues are now
distributed uniformly over an $S^5/\mathbb{Z}_M$ fibered over this
interval, with the radius of the $S^5/\mathbb{Z}_M$ at any point
$\theta_i$ in the interval being proportional to the density of
Polyakov loop eigenvalues at $\theta_i$ (for fixed $TR$). The
$S^5/\mathbb{Z}_M$ thus shrinks to zero radius at the endpoints
$\pm \theta_0$ of the interval: the topology of the joint
eigenvalue distribution is an $S^6/\mathbb{Z}_M$ where the
$\mathbb{Z}_M$ is understood to act on the $S^5$ transverse to an
$S^1$ diameter. Thus, the Hagedorn phase transition manifests
itself in the general case of non-zero scalar VEV's as a change in
the topology of the joint eigenvalue distribution
$S^5/\mathbb{Z}_M \times S^1 \longrightarrow S^6/\mathbb{Z}_M$.

In order to understand how the $S^6/\mathbb{Z}_M$ eigenvalue
distribution may be realized in the dual AdS spacetime we first
need to consider the $S^1$ part of the low-temperature
distribution $S^1 \times S^5/\mathbb{Z}_M$. The eigenvalues of the
Wilson line wound around the thermal circle give the positions of
D2-branes\footnote{The D2-branes here are T-dual to the original
D3-branes.} on the T-dual of the thermal circle in thermal
$AdS_5$. As the temperature is raised higher and higher beyond
$T_H$, the Polyakov loop eigenvalues become localized to smaller
and smaller intervals. On the AdS side one therefore finds a
localized D2-brane configuration. It was noted in
\cite{Gursoy:2007np} that a similar localization of D2-branes on a
\emph{spatial} circle, at finite temperature, was investigated in
\cite{Barbon:2004dd} where it was observed to produce a
near-horizon geometry containing a non-contractible $S^6$.
Moreover, it was predicted in \cite{Barbon:2004dd} from
supergravity that a $S^1 \times S^5 \to S^6$ topological
transition of a Gregory-Laflamme type should take place. In the
present case, where the dual spacetime is $AdS_5 \times
S^5/\mathbb{Z}_M$, we expect the appearance of an
$S^6/\mathbb{Z}_M$ in the near-horizon geometry of the localized
configuration of D2-branes on the T-dual of the thermal circle.

We now address the important question regarding the stability of
the saddle points (\ref{eq:low-T-saddle-point}) and
(\ref{eq:JED_factorizes}) against off-diagonal fluctuations. As
one may read off from (\ref{eq:quadratic-part-of-bos.-action}),
the mass of the $ij$ entry of a fluctuating scalar field is
$\sqrt{R^{-2} + \big( (\varphi_a)_{ii} - (\varphi_a)_{jj}
\big)^2}$ (for notational convenience we here use $\varphi_a$
which are real-valued scalar fields related to the complex scalar
fields by (\ref{eq:defofABFi})-(\ref{eq:defofABFibar})). In the
saddle point (\ref{eq:low-T-saddle-point}), one finds from
(\ref{eq:radius_of_low-T_S5}) that
\begin{equation}
\big( (\varphi_a)_{ii} - (\varphi_a)_{jj} \big)^2 \sim
\frac{\lambda^2}{R^2}
\end{equation}
and so the ratio of masses of an off-diagonal fluctuation to a
diagonal fluctuation is
$\frac{m_\mathrm{off-diag}}{m_\mathrm{diag}} \sim \sqrt{1 +
\mathcal{O}(\lambda^2)}$. For small $\lambda$ this ratio is very
close to 1. A priori it thus appears possible for the fluctuating
fields to have off-diagonal elements in this background, causing
the background to be unstable.

Despite this we believe that the sector of constant and
`commuting' VEV's is interesting since it may have a connection
with the dominant saddle points at strong 't Hooft coupling (where
the assumption $[\varphi_a, \varphi_b] = 0$ seems natural). It is
also worth remarking that at zero temperature and strong 't Hooft
coupling one finds an $S^5/\mathbb{Z}_M$ distribution of the
scalar VEV eigenvalues \cite{Berenstein:2006yy}. In a different
vein, the topological phase transition $S^5/\mathbb{Z}_M \times
S^1 \longrightarrow S^6/\mathbb{Z}_M$ provides a natural extension
of the Hagedorn/deconfinement phase transition for the Polyakov
loop eigenvalues studied in Section 4, and it is a tantalizing
question whether it extends to a phase transition of the full
gauge theory at weak 't Hooft coupling.

Finally it should be noted that the effective action
(\ref{eq:quantum-effective-action(total)}) has other saddle points
in which off-diagonal fluctuations can have parametrically large
masses. (Indeed, compare with \cite{Hollowood:2006xb} where these
saddle points were associated with the Gregory-Laflamme
instability in the gravity dual theory.) These saddle points are
therefore guaranteed to be stable backgrounds. In particular, the
effective action (\ref{eq:quantum-effective-action(total)}) has
interesting physics beyond the saddle points studied in this
section.

\subsubsection*{Generalization to other $\mathbb{Z}_M$ orbifold
field theories}

The computations in this section immediately carry over to the
more general $\mathbb{Z}_M$ orbifold field theories considered in
Section 5.3. In this paragraph we remark on the theory defined by
letting the action of $\mathbb{Z}_M$ be that of
(\ref{eq:orbifoldaction}) with $\omega$ replaced by $\omega^p$ for
some fixed $p \in \mathbb{Z}$. The quantum effective action of the
corresponding field theory is obtained from that of
$\mathcal{N}=2$ quiver gauge theory by defining $v_{i,j;
\hspace{0.7mm} n,m}$ to be as given in
(\ref{eq:VEV-structure-of-generalized-QEA}). The minima of this
effective action are found by making the orbifold identifications
\begin{equation}
a_{i,(i+1)} \cong \omega^{-p} a_{i,(i+1)} \: , \hspace{1.2cm}
b_{(i+1),i} \cong \omega^p \hspace{0.5mm} b_{(i+1),i} \: ,
\hspace{1.2cm} \phi_i \cong \phi_i \: .
\label{eq:generalized-orbifolding}
\end{equation}
The resulting expression for the effective action is then
precisely the same as in the case of $\mathcal{N}=2$ quiver gauge
theory treated in this section, and the conclusions carry directly
over. In particular, having made the orbifold identifications
(\ref{eq:generalized-orbifolding}), one finds the low-temperature
joint eigenvalue distribution $S^5 \times S^1$ and the
high-temperature distribution $S^6$. Alternatively, the joint
eigenvalue distributions are $S^5/\mathbb{Z}_M \times S^1$ and
$S^6/\mathbb{Z}_M$, respectively, where the action of
$\mathbb{Z}_M$ is precisely the orbifold action defining the
$\mathbb{Z}_M$ orbifold theory. It is important to note that the
orbifold identifications (\ref{eq:generalized-orbifolding})
correspond uniquely to the quantum effective action of the field
theory. Indeed, assume that we make the identifications
(\ref{eq:generalized-orbifolding}) with some $\omega^q$ replacing
$\omega^p$. In order for the quantum effective action to reduce to
an expression involving norms on $\mathbb{C}^3$ we must require
$\omega^q$ to have the same order as $\omega^p$. That is, we must
have $\forall M \in \mathbb{N}: \gcd(q,M) = \gcd(p,M)$ which
implies $q = p$. Identifying the above $S^5/\mathbb{Z}_M$
distribution with the $S^5/\mathbb{Z}_M$ part of the
holographically dual $AdS_5 \times S^5/\mathbb{Z}_M$ spacetime,
this shows in particular that, within this class of $\mathbb{Z}_M$
orbifold field theories, the geometry of the dual AdS spacetime is
mirrored in the structure of the quantum effective action in a
precise way.

\section{Discussion and conclusions}

In this paper we have investigated different aspects of the phase
structure of $\mathcal{N}=2$ $U(N)^M$ quiver gauge theories. We
have set up a matrix model for $\mathcal{N}=2$ quiver gauge
theories on $S^1 \times S^3$ with chemical potentials conjugate to
the $R$-charges. We then found the stable saddle points of the
model as a function of temperature and chemical potentials. More
specifically, we identified a low and a high temperature phase
separated by a threshold temperature $T_H(\mu)$ which marks a
Hagedorn/deconfinement phase transition. The condition of
stability of the low-temperature saddle point was translated into
a phase diagram of $\mathcal{N}=2$ quiver gauge theory as a
function of both temperature and chemical potentials. We observed
that in regions of small temperature and near-critical chemical
potential the Hilbert space of gauge invariant operators truncates
to the $SU(2)$ subsector, or to a larger subsector whose symmetry
group has yet to be determined. More specifically, we found the
$SU(2)$ subsector when the chemical potential corresponding to the
$SU(2)_R$ factor of the $R$-symmetry group $SU(2)_R \times U(1)_R$
is turned on, whereas the larger subsector emerged from turning on
both chemical potentials and setting them equal.

We then developed the matrix model of $\mathcal{N}=2$ quiver gauge
theory in a different direction, allowing non-zero VEV's for the
scalar fields, but setting the $R$-symmetry chemical potentials to
zero. We did this by computing a 1-loop effective potential for
constant and ``commuting'' VEV's, valid at weak 't Hooft coupling
and in the temperature range $0 \leq TR \ll \lambda^{-1/2}$.  We
furthermore obtained the effective potential for more general
$\mathbb{Z}_M$ orbifold field theories by an immediate
generalization. Then we found the equilibrium configurations of
the eigenvalues of the Polyakov loop and the scalar VEV's. The
eigenvalues of the scalar VEV's localize to a hypersurface in
$\mathbb{C}^3$ due to a repulsive part of the effective potential
of a Vandermonde type, originating from the quantum corrections.
We found that at the Hagedorn temperature the topology of the
joint distribution of the eigenvalues undergoes a phase transition
$S^1 \times S^5/\mathbb{Z}_M \to S^6/\mathbb{Z}_M$. Finally, we
identified the $S^5/\mathbb{Z}_M$ part of the low-temperature
eigenvalue distribution as the emergence of the $S^5/\mathbb{Z}_M$
part of the holographically dual geometry $AdS_5 \times
S^5/\mathbb{Z}_M$. It should be noted, though, that the latter
is a dominant geometrical saddle at strong 't Hooft coupling while the
``commuting'' saddle found from the effective potential at weak 't Hooft coupling
is not an absolute minimum \cite{Aharony:2007rj, Gursoy:2007np}.
Extrapolating this identification to high temperatures,
we furthermore note that the dual spacetime interpretation of
the high-temperature $S^6/\mathbb{Z}_M$ is at present not entirely clear.
We have also generalized the analysis to a class of $\mathbb{Z}_M$ orbifold
field theories, thereby finding that the geometry of the dual AdS
spacetime is similarly mirrored in the structure of the quantum effective
action in a precise way.

There are several interesting future directions to pursue. It
would be interesting to investigate other vacua of $\mathcal{N}=2$
quiver gauge theory which preserve less $R$-symmetry. In
particular, such vacua could prove important when the matrix model
with non-zero scalar VEV's developed in this paper is extended to
include $R$-symmetry chemical potentials.

One could also consider the gravity duals of the phase transitions
studied in this paper. In particular, the effective potential we
computed in Section 5 can be used to study the manifestation of
Gregory-Laflamme instability from the weakly coupled gauge theory
point of view \cite{Hollowood:2006xb}. This would proceed along
the lines of \cite{Hollowood:2006xb} where, above a critical
temperature $T_c \gg T_H$, the effective potential computed for
$\mathcal{N}=4$ SYM theory was observed to develop new unstable
directions along the scalar directions accompanied by new saddle
points which only preserve an $SO(5)$ subgroup of the global
$SO(6)$ isometry group. This phenomenon was identified as the weak
coupling version of the Gregory-Laflamme localization instability
of the small $AdS_5$ black hole in the gravity dual of the
strongly coupled gauge theory.

Furthermore, the results obtained in this paper can be applied to
compute the Polyakov-Maldacena loop \cite{Maldacena:1998im} at
weak coupling. It can also be computed at strong coupling
following \cite{Hartnoll:2006hr}, so this is an interesting object
to compare at weak and strong coupling.

It would be very interesting to study the subsectors of the
Hilbert space of gauge invariant operators that we identified in
Section 4.3 in further detail, in particular to determine the
symmetry group of the subsector corresponding to turning on both
chemical potentials and setting them equal. A further point of
particular interest would be to examine whether these subsectors
are closed under the action of the full dilatation operator in
analogy with \cite{Harmark:2006di}. More generally, these results
could prove useful to further investigate the corresponding spin
chain for the $\mathcal{N}=2$ quiver gauge theory.

Another direction to pursue would be to examine, following
\cite{Harmark:2006ta, Harmark:2006ie}, whether the Hagedorn
temperature of $\mathcal{N}=2$ quiver gauge theory in the
near-critical limit combined with the triple scaling limit of
\cite{Bertolini:2002nr} can be matched to the Hagedorn temperature
of Type IIB string theory on a pp-wave with a compact spacelike
circle. This would involve computing the spectrum of a certain
subsector of the $SU(2)$ sector of gauge invariant operators which
are dual to strings that wind about and/or have oscillators in the
compact direction. This might require finding novel Bethe
ans\"atze since the ground states of the spin chain governing the
truncated $SU(2)$ sector appear to be inherently different from,
say, ferromagnetic ground states.

\section*{Acknowledgements}

We would like to thank Poul Damgaard, Charlotte Kristjansen, Herbert Neuberger,
Marta Orselli and especially Troels Harmark and Paolo Di Vecchia for useful
discussions and comments. KJL wishes to thank Stefan Rozental and
Hanna Kobylinski Rozental's Foundation for financial support. The
work of NO is partially supported by the European Community's
Human Potential Programme under contract MRTN-CT-2004-005104
`Constituents, fundamental forces and symmetries of the universe'.

\appendix

\section{Detailed description of $\mathcal{N}=2$ quiver gauge theory}

This appendix is intended to give a detailed description of
$\mathcal{N}=2$ quiver gauge theory, including details which the
authors have not found elsewhere in the literature.

\subsection{Relation to $\mathcal{N}=4$ SYM theory}

In this section we give a detailed description of how
$\mathcal{N}=2 \:\: U(N)^M$ quiver gauge theory can be obtained by
applying a $\mathbb{Z}_M$ projection to $\mathcal{N}=4 \:\: U(NM)$
SYM theory.

Consider Type IIB string theory and introduce a stack of $NM$
coincident D3-branes into the 10-dimensional (initially flat)
spacetime. It is well known that the low-energy effective field
theory of open strings with endpoints attached to the D3-branes is
4-dimensional $\mathcal{N}=4$ SYM theory with gauge group $U(NM)$.
The space transverse to the world volume of the D3-branes is
$\mathbb{R}^6 \cong \mathbb{C}^3$ which has the isometry group
$SO(6)$. Now we consider the action of the subgroup $\mathbb{Z}_M$
on $\mathbb{C}^3$ given by
\begin{equation}
(z_1, z_2, z_3) \:\, \longrightarrow \:\, (z_1, \omega^{-1} z_2,
\omega z_3) \: , \qquad \qquad \omega \equiv e^{2\pi i/M} \: .
\label{eq:orbifoldaction}
\end{equation}
The group $\mathbb{Z}_M$ is called the \emph{orbifold group}. We
will denote the resulting quotient of $\mathbb{C}^3$ by
$\mathbb{C}^3/\mathbb{Z}_M$ where it is implied that the action of
$\mathbb{Z}_M$ on $\mathbb{C}^3$ is always that given in Eq.
(\ref{eq:orbifoldaction}).

Consider now open strings living on the stack of D3-branes where
the transverse space is $\mathbb{C}^3/\mathbb{Z}_M$. The
low-energy effective field theory is no longer $\mathcal{N}=4 \:\:
U(NM)$ SYM theory. This is because associated with the orbifold
group action (\ref{eq:orbifoldaction}) on the coordinates of
$\mathbb{C}^3$ there is an orbifold group action on the scalar
fields and their superpartners (to be defined below), and we must
require that all quantum fields of $\mathcal{N}=4$ SYM theory be
invariant under this action. The gauge theory obtained from
$\mathcal{N}=4$ SYM theory by truncating the Hilbert spaces of
quantum fields to $\mathbb{Z}_M$-invariant fields is called
\emph{$\mathcal{N}=2$ quiver gauge theory}.

The $R$-symmetry group of $\mathcal{N}=2$ quiver gauge theory is
$SU(2)_R \times U(1)_R$. This is shown explicitly in Appendix A.2
where the Lagrangian density of the quiver gauge theory is
expressed in terms of $SU(2)_R \times U(1)_R$ invariants. The
quiver gauge theory thus indeed has $\mathcal{N}=2$ supersymmetry.

The orbifold group action (\ref{eq:orbifoldaction}) breaks the
gauge group $U(NM)$ of the $\mathcal{N}=4$ theory into
\begin{equation}
U(N)^{(1)} \times U(N)^{(2)} \times \cdots \times U(N)^{(M)}
\end{equation}
which is thus the gauge group of $\mathcal{N}=2$ quiver gauge
theory. We can see this as a manifestation of the fact that the
quiver gauge theory is a low-energy effective field theory of open
strings. Indeed, each of the $M$ copies of
$\mathbb{C}^3/\mathbb{Z}_M$ embedded in $\mathbb{C}^3$ will
contain $N$ coincident D3-branes, and an open string can attach
its endpoints to any of the stacks. Finally, to conclude the
enumeration of the symmetries of $\mathcal{N}=2$ quiver gauge
theory, we note that it is known to be a conformally invariant
theory like the parent $\mathcal{N}=4$ SYM theory
\cite{Kachru:1998ys}.

In order to define the action of the orbifold group $\mathbb{Z}_M$
on the $\mathcal{N}=4$ SYM fields we first set up some notation.
First $\mathbb{Z}_M$ is embedded into $U(NM)$ by defining the
twist matrix $\gamma \equiv \mathrm{diag}(1,\omega, \ldots,
\omega^{M-1})$ and mapping $\mathbb{Z}_M \ni k \longmapsto
\gamma^k \in U(NM)$. (Note that the entries $\omega^j$ of $\gamma$
are really $N \times N$ matrices.)\footnote{Note that this
representation of $\mathbb{Z}_M$ satisfies $\Tr \gamma^k = 0$ for
all $k \in \mathbb{Z}_M \setminus \{ 0 \}.$ As pointed out in Ref.
\cite{Bershadsky:1998mb}, this is needed for consistency (the
cancellation of one-loop open string tadpole diagrams).} The
action of $\mathbb{Z}_M$ on the $\mathcal{N}=4$ SYM fields is then
\begin{equation}
\phi \longrightarrow (\gamma^{k})^\dag \hspace{0.4mm} (\rho \cdot
\phi) \, \gamma^k
\end{equation}
where $\rho \cdot \phi$ equals a phase times the field $\phi$. For
the scalar fields the phase is determined by their identifications
with the $z_1, z_2$ and $z_3$ directions in $\mathbb{C}^3$ and
comparing with (\ref{eq:orbifoldaction}). For the gauge field the
phase is 1. For the spinor fields the phase equals that of their
bosonic superpartner. Thus the condition for the $\mathcal{N}=4$
SYM fields $\phi$ to be invariant under the action of
$\mathbb{Z}_M$ is
\begin{equation}\label{eq:ZM-invarianceforfields}
\phi = \gamma^\dag \hspace{0.4mm} (\rho \cdot \phi) \, \gamma \: .
\end{equation}
In the following we will obtain the Lagrangian density of
$\mathcal{N}=2 \:\: U(N)^M$ quiver gauge theory by rewriting the
$\mathcal{N}=4 \:\: U(NM) $ SYM Lagrangian density and require
that all the fields satisfy the $\mathbb{Z}_M$-invariance
condition (\ref{eq:ZM-invarianceforfields}).

We now consider $\mathcal{N}=4$ $U(NM)$ SYM theory on $\mathbb{R}
\times S^3$ where the radius of $S^3$ is denoted by $R$. The
scalar fields will couple conformally to the curvature of the
$S^3$ through a quadratic term in the action. In the
decompactification limit $R \to \infty$ this term will vanish. The
action of $\mathcal{N}=4 \:\: U(NM)$ SYM theory on $\mathbb{R}
\times S^3$ equipped with a metric of Euclidean signature reads
\begin{eqnarray}
S^{\mathcal{N}=4} &=& \int d^4 x \: \Tr \left( \frac{1}{4} F_{\mu
\nu} F_{\mu \nu} + \frac{1}{2}(D_\mu \phi^i)(D_\mu \phi^i) +
\frac{1}{2}R^{-2}
\phi^i \phi^i - \frac{1}{4} g^2 \big[\phi^i, \phi^j \big] \big[\phi^i, \phi^j\big] \right. \nonumber \\
&\phantom{=}& \phantom{\int d^4 x \: \Tr \Bigg(} \left. \quad + \,
\frac{i}{2} \overline{\psi_p} \gamma_\mu D_\mu \psi_p -
\frac{g}{2} \overline{\psi_p} \big[(\alpha^k_{pq} \phi^{2k-1} +
i\beta^k_{pq} \gamma_5 \phi^{2k}),\psi_q \big] \right)
\label{eq:N=4action}
\end{eqnarray}
where $F_{\mu \nu} = \partial_\mu A_\nu - \partial_\nu A_\mu +
ig[A_\mu, A_\nu]$ and $D_\mu = \partial_\mu + ig[A_\mu, \, \cdot
\, ].$ The traces are taken over the gauge indices. The indices
have the ranges $\mu,\nu = 0,\ldots,3$; $i,j=1,\ldots,6$;
$p,q=1,\ldots,4$ and $k=1,\ldots,3$. Here $\phi^i$ are six real
scalar fields and $\psi_p$ are four 4-component Majorana spinors.
Moreover, $\gamma_\mu$ are the 4-dimensional $4 \times 4$ gamma
matrices and $\alpha^k$ and $\beta^k$ are $4 \times 4$ matrices
satisfying the relations
\begin{equation}
\{ \alpha^k, \alpha^l \} = -2 \delta^{kl} \mathbf{1}_4 \: , \qquad
\{ \beta^k, \beta^l \} = -2 \delta^{kl} \mathbf{1}_4 \: , \qquad
[\alpha^k, \beta^l] = 0 \: .
\label{eq:(anti)commutation_for_alpha_beta}
\end{equation}
Explicit representations can be given as
\begin{equation}\label{eq:alphamatrices}
\alpha^1 = \left( \hspace{-0.05cm} \begin{array}{cc} 0 & \sigma_1 \\
-\sigma_1 & 0 \end{array} \right) \: , \qquad \alpha^2 = \left( \begin{array}{cc} 0 & -\sigma_3 \\
\sigma_3 & 0 \end{array} \right) \: , \qquad \alpha^3 = \left( \begin{array}{cc} i\sigma_2 & 0 \\
0 & i\sigma_2 \end{array} \right) \: ,
\end{equation}
\begin{equation}\label{eq:betamatrices}
\beta^1 = \left( \begin{array}{cc} 0 & i \sigma_2 \\
i \sigma_2 & 0 \end{array} \right) \: , \qquad \beta^2
= \left( \hspace{-0.08cm} \begin{array}{cc} 0 & \mathbf{1}_2 \\
-\mathbf{1}_2 & 0 \end{array} \right) \: ,
\qquad \beta^3 = \left( \hspace{-0.08cm} \begin{array}{cc} -i\sigma_2 & 0 \\
0 & i\sigma_2 \end{array} \right) \: .
\end{equation}

\subsubsection*{The bosonic part of the quiver action}

To put the action of $\mathcal{N}=4$ SYM theory in a form suitable
for performing the orbifold projection we now define three complex
scalar fields
\begin{equation}
A = \frac{1}{\sqrt{2}}(\phi^1 + i\phi^2)\:, \quad \quad B =
\frac{1}{\sqrt{2}}(\phi^3 + i\phi^4)\:, \quad \quad \Phi =
\frac{1}{\sqrt{2}}(\phi^5 + i\phi^6)\: . \label{eq:defofABFi}
\end{equation}
The fields $\phi^i$ are Hermitian (since they transform in the
adjoint representation of the gauge group $U(NM)$), so by
Hermitian conjugation of (\ref{eq:defofABFi}) we find
\begin{equation}
\overline{A} = \frac{1}{\sqrt{2}}(\phi^1 - i\phi^2)\:, \quad \quad
\overline{B} = \frac{1}{\sqrt{2}}(\phi^3 - i\phi^4)\:, \quad \quad
\overline{\Phi} = \frac{1}{\sqrt{2}}(\phi^5 - i\phi^6)\: .
\label{eq:defofABFibar}
\end{equation}
The scalar field part of the $\mathcal{N}=4$ SYM Lagrangian
density written in terms of these fields takes the form
\begin{eqnarray}
\mathcal{L}_\mathrm{scalar}^{\mathcal{N}=4} &=& \Tr \left(
\frac{1}{2} (D_\mu \phi^i) (D_\mu \phi^i) + \frac{1}{2} R^{-2}
\phi^i \phi^i - \frac{1}{4} g^2 \big[\phi^i, \phi^j\big] \big[\phi^i, \phi^j\big] \right) \nonumber \\
&=& \Tr \Big( \hspace{0.4mm} D_\mu A \hspace{0.4mm} D_\mu
\overline{A} + D_\mu \overline{B} \hspace{0.4mm} D_\mu B + D_\mu
\Phi \hspace{0.4mm}
D_\mu \overline{\Phi} \hspace{0.4mm} \Big) \nonumber \\
&\phantom{=}& \: + \, R^{-2} \Tr \Big( A \overline{A} +
\overline{B} B + \Phi \overline{\Phi} \Big) +
\mathcal{L}_D^{\mathcal{N}=4} + \mathcal{L}_F^{\mathcal{N}=4}
\label{eq:quiveraction(scalar,nf1)}
\end{eqnarray}
where the $D$ and $F$ terms are, respectively,
\begin{eqnarray}
\mathcal{L}_D^{\mathcal{N}=4} &=& \frac{1}{2} g^2 \Tr \Big([A,
\overline{A}] + [B,\overline{B}] + [\Phi, \overline{\Phi}] \Big)^2 \label{eq:quiveraction(scalar,nf2)} \\
\mathcal{L}_F^{\mathcal{N}=4} &=& -2 g^2 \Tr \Big([A, B]
\hspace{0.3mm} [\overline{A}, \overline{B}] + [A, \Phi]
\hspace{0.3mm} [\overline{A}, \overline{\Phi}] + [B,\Phi]
\hspace{0.3mm} [\overline{B},\overline{\Phi}]\Big) \: .
\label{eq:quiveraction(scalar,nf3)}
\end{eqnarray}
The scalar fields $\Phi, A$ and $B$ can be identified with the
$z_1, z_2$ and $z_3$ directions of the $\mathbb{C}^3$ (because
they are the Goldstone bosons associated with breaking the
translational invariance in the directions transverse to the
D3-branes), so we have the orbifold group action $\rho :
(\Phi,A,B) \mapsto (\Phi, \omega^{-1} A, \omega B)$, and the
condition for these fields to be invariant under the
$\mathbb{Z}_M$-transformation is then
\begin{equation}
\gamma^\dag \Phi \gamma = \Phi \: , \quad \quad \gamma^\dag A
\gamma = \omega A \: , \quad \quad \gamma^\dag B \gamma =
\omega^{-1} B \: .
\label{eq:Z_M-invariance-condition-for-bos-fields}
\end{equation}
One easily checks that these conditions are satisfied by splitting
the $NM \times NM$ matrix fields of the $\mathcal{N}=4 \:\: U(NM)$
SYM theory up into $M \times M$ block matrices whose entries are
$N \times N$ matrices:
\begin{equation}\label{eq:projinvariantAmuA}
A_\mu = \left( \begin{array}{cccc} A_{\mu 1} & & & \\
& A_{\mu 2} & & \\ & & \ddots & \\ & & & A_{\mu M} \end{array}
\right) \: , \qquad \quad A = \left( \begin{array}{ccccc} 0 & A_{1,2} & & & \\
& 0 & A_{2,3} & & \\ & & \ddots & \ddots & \\ & & & 0 & A_{(M-1),M} \\
A_{M,1} & & & & 0 \end{array} \right) \: ,
\end{equation}
\begin{equation}\label{eq:projinvariantBFi}
B = \left( \begin{array}{ccccc} 0 & & & & B_{1,M} \\
B_{2,1} & 0 & & & \\ & B_{3,2} & \ddots & & \\ & & \ddots & 0 & \\
& & & B_{M,(M-1)} & 0
\end{array} \right) \: , \qquad \quad \Phi = \left( \begin{array}{cccc} \Phi_1 & & & \\
& \Phi_2 & & \\ & & \ddots & \\ & & & \Phi_M \end{array} \right)
\: .
\end{equation}
Here $A_{\mu i}, A_{i,(i+1)}, B_{(i+1),i}$ and $\Phi_i$ are $N
\times N$ matrices (where $i=1,\ldots,M$ and we identify $i \simeq
i + M$). Inserting the $\mathbb{Z}_M$-invariant forms of $A_\mu,
A, B$ and $\Phi$ given in Eqs.
(\ref{eq:projinvariantAmuA})-(\ref{eq:projinvariantBFi}) into
(\ref{eq:quiveraction(scalar,nf1)})-(\ref{eq:quiveraction(scalar,nf3)})
the scalar field part of the $\mathcal{N}=2$ quiver gauge theory
Lagrangian density reads
\begin{eqnarray}
\mathcal{L}_\mathrm{scalar} &=&  \sum_{i=1}^M \Bigg\{
\hspace{0.4mm} \Tr \Big[ \Big( \partial_\mu A_{i,(i+1)} + ig
A_{\mu i} A_{i,(i+1)} - ig A_{i,(i+1)} A_{\mu (i+1)} \Big) \nonumber \phantom{aaaaaaaaaaaaaaaa}\\
&\phantom{=}& \hspace{1.8cm} \times \, \Big(
\partial_\mu \overline{ A_{i,(i+1)} } + ig A_{\mu (i+1)}
\overline{A_{i,(i+1)}} - ig \overline{ A_{i,(i+1)} } A_{\mu i} \Big) \Big] \nonumber \\
&\phantom{=}& + \, \Tr \Big[ \Big( \partial_\mu B_{(i+1),i} +
ig A_{\mu (i+1)} B_{(i+1),i} - ig B_{(i+1),i} A_{\mu i} \Big) \nonumber \\
&\phantom{=}& \hspace{1.6cm} \times \, \Big(
\partial_\mu \overline{B_{(i+1),i}} + ig A_{\mu i} \overline{B_{(i+1),i}} -
ig \overline{B_{(i+1),i}} A_{\mu (i+1)} \Big) \Big] \nonumber \\
&\phantom{=}& + \, \Tr \Big[ \Big(
\partial_\mu \Phi_i + ig [A_{\mu i},\Phi_i] \Big) \Big(
\partial_\mu \overline{\Phi_i} + ig [A_{\mu i},\overline{\Phi_i}] \Big) \Big] \nonumber \\
&\phantom{=}& + \: R^{-2} \Tr \Big( A_{i,(i+1)}
\overline{A_{i,(i+1)}} + \overline{B_{(i+1),i}} B_{(i+1),i} +
\Phi_i \overline{\Phi_i} \Big) \nonumber
\end{eqnarray}
\begin{eqnarray}
&\phantom{=}& + \: \frac{1}{2} g^2 \Tr \Big[ \Big( A_{i,(i+1)}
\overline{A_{i,(i+1)}} - \overline{A_{(i-1),i}} A_{(i-1),i} \nonumber \\
&\phantom{=}& \hspace{2.2cm} + \, B_{i,(i-1)}
\overline{B_{i,(i-1)}} - \overline{B_{(i+1),i}} B_{(i+1),i} +
[\Phi_i,\overline{\Phi_i}] \Big)^2 \Big] \nonumber \\
&\phantom{=}& - \: 2 g^2 \Tr \Big[ \Big( A_{i,(i+1)}
B_{(i+1),i} - B_{i,(i-1)} A_{(i-1),i} \Big) \nonumber \\
&\phantom{=}& \hspace{2.1cm} \times \, \Big(
\overline{A_{(i-1),i}} \: \: \overline{B_{i,(i-1)}} -
\overline{B_{(i+1),i}} \: \: \overline{A_{i,(i+1)}} \Big) \Big] \nonumber \\
&\phantom{=}& - \: 2 g^2 \Tr \Big[ \Big( A_{i,(i+1)} \Phi_{i+1} -
\Phi_i A_{i,(i+1)} \Big) \Big( \overline{A_{i,(i+1)}} \: \:
\overline{\Phi_i} - \overline{\Phi_{i+1}} \: \:
\overline{A_{i,(i+1)}} \Big) \Big] \nonumber \\
&\phantom{=}& - \: 2 g^2 \Tr \Big[ \Big( B_{(i+1),i} \Phi_i -
\Phi_{i+1} B_{(i+1),i} \Big) \Big(\overline{B_{(i+1),i}} \: \:
\overline{\Phi_{i+1}} - \overline{\Phi_i} \: \:
\overline{B_{(i+1),i}} \Big) \Big] \Bigg\} \: .
\label{eq:quiveraction(scalar,f)}
\end{eqnarray}

Inserting the form of $A_\mu$ given in
(\ref{eq:projinvariantAmuA}) into (\ref{eq:N=4action}), the gauge
field part of the $\mathcal{N}=2$ quiver gauge theory Lagrangian
density reads
\begin{equation}\label{eq:quiveraction(gauge,f)}
\mathcal{L}_\mathrm{gauge} = \frac{1}{4} \sum_{i=1}^M
\hspace{0.4mm} \Tr F_{\mu \nu}^i F_{\mu \nu}^i
\end{equation}
where of course $F_{\mu \nu}^i = \partial_\mu A_\nu^i -
\partial_\nu A_\mu^i + ig[A_\mu^i, A_\nu^i]$.

\subsubsection*{The fermionic part of the quiver action}

The fermionic part of the $\mathcal{N}=4$ SYM Lagrangian density
reads
\begin{equation}
\mathcal{L}^{\mathcal{N}=4}_\mathrm{ferm} = \Tr \left( \frac{i}{2}
\hspace{0.5mm} \overline{\psi_p} \gamma_\mu D_\mu \psi_p -
\frac{g}{2} \overline{\psi_p} \big[(\alpha^k_{pq} \phi^{2k-1} +
i\beta^k_{pq} \gamma_5 \phi^{2k}),\psi_q \big] \right)
\label{eq:fermionicpartofN=4Lagrangian}
\end{equation}
where the gamma matrices are given by
\begin{eqnarray}
\gamma_\mu &\equiv& \left( \hspace{-0.1cm} \begin{array}{cc} 0 & \tau_\mu \\
\overline{\tau}_\mu & 0 \end{array} \hspace{-0.1cm} \right) \: ,
\hspace{1.84cm} \gamma_5 \hspace{0.05cm} \equiv \hspace{0.05cm}
\gamma_0 \gamma_1 \gamma_2 \gamma_3 = \left( \hspace{-0.1cm}
\begin{array}{cc} \mathbf{1} & 0 \\ 0 &
-\mathbf{1} \end{array} \hspace{-0.1cm} \right) \label{eq:gammamatrices1}\\
\tau_\mu &\equiv&  (1, \hspace{0.5mm} i\boldsymbol{\sigma}) \: ,
\hspace{2.18cm} \overline{\tau}_\mu \hspace{0.05cm} \equiv
\hspace{0.05cm} (1, \hspace{0.3mm} -i\boldsymbol{\sigma})
\phantom{\Big)} \label{eq:gammamatrices2}
\end{eqnarray}
and representations of $\alpha^k$ and $\beta^k$ are given in Eqs.
(\ref{eq:alphamatrices}) and (\ref{eq:betamatrices}),
respectively. The fields $\psi_p$, $p = 1, \ldots, 4$ are
4-component Majorana spinors which can be decomposed in terms of
2-component Weyl spinors as follows
\begin{equation}
(\psi_p)^a \hspace{0.05cm} \equiv \hspace{0.05cm}
\left( \hspace{-0.1cm} \begin{array}{c} (\lambda_p)_\alpha \\
(\overline{\lambda_p})^{\dot{\alpha}} \end{array} \hspace{-0.1cm}
\right) \: , \hspace{1.5cm} (\overline{\psi_p})_a \hspace{0.05cm}
\equiv \hspace{0.05cm} \left( \hspace{-0.1cm} \begin{array}{c}
(\lambda_p)^\alpha \\ (\overline{\lambda_p})_{\dot{\alpha}}
\end{array} \hspace{-0.1cm} \right) \label{eq:MajoranadecomposetoWeyl}
\end{equation}
where $a=1,\ldots,4$ is the spinor index on $\psi_p$. The Majorana
spinors are related to their conjugates through the Majorana
condition
\begin{equation}
\psi_p = C \hspace{0.4mm} \overline{\psi_p}
\end{equation}
where the Majorana conjugation matrix is $C = \textstyle{\left(
\hspace{-0.1cm} \begin{array}{cc} \epsilon_{\alpha \beta} & 0 \\ 0
& \epsilon^{\dot{\alpha} \dot{\beta}} \end{array} \hspace{-0.1cm}
\right)}$ with $\epsilon_{12} = -\epsilon_{21} = -1.$

Combining Eqs. (\ref{eq:MajoranadecomposetoWeyl}) and
(\ref{eq:gammamatrices1})-(\ref{eq:gammamatrices2}) one finds
\begin{equation}\label{eq:decomposekineticactioninWeyls}
\frac{1}{2} \overline{\psi_p} \gamma_\mu D_\mu \psi_p =
(\lambda_p)^\alpha (\tau_\mu)_{\alpha \dot{\beta}}
\stackrel{\leftrightarrow}{D}_\mu
(\overline{\lambda_p})^{\dot{\beta}} \: .
\end{equation}
Here the operator $\stackrel{\leftrightarrow}{D}_\mu$ is defined
by $\chi_p \hspace{-0.9mm} \stackrel{\leftrightarrow}{D}_\mu
\hspace{-0.9mm} \chi_q \equiv \frac{1}{2} \big( \chi_p D_\mu
\chi_q - (D_\mu \chi_p) \, \chi_q \big)$.

It will be useful for exhibiting the $R$-symmetry of the quiver
gauge theory to express the fermionic Lagrangian density in terms
of the following Weyl spinors
\begin{equation}
\chi_A \equiv \overline{\lambda_1} \: , \qquad \chi_B \equiv
\overline{\lambda_2} \: , \qquad \psi \equiv \overline{\lambda_3}
\: , \qquad \psi_\Phi \equiv \lambda_4 \: .
\label{eq:defquiverspinors(nf)}
\end{equation}
Here $\chi_A, \chi_B, \psi, \psi_\Phi$ are the respective
superpartners of $A, B, A_\mu, \Phi$. Note here that the bar used
over the spinors in (\ref{eq:defquiverspinors(nf)}) is understood
to mean the \emph{Hermitian conjugate} whereas the bar over the
$\lambda_p$ in (\ref{eq:decomposekineticactioninWeyls}) denotes
the usual conjugate of Weyl spinors. Explicitly, letting $\alpha =
1,2$ be the spinor index and letting $m,n$ be the gauge indices,
\begin{equation}
(\lambda_1)_{\alpha,mn} \equiv (\chi_A)^*_{\alpha, nm} =
(\overline{\chi_A})_{\alpha,mn}
\end{equation}
and
\begin{equation}
(\overline{\lambda_1})_{\dot{\alpha},mn} = \left(
(\lambda_1)^*_{\alpha,mn}\right)^T = (\lambda_1)^*_{\alpha,nm} =
(\chi_A)_{\alpha,mn}
\end{equation}
and analogously for $\chi_B, \psi_\Phi$ and $\psi$. In particular,
note that all the Weyl spinors $\chi_A, \chi_B, \psi_\Phi$ and
$\psi$ have undotted indices.

Inserting the definitions (\ref{eq:defquiverspinors(nf)}) into the
decomposition (\ref{eq:decomposekineticactioninWeyls}) we can
write the kinetic part of the fermionic $\mathcal{N}=4$ SYM
Lagrangian density (\ref{eq:fermionicpartofN=4Lagrangian}) in the
form
\begin{eqnarray}
\mathcal{L}_\mathrm{ferm}^{\mathcal{N}=4,\mathrm{kin}} &=&
\frac{i}{2} \Tr \Big( \overline{\psi_p} \gamma_\mu D_\mu \psi_p \Big) \nonumber \\
&=& i \Tr \Big( \overline{\chi_A} \, \tau_\mu \hspace{-0.9mm}
\stackrel{\leftrightarrow}{D}_\mu \hspace{-0.9mm} \chi_A +
\overline{\chi_B} \, \tau_\mu \hspace{-0.9mm}
\stackrel{\leftrightarrow}{D}_\mu \hspace{-0.9mm} \chi_B +
\overline{\psi} \, \tau_\mu \hspace{-0.9mm}
\stackrel{\leftrightarrow}{D}_\mu \hspace{-0.9mm} \psi + \psi_\Phi
\, \tau_\mu \hspace{-0.9mm} \stackrel{\leftrightarrow}{D}_\mu
\hspace{-0.9mm} \overline{\psi_\Phi} \Big) \: . \phantom{aaaaaa}
\end{eqnarray}
In order to find the potential part of the fermionic
$\mathcal{N}=2$ quiver gauge theory Lagrangian density we first
rewrite the analogous part of the $\mathcal{N}=4$ SYM Lagrangian
density (\ref{eq:fermionicpartofN=4Lagrangian}). By inserting the
explicit forms of the $\alpha^k, \beta^k$ matrices given in Eqs.
(\ref{eq:alphamatrices})-(\ref{eq:betamatrices}) into
(\ref{eq:fermionicpartofN=4Lagrangian}) and then decomposing the
4-component Majorana spinors into 2-component Weyl spinors
according to (\ref{eq:MajoranadecomposetoWeyl}) and finally making
the substitutions (\ref{eq:defquiverspinors(nf)}), the
$\mathcal{N}=4$ SYM theory result may be expressed as
\begin{eqnarray}
\mathcal{L}_\mathrm{ferm}^{\mathcal{N}=4} &=& i \Tr \Big(
\overline{\chi_A} \, \tau_\mu \hspace{-0.9mm}
\stackrel{\leftrightarrow}{D}_\mu \hspace{-0.9mm} \chi_A +
\overline{\chi_B} \, \tau_\mu \hspace{-0.9mm}
\stackrel{\leftrightarrow}{D}_\mu \hspace{-0.9mm} \chi_B +
\overline{\psi} \, \tau_\mu \hspace{-0.9mm}
\stackrel{\leftrightarrow}{D}_\mu \hspace{-0.9mm} \psi + \psi_\Phi
\, \tau_\mu \hspace{-0.9mm} \stackrel{\leftrightarrow}{D}_\mu
\hspace{-0.9mm} \overline{\psi_\Phi} \Big) \phantom{aaaaaaaa} \nonumber \\
&\phantom{=}& + \hspace{0.5mm} \frac{g}{\sqrt{2}} \hspace{0.3mm}
\Tr \Big( \hspace{0.3mm} \overline{\chi_A} \hspace{0.3mm} \big(
[A,\psi_\Phi] - [\overline{B},\overline{\psi}] \big)
\hspace{0.5mm} + \hspace{0.5mm} \overline{\chi_B} \hspace{0.3mm}
\big( [\overline{A},\overline{\psi}] + [B,\psi_\Phi] \big)
\hspace{0.5mm} \nonumber \\
&\phantom{=}& \hspace{1.9cm} - \hspace{0.5mm} \overline{\psi}
\hspace{0.2mm} \big( [\overline{A},\overline{\chi_B}] -
[\overline{B},\overline{\chi_A}] \big) - \hspace{0.1cm} \psi_\Phi
\hspace{0.1mm} \big( [A,\overline{\chi_A}] + [B,\overline{\chi_B}]
\big) \nonumber \\
&\phantom{=}& \hspace{1.9cm} + \hspace{0.5mm} \chi_A \big(
[\overline{A},\overline{\psi_\Phi}] - [B,\psi] \big)
\hspace{0.5mm} + \hspace{0.5mm} \chi_B \big(
[A,\psi] + [\overline{B},\overline{\psi_\Phi}] \big) \nonumber \\
&\phantom{=}& \hspace{1.9cm} - \hspace{0.1cm} \psi \big(
[A,\chi_B] - [B,\chi_A] \big) \hspace{0.5mm} - \hspace{0.5mm}
\overline{\psi_\Phi} \hspace{0.1mm} \big( [\overline{A},\chi_A] +
[\overline{B},\chi_B] \big) \nonumber \phantom{\Big)} \\
&\phantom{=}& \hspace{1.9cm} + \hspace{0.1cm} \overline{\chi_A}
\hspace{0.4mm} [\overline{\Phi}, \overline{\chi_B}] \hspace{0.5mm}
- \hspace{0.5mm} \overline{\chi_B} \hspace{0.4mm}
[\overline{\Phi},\overline{\chi_A}] \hspace{0.5mm} +
\hspace{0.5mm} \overline{\psi} \hspace{0.4mm} [\Phi, \psi_\Phi]
\hspace{0.5mm} - \hspace{0.5mm} \psi_\Phi \hspace{0.25mm} [\Phi, \overline{\psi}] \nonumber \\
&\phantom{=}& \hspace{1.9cm} + \hspace{0.1cm} \chi_A
\hspace{0.3mm} [\Phi, \chi_B] \hspace{0.5mm} - \hspace{0.5mm}
\chi_B \hspace{0.3mm} [\Phi, \chi_A] \hspace{0.5mm} +
\hspace{0.5mm} \psi \hspace{0.2mm} [\overline{\Phi},
\overline{\psi_\Phi}] \hspace{0.5mm} - \hspace{0.5mm}
\overline{\psi_\Phi} \hspace{0.4mm} [\overline{\Phi}, \psi]
\hspace{1mm} \Big) \: . \label{eq:quiveraction(ferm,nf)}
\end{eqnarray}
The Weyl spinor fields $\chi_A, \chi_B, \psi_\Phi, \psi$ are the
respective superpartners of $A, B, \Phi, A_\mu$. Therefore they
must satisfy the $\mathbb{Z}_M$-invariance conditions
\begin{equation}
\gamma^\dag \chi_A \gamma = \omega \chi_A \: , \quad \quad
\gamma^\dag \chi_B \gamma = \omega^{-1} \chi_B \: , \quad \quad
\gamma^\dag \psi_\Phi \gamma = \psi_\Phi \: , \quad \quad
\gamma^\dag \psi \gamma = \psi \: .
\label{eq:Z_M-invariance-condition-for-ferm-fields}
\end{equation}
One easily checks that these conditions are satisfied by splitting
the $NM \times NM$ matrix fields of the $\mathcal{N}=4 \:\: U(NM)$
SYM theory up into $M \times M$ block matrices whose entries are
$N \times N$ matrices:
\begin{equation}\label{eq:projinvariantpsichiA}
\psi = \left( \begin{array}{cccc} \psi_1 & & & \\
& \psi_2 & & \\ & & \ddots & \\ & & & \psi_M \end{array} \right)
\: , \qquad \quad \chi_A = \left( \begin{array}{ccccc} 0 & \chi_{A,1} & & & \\
& 0 & \chi_{A,2} & & \\ & & \ddots & \ddots & \\ & & & 0 & \chi_{A,M-1} \\
\chi_{A,M} & & & & 0
\end{array} \right) \: ,
\end{equation}
\begin{equation}\label{eq:projinvariantchiBpsiPhi}
\chi_B = \left( \begin{array}{ccccc} 0 & & & & \chi_{B,M} \\
\chi_{B,1} & 0 & & & \\ & \chi_{B,2} & \ddots & & \\ & & \ddots & 0 & \\
& & & \chi_{B,M-1} & 0
\end{array} \right) \: , \qquad
\psi_\Phi = \left( \begin{array}{cccc} \psi_{\Phi,1} & & & \\
& \psi_{\Phi,2} & & \\ & & \ddots & \\ & & & \psi_{\Phi,M}
\end{array} \right) \: .
\end{equation}
Here $\psi_i, \chi_{A,i}, \chi_{B,i}$ and $\psi_{\Phi,i}$ are $N
\times N$ matrices (where $i=1,\ldots,M$ and we identify $i \simeq
i + M$). Inserting the $\mathbb{Z}_M$-invariant forms of $\psi,
\chi_A, \chi_B$ and $\psi_\Phi$ given in Eqs.
(\ref{eq:projinvariantpsichiA})-(\ref{eq:projinvariantchiBpsiPhi})
into (\ref{eq:quiveraction(ferm,nf)}), the spinor field part of
the $\mathcal{N}=2$ quiver gauge theory Lagrangian density reads
(summation over $i = 1, \ldots, M$ implied)
\begin{eqnarray}
\mathcal{L}_\mathrm{ferm} &=& i \Tr \Big( \overline{\chi_{A,i}} \,
\tau_\mu \hspace{-0.9mm} \stackrel{\leftrightarrow}{D}_\mu
\hspace{-0.9mm} \chi_{A,i} + \overline{\chi_{B,i}} \, \tau_\mu
\hspace{-0.9mm} \stackrel{\leftrightarrow}{D}_\mu \hspace{-0.9mm}
\chi_{B,i} + \overline{\psi_i} \, \tau_\mu \hspace{-0.9mm}
\stackrel{\leftrightarrow}{D}_\mu \hspace{-0.9mm} \psi_i +
\psi_{\Phi,i} \, \tau_\mu \hspace{-0.9mm}
\stackrel{\leftrightarrow}{D}_\mu
\hspace{-0.9mm} \overline{\psi_{\Phi,i}} \Big) \nonumber \phantom{aaaaaaaaa} \\
&\phantom{=}& \hspace{-1.7cm} + \hspace{0.2mm} \frac{g}{\sqrt{2}}
\hspace{0.1mm} \Tr \Big( \overline{\chi_{A,i}} \hspace{0.2mm}
A_{i,(i+1)} \hspace{0.2mm} \psi_{\Phi,(i+1)} -
\overline{\chi_{A,i}} \hspace{0.2mm} \psi_{\Phi,i} \hspace{0.2mm}
A_{i,(i+1)} - \overline{\chi_{A,i}} \hspace{0.2mm}
\overline{B_{(i+1),i}} \hspace{0.2mm} \overline{\psi_{i+1}} +
\overline{\chi_{A,i}} \hspace{0.2mm}
\overline{\psi_i} \hspace{0.2mm} \overline{B_{(i+1),i}} \nonumber \\
&\phantom{=}& \hspace{-0.08mm} + \hspace{1mm}
\overline{\chi_{B,i}} \hspace{0.2mm} \overline{A_{i,(i+1)}}
\hspace{0.2mm} \overline{\psi_i} - \overline{\chi_{B,i}}
\hspace{0.2mm} \overline{\psi_{i+1}} \hspace{0.2mm}
\overline{A_{i,(i+1)}} + \overline{\chi_{B,i}} \hspace{0.2mm}
B_{(i+1),i} \hspace{0.2mm} \psi_{\Phi,i}
- \overline{\chi_{B,i}} \hspace{0.2mm} \psi_{\Phi,(i+1)} \hspace{0.2mm} B_{(i+1),i} \nonumber \\
&\phantom{=}&  \hspace{-0.08mm} - \overline{\psi_{i+1}}
\hspace{0.2mm} \overline{A_{i,(i+1)}} \hspace{0.2mm}
\overline{\chi_{B,i}} + \overline{\psi_i} \hspace{0.2mm}
\overline{\chi_{B,i}} \hspace{0.2mm} \overline{A_{i,(i+1)}} +
\overline{\psi_i} \hspace{0.2mm} \overline{B_{(i+1),i}}
\hspace{0.2mm} \overline{\chi_{A,i}} - \overline{\psi_{i+1}}
\hspace{0.2mm} \overline{\chi_{A,i}} \hspace{0.2mm}
\overline{B_{(i+1),i}} \nonumber \\
&\phantom{=}& \hspace{-0.08mm} - \hspace{1mm} \psi_{\Phi,i}
\hspace{0.2mm} A_{i,(i+1)} \hspace{0.2mm} \overline{\chi_{A,i}} +
\psi_{\Phi,(i+1)} \hspace{0.2mm} \overline{\chi_{A,i}}
\hspace{0.2mm} A_{i,(i+1)} - \psi_{\Phi,(i+1)} \hspace{0.2mm}
B_{(i+1),i} \hspace{0.2mm} \overline{\chi_{B,i}} + \psi_{\Phi,i}
\hspace{0.2mm} \overline{\chi_{B,i}} \hspace{0.2mm}
B_{(i+1),i} \nonumber \\
&\phantom{=}& \hspace{-0.08mm} + \hspace{1mm} \chi_{A,i}
\hspace{0.2mm} \overline{A_{i,(i+1)}} \hspace{0.2mm}
\overline{\psi_{\Phi,i}} - \chi_{A,i} \hspace{0.2mm}
\overline{\psi_{\Phi,(i+1)}} \hspace{0.2mm} \overline{A_{i,(i+1)}}
- \chi_{A,i} \hspace{0.2mm} B_{(i+1),i} \hspace{0.2mm} \psi_i +
\chi_{A,i} \hspace{0.2mm} \psi_{i+1} \hspace{0.2mm}
B_{(i+1),i}\nonumber \\
&\phantom{=}& \hspace{-0.08mm} + \hspace{1mm} \chi_{B,i}
\hspace{0.2mm} A_{i,(i+1)} \hspace{0.2mm} \psi_{i+1} - \chi_{B,i}
\hspace{0.2mm} \psi_i \hspace{0.2mm} A_{i,(i+1)} + \chi_{B,i}
\hspace{0.2mm} \overline{B_{(i+1),i}} \hspace{0.2mm}
\overline{\psi_{\Phi,(i+1)}} - \chi_{B,i} \hspace{0.2mm}
\overline{\psi_{\Phi,i}} \hspace{0.2mm} \overline{B_{(i+1),i}} \nonumber \\
&\phantom{=}& \hspace{-0.08mm} - \psi_i \hspace{0.2mm} A_{i,(i+1)}
\hspace{0.2mm} \chi_{B,i} + \psi_{i+1} \hspace{0.2mm} \chi_{B,i}
\hspace{0.2mm} A_{i,(i+1)} + \psi_{i+1} \hspace{0.2mm} B_{(i+1),i}
\hspace{0.2mm} \chi_{A,i} - \psi_i \hspace{0.2mm} \chi_{A,i}
\hspace{0.2mm} B_{(i+1),i} \nonumber \\
&\phantom{=}& \hspace{-0.08mm} - \hspace{1mm}
\overline{\psi_{\Phi,(i+1)}} \hspace{0.2mm} \overline{A_{i,(i+1)}}
\hspace{0.2mm} \chi_{A,i} + \overline{\psi_{\Phi,i}}
\hspace{0.2mm} \chi_{A,i} \hspace{0.2mm} \overline{A_{i,(i+1)}} -
\overline{\psi_{\Phi,i}} \hspace{0.2mm} \overline{B_{(i+1),i}}
\hspace{0.2mm} \chi_{B,i} + \overline{\psi_{\Phi,(i+1)}}
\hspace{0.2mm} \chi_{B,i} \hspace{0.2mm}
\overline{B_{(i+1),i}} \nonumber \\
&\phantom{=}& \hspace{-0.08mm} + \hspace{1mm}
\overline{\chi_{A,i}} \hspace{0.2mm} \overline{\Phi_i}
\hspace{0.2mm} \overline{\chi_{B,i}} - \overline{\chi_{A,i}}
\hspace{0.2mm} \overline{\chi_{B,i}} \hspace{0.2mm}
\overline{\Phi_{i+1}} - \overline{\chi_{B,i}} \hspace{0.2mm}
\overline{\Phi_{i+1}} \hspace{0.2mm} \overline{\chi_{A,i}} +
\overline{\chi_{B,i}} \hspace{0.2mm} \overline{\chi_{A,i}}
\hspace{0.2mm} \overline{\Phi_i} \nonumber \\
&\phantom{=}& \hspace{-0.08mm} + \hspace{1mm} \chi_{A,i}
\hspace{0.2mm} \Phi_{i+1} \hspace{0.2mm} \chi_{B,i} - \chi_{A,i}
\hspace{0.2mm} \chi_{B,i} \hspace{0.2mm} \Phi_i - \chi_{B,i}
\hspace{0.2mm} \Phi_i \hspace{0.2mm} \chi_{A,i} + \chi_{B,i}
\hspace{0.2mm} \chi_{A,i} \hspace{0.2mm} \Phi_{i+1} \nonumber \\
&\phantom{=}& \hspace{-0.08mm} + \hspace{1mm} \overline{\psi_i}
\hspace{0.2mm} \big[ \Phi_i, \psi_{\Phi,i} \big] - \psi_{\Phi,i}
\hspace{0.2mm} \big[ \Phi_i, \overline{\psi_i} \big] + \psi_i
\hspace{0.2mm} \big[ \overline{\Phi_i}, \overline{\psi_{\Phi,i}}
\big] - \overline{\psi_{\Phi,i}} \hspace{0.2mm} \big[
\overline{\Phi_i}, \psi_i \big] \Big) \: .
\label{eq:quiveraction(spinor,f)}
\end{eqnarray}
We conclude that the Lagrangian density of $\mathcal{N}=2$
$U(N)^M$ quiver gauge theory is
\begin{equation}
\mathcal{L} = \mathcal{L}_\mathrm{scalar} +
\mathcal{L}_\mathrm{gauge} + \mathcal{L}_\mathrm{ferm}
\label{eq:quiverLagrangian}
\end{equation}
where $\mathcal{L}_\mathrm{scalar}, \mathcal{L}_\mathrm{gauge}$
and $\mathcal{L}_\mathrm{ferm}$ are given in Eqs.
(\ref{eq:quiveraction(scalar,f)}),
(\ref{eq:quiveraction(gauge,f)}) and
(\ref{eq:quiveraction(spinor,f)}), respectively.

\subsection{$R$-symmetry}

The Lagrangian density of $\mathcal{N}=2$ quiver gauge theory
(given in Eqs. (\ref{eq:quiverLagrangian}),
(\ref{eq:quiveraction(scalar,f)}),
(\ref{eq:quiveraction(gauge,f)}) and
(\ref{eq:quiveraction(spinor,f)})) is invariant under global
$SU(2)_R \times U(1)_R$ transformations. The $U(1)_R$ factor of
the $R$-symmetry group acts on the fields as
\begin{equation}
A_{i,(i+1)} \: \longrightarrow \: A_{i,(i+1)} \: \: , \quad \qquad
B_{(i+1),i} \: \longrightarrow \: B_{(i+1),i} \: \: , \quad \qquad
\Phi_i \: \longrightarrow \: e^{i\zeta} \Phi_i
\end{equation}
\begin{equation}
\chi_{A,i} \: \longrightarrow \: e^{-i\zeta/2} \chi_{A,i} \: \: ,
\quad \qquad \chi_{B,i} \: \longrightarrow \: e^{-i\zeta/2}
\chi_{B,i}
\end{equation}
\begin{equation}
\psi_i \: \longrightarrow \: e^{i\zeta/2} \psi_i \: \: , \quad
\qquad \psi_{\Phi,i} \: \longrightarrow \: e^{-i\zeta/2}
\psi_{\Phi,i} \: .
\end{equation}
The $U(1)_R$ transformations of the Hermitian conjugate fields are
obtained by flipping $\zeta \to -\zeta$. The Lagrangian density is
manifestly invariant under the $U(1)_R$ transformation.

We now move to consider the $SU(2)_R$ transformations. Define the
2-component spinors
\begin{equation}
(\lambda_i)_a \: \equiv \: \left( \begin{array}{c} A_{i,(i+1)} \\
\overline{B_{(i+1),i}}
\end{array} \right) \: , \qquad \qquad
(\overline{\lambda_i})^a \: \equiv \: \left( \begin{array}{c} \overline{A_{i,(i+1)}} \\
B_{(i+1),i}
\end{array} \right) \: .
\end{equation}
Under $\sigma \in SU(2)_R$ these spinors have the transformations
\begin{eqnarray}
(\lambda_i)_a  &\longrightarrow&  \sigma_a^{\phantom{a}b} \, (\lambda_i)_b \\
(\overline{\lambda_i})^a  &\longrightarrow&
(\overline{\lambda_i})^b \, \overline{\sigma}_b^{\phantom{b}a} \:
.
\end{eqnarray}
Note that $(\overline{\lambda_i})_a = \epsilon_{ab}
(\overline{\lambda_i})^b$ has the transformation
\begin{equation}
(\overline{\lambda_i})_a \: \: \longrightarrow \: \: \epsilon_{ab}
\, \overline{\sigma}_c^{\phantom{c}b} \, \epsilon^{dc} \,
(\overline{\lambda_i})_d = \sigma_a^{\phantom{a}d}
(\overline{\lambda_i})_d
\end{equation}
where the equality follows by using $\sigma \in SU(2)_R$. Thus,
$(\lambda_i)_a$ and $(\overline{\lambda_i})_a$ are $SU(2)_R$
doublets. To exhibit the $SU(2)_R$ invariance of the Lagrangian
density we define $SU(2)_R$ invariants such as
\begin{equation}
(\lambda_i)_a \, (\overline{\lambda_i})^a = -\epsilon^{ab}
(\lambda_i)_a (\overline{\lambda_i})_b = - A_{i,(i+1)}
\overline{A_{i,(i+1)}} - \overline{B_{(i+1),i}} B_{(i+1),i}
\end{equation}
and write the Lagrangian density in terms of these. For
$\mathcal{N}=2$ quiver gauge theory the bifundamental scalars and
the adjoint fermions are organized into  $SU(2)_R$ doublets as
follows
\begin{equation}
(\lambda_i)_a \: \equiv \: \left( \begin{array}{c} A_{i,(i+1)} \\
\overline{B_{(i+1),i}} \end{array} \right) \: , \qquad \qquad
(\overline{\lambda_i})_a \: \equiv \: \left( \begin{array}{c} - B_{(i+1),i} \\
\overline{A_{i,(i+1)}} \end{array} \right)
\label{eq:SU(2)doublets(scalar)}
\end{equation}
\begin{equation}
(\chi_i)_a \: \equiv \: \left( \begin{array}{c} \overline{\psi_i} \\
\psi_{\Phi,i} \end{array} \right) \: , \qquad \qquad
(\overline{\chi_i})_a \: \equiv \: \left( \begin{array}{c} -\overline{\psi_{\Phi,i}} \\
\psi_i \end{array} \right) \label{eq:SU(2)doublets(spinor)} \: .
\end{equation}
The scalar field Lagrangian density written in terms of the
$SU(2)_R$ doublets takes the following form\footnote{Note that the
term $R^{-2} \Tr \big( \epsilon^{ab} (\lambda_i)_a
(\overline{\lambda_i})_b + \Phi_i \overline{\Phi_i} \big)$
describing the conformal coupling of the scalar fields to the
curvature has been omitted here.}

\begin{eqnarray}
\mathcal{L}_\mathrm{scalar} &=& \sum_{i=1}^M \Big[ \Tr
\Big(\epsilon^{ab} (D_\mu \lambda_i)_a (D_\mu
\overline{\lambda_i})_b + D_\mu \Phi_i D_\mu \overline{\Phi_i}
\Big) \nonumber \\
&\phantom{=}& \hspace{0.9cm} + \, \frac{1}{2}g^2 \Tr \Big(
\epsilon^{ab} (\lambda_i)_a (\overline{\lambda_i})_b -
\epsilon^{ab} (\overline{\lambda_{i-1}})_a (\lambda_{i-1})_b +
[\Phi_i,\overline{\Phi_i}] \Big)^2 \nonumber \\
&\phantom{=}& \hspace{0.9cm} - \, 2g^2 \Tr \Big( \epsilon^{ab}
(\lambda_i)_a (\overline{\lambda_i})_b - \epsilon^{ab}
(\overline{\lambda_{i-1}})_a (\lambda_{i-1})_b \Big)^2 \nonumber \\
&\phantom{=}& \hspace{0.9cm} - \, 2g^2 \Tr \Big(
\epsilon^{ab}(\lambda_i)_a (\overline{\lambda_i})_b \epsilon^{cd}
(\overline{\lambda_{i-1}})_c (\lambda_{i-1})_d +
\epsilon^{ab}(\lambda_{i-1})_a (\lambda_i)_b \epsilon^{cd}
(\overline{\lambda_i})_c (\overline{\lambda_{i-1}})_d \Big) \nonumber \\
&\phantom{=}& \hspace{0.9cm} + \, 2g^2 \Tr \Big( \epsilon^{ab}
(\overline{\lambda_i})_a (\lambda_i)_b \epsilon^{cd}
(\overline{\lambda_i})_c (\lambda_i)_d + \epsilon^{ab}
(\lambda_i)_a (\overline{\lambda_i})_b
\epsilon^{cd} (\lambda_i)_c (\overline{\lambda_i})_d \Big) \nonumber \\
&\phantom{=}& \hspace{0.9cm} - \, 2g^2 \Tr \Big( \epsilon^{ab}
(\lambda_i)_a \Phi_{i+1} (\overline{\lambda_i})_b
\overline{\Phi_i} + \epsilon^{ab} (\lambda_i)_a
\overline{\Phi_{i+1}} (\overline{\lambda_i})_b
\Phi_i \Big) \nonumber \\
&\phantom{=}& \hspace{0.9cm} + \, 2g^2 \Tr \Big( \epsilon^{ab}
(\lambda_i)_a \overline{\Phi_{i+1}} \Phi_{i+1}
(\overline{\lambda_i})_b + \epsilon^{ab} (\lambda_i)_a
(\overline{\lambda_i})_b \overline{\Phi_i} \Phi_i \Big) \Big] \: .
\end{eqnarray}
The spinor field Lagrangian density written in terms of the
$SU(2)_R$ doublets takes the following form
\begin{eqnarray}
\mathcal{L}_\mathrm{ferm} &=& \sum_{i=1}^M \Big[ \, i \Tr \Big(
\overline{\chi_{A,i}} \, \tau_\mu \hspace{-0.9mm}
\stackrel{\leftrightarrow}{D}_\mu \hspace{-0.9mm} \chi_{A,i} +
\overline{\chi_{B,i}} \, \tau_\mu \hspace{-0.9mm}
\stackrel{\leftrightarrow}{D}_\mu \hspace{-0.9mm} \chi_{B,i} +
\epsilon^{cd}(\chi_i)_c (\tau_\mu \hspace{-0.9mm}
\stackrel{\leftrightarrow}{D}_\mu \hspace{-0.9mm}
\overline{\chi_i})_d \Big) \nonumber \\
&\phantom{=}& \hspace{1cm} + \frac{g}{\sqrt{2}} \Tr \Big(
\epsilon^{cd} \big\{ \chi_{A,i} \hspace{0.2mm}
(\overline{\lambda_i})_c , \hspace{0.5mm} (\overline{\chi_i})_d
\big\} \hspace{0.8mm} + \hspace{0.5mm} \epsilon^{cd} \big\{
\chi_{A,i}, \hspace{0.5mm} (\overline{\chi_{i+1}})_c
\hspace{0.2mm} (\overline{\lambda_i})_d \big\} \nonumber \\
&\phantom{=}& \hspace{2.7cm} + \hspace{1mm} \epsilon^{cd} \big\{
\hspace{0.2mm} \overline{\chi_{A,i}} \hspace{0.5mm} (\lambda_i)_c
, \hspace{0.6mm} (\chi_{i+1})_d \big\} \hspace{0.6mm} +
\hspace{0.5mm} \epsilon^{cd}\big\{ \hspace{0.2mm}
\overline{\chi_{A,i}}, \hspace{0.6mm} (\chi_i)_c \hspace{0.2mm} (\lambda_i)_d \big\} \nonumber \\
&\phantom{=}& \hspace{2.7cm} + \hspace{1mm} \epsilon^{cd} \big\{
\chi_{B,i} \hspace{0.2mm} (\lambda_i)_c, \hspace{0.6mm}
(\overline{\chi_{i+1}})_d \big\} \hspace{0.8mm} + \hspace{0.6mm}
\epsilon^{cd} \big\{ \chi_{B,i}, \hspace{0.6mm}
(\overline{\chi_i})_c \hspace{0.2mm}
(\lambda_i)_d \big\} \nonumber \phantom{\Big)} \\
&\phantom{=}& \hspace{2.7cm} - \hspace{1mm} \epsilon^{cd} \big\{
\hspace{0.2mm} \overline{\chi_{B,i}} \hspace{0.6mm}
(\overline{\lambda_i})_c, \hspace{0.5mm} (\chi_i)_d \big\}
\hspace{0.6mm} - \hspace{0.5mm} \epsilon^{cd} \big\{
\hspace{0.2mm} \overline{\chi_{B,i}} , \hspace{0.5mm}
(\chi_{i+1})_c \hspace{0.2mm} (\overline{\lambda_i})_d \big\} \nonumber \phantom{\big)} \\
&\phantom{=}& \hspace{2.7cm} + \hspace{1mm} \epsilon^{cd} \big\{
(\chi_i)_c \hspace{0.2mm} \Phi_i, \hspace{0.5mm} (\chi_i)_d \big\}
\hspace{0.6mm} + \hspace{0.6mm} \epsilon^{cd} \big\{
(\overline{\chi_i})_c \hspace{0.4mm} \overline{\Phi_i},
\hspace{0.5mm} (\overline{\chi_i})_d \big\} \nonumber \\
&\phantom{=}& \hspace{2.7cm} + \hspace{1mm} \big\{ \chi_{A,i}
\hspace{0.2mm} \Phi_{i+1}, \hspace{0.6mm} \chi_{B,i} \big\}
\hspace{0.8mm} + \hspace{0.3mm} \big\{ \overline{\chi_{A,i}}
\hspace{0.5mm} \overline{\Phi_i}, \hspace{0.6mm}  \overline{\chi_{B,i}} \big\} \nonumber \\
&\phantom{=}& \hspace{2.7cm} - \hspace{1mm} \big\{ \chi_{B,i}
\hspace{0.2mm} \Phi_i , \hspace{0.6mm} \chi_{A,i} \big\}
\hspace{0.8mm} - \hspace{0.3mm} \big\{ \overline{\chi_{B,i}}
\hspace{0.6mm} \overline{\Phi_{i+1}} , \hspace{0.6mm}
\overline{\chi_{A,i}} \big\} \Big) \Big] \: .
\end{eqnarray}
\vspace{0.1cm}

\noindent These results are conveniently summarized in Table A
which lists the $R$-charges of all the fields in $\mathcal{N}=2$
quiver gauge theory.

\begin{center}
\vspace{0.1cm}
\begin{tabular}{|c|c|c|c|c|c|c|c|c|} \hline & \, $A_{i,(i+1)}$
\, & \, $B_{(i+1),i}$ \, & \, $\Phi_i$ \, & \, $A_{\mu i}$ \, & \,
$\chi_{A,i}$ \, & \, $\chi_{B,i}$ \, & \, $\psi_{\Phi,i}$ \, & \, $\psi_i$ \, \\
\hline \: $U(1)_R$ \phantom{\Big[} & 0 & 0 & 1 & 0 &
$-\frac{1}{2}$ & $ -\frac{1}{2}$ & $-\frac{1}{2}$ & $\frac{1}{2}$ \\
\hline \: $SU(2)_R$ \phantom{\Big[} & $\frac{1}{2}$ &
$\frac{1}{2}$ & 0 & 0 & 0 &
0 & $-\frac{1}{2}$ & $-\frac{1}{2}$ \\
\hline
\end{tabular}
\\
\vspace{0.5cm} {\small{\textbf{Table A.} \textit{$R$-charges for
the bosonic and fermionic fields}}}
\end{center}
Here the generators of $\mathfrak{su}(2)_R$ are taken in the
fundamental representation and chosen as $\frac{1}{2} (\sigma_x,
\sigma_y, \sigma_z)$. The $R$-charges of the corresponding
Hermitian conjugate fields are obtained by simply changing the
signs of the $U(1)_R$ and $SU(2)_R$ charges.

\section{Bosonic and fermionic fluctuation determinants}

In this appendix we present some technical details of the
computation of the 1-loop quantum effective action given in
Section 5. More specifically, we explain here how to evaluate the
fluctuation determinants arising from path integrating over the
fluctuating fields.

\subsection{Bosonic case}

The fluctuation operators $\Box_g^{mn}, \Box_\mathbf{A}^{mn},
\Box_\mathbf{B}^{mn}$ and $\Box_\mathbf{\Phi}^{mn}$ in Eqs.
(\ref{eq:quadratic-part-of-bos.-action})-(\ref{eq:formal_bosonic_QEA})
are given as below.
{\small{\begin{equation} (\Box_g^{mn})_{ij} =
\left\{
\begin{array}{ll} - 2 \, \Big( (\overline{a_{(i-1),i}})_{nn}
\hspace{0.4mm} (a_{(i-1),i})_{mm} + (b_{i,(i-1)})_{nn}
\hspace{0.4mm} (\overline{b_{i,(i-1)}})_{mm}
\Big) & \hspace{0.6cm} \mathrm{for} \:\, j = i - 1 \vspace{0.45cm} \\
- \partial^2 - 2i(\alpha_i^n - \alpha_i^m) \hspace{0.4mm}
\partial_0 + (\alpha_i^n - \alpha_i^m)^2 & \vspace{0.05cm} \\
\hspace{0.3cm} + 2 \hspace{1.0mm} \Big( (a_{i,(i+1)})_{nn}
\hspace{0.4mm} (\overline{a_{i,(i+1)}})_{nn}
+ (a_{(i-1),i})_{mm} \hspace{0.4mm} (\overline{a_{(i-1),i}})_{mm} & \vspace{0.15cm} \\
\hspace{1.2cm} + \, (b_{i,(i-1)})_{nn} \hspace{0.4mm}
(\overline{b_{i,(i-1)}})_{nn} + (b_{(i+1),i})_{mm} \hspace{0.4mm}
(\overline{b_{(i+1),i}})_{mm} & \vspace{0.15cm} \\
\hspace{1.2cm} + \, \big( (\phi_i)_{nn} - (\phi_i)_{mm} \big)
\hspace{0.5mm} \big( (\overline{\phi_i})_{nn}
- (\overline{\phi_i})_{mm}\big) \Big) & \hspace{0.6cm} \mathrm{for} \:\, j = i \vspace{0.45cm} \\
- 2 \, \Big( (a_{i,(i+1)})_{nn} \hspace{0.4mm}
(\overline{a_{i,(i+1)}})_{mm} + (\overline{b_{(i+1),i}})_{nn}
\hspace{0.4mm} (b_{(i+1),i})_{mm} \Big) & \hspace{0.6cm}
\mathrm{for} \:\, j = i + 1
\end{array}\right. \label{eq:fluctuation-operator(Box-g)}
\end{equation}}}
and
{\small{\begin{equation} (\Box_\mathbf{A}^{mn})_{ij} = \left\{
\begin{array}{ll} -2 \, \Big( (\overline{a_{i,(i+1)}})_{nn}
\hspace{0.4mm} (a_{(i-1),i})_{mm} + (b_{(i+1),i})_{nn}
\hspace{0.4mm} (\overline{b_{i,(i-1)}})_{mm}
\Big) & \hspace{-0.7mm} \mathrm{for} \:\, j = i - 1 \vspace{0.45cm} \\
-\partial^2 - 2i(\alpha_{i+1}^n -\alpha_i^m) \hspace{0.4mm}
\partial_0 + (\alpha_{i+1}^n - \alpha_i^m)^2 + R^{-2} & \vspace{0.05cm} \\
\hspace{0.3cm} + \hspace{1.5mm} 2 \hspace{0.4mm} \Big(
(a_{i,(i+1)})_{nn} \hspace{0.4mm} (\overline{a_{i,(i+1)}})_{nn}
+ (a_{i,(i+1)})_{mm} \hspace{0.4mm} (\overline{a_{i,(i+1)}})_{mm} & \vspace{0.15cm} \\
\hspace{1.2cm} + \, (b_{(i+1),i})_{nn} \hspace{0.4mm}
(\overline{b_{(i+1),i}})_{nn} + (b_{(i+1),i})_{mm} \hspace{0.4mm}
(\overline{b_{(i+1),i}})_{mm} & \vspace{0.15cm} \\
\hspace{1.2cm} + \, \big( (\phi_{i+1})_{nn} - (\phi_i)_{mm} \big)
\hspace{0.5mm} \big( (\overline{\phi_{i+1}})_{nn}
- (\overline{\phi_i})_{mm}\big) \Big) & \hspace{-0.7mm} \mathrm{for} \:\, j = i \vspace{0.45cm} \\
-2 \, \Big( (a_{(i+1),(i+2)})_{nn} \hspace{0.4mm}
(\overline{a_{i,(i+1)}})_{mm} + (\overline{b_{(i+2),(i+1)}})_{nn}
\hspace{0.4mm} (b_{(i+1),i})_{mm} \Big) & \hspace{-0.7mm}
\mathrm{for} \:\, j = i + 1
\end{array}\right. \label{eq:fluctuation-operator(Box-A)}
\end{equation}}}
and
{\small{\begin{equation} (\Box_\mathbf{B}^{mn})_{ij} = \left\{
\begin{array}{ll} -2 \, \Big( (\overline{a_{(i-1),i}})_{nn}
\hspace{0.4mm} (a_{i,(i+1)})_{mm} + (b_{i,(i-1)})_{nn}
\hspace{0.4mm} (\overline{b_{(i+1),i}})_{mm}
\Big) & \hspace{-0.7mm} \mathrm{for} \:\, j = i - 1 \vspace{0.45cm} \\
-\partial^2 - 2i(\alpha_i^n -\alpha_{i+1}^m) \hspace{0.4mm}
\partial_0 + (\alpha_i^n - \alpha_{i+1}^m)^2 + R^{-2} & \vspace{0.05cm} \\
\hspace{0.3cm} + \hspace{1.5mm} 2 \hspace{0.4mm} \Big(
(a_{i,(i+1)})_{nn} \hspace{0.4mm} (\overline{a_{i,(i+1)}})_{nn}
+ (a_{i,(i+1)})_{mm} \hspace{0.4mm} (\overline{a_{i,(i+1)}})_{mm} & \vspace{0.15cm} \\
\hspace{1.2cm} + \, (b_{(i+1),i})_{nn} \hspace{0.4mm}
(\overline{b_{(i+1),i}})_{nn} + (b_{(i+1),i})_{mm} \hspace{0.4mm}
(\overline{b_{(i+1),i}})_{mm} & \vspace{0.15cm} \\
\hspace{1.2cm} + \, \big( (\phi_i)_{nn} - (\phi_{i+1})_{mm} \big)
\hspace{0.5mm} \big( (\overline{\phi_i})_{nn}
- (\overline{\phi_{i+1}})_{mm}\big) \Big) & \hspace{-0.7mm} \mathrm{for} \:\, j = i \vspace{0.45cm} \\
-2 \, \Big( (a_{i,(i+1)})_{nn} \hspace{0.4mm}
(\overline{a_{(i+1),(i+2)}})_{mm} + (\overline{b_{(i+1),i}})_{nn}
\hspace{0.4mm} (b_{(i+2),(i+1)})_{mm} \Big) & \hspace{-0.7mm}
\mathrm{for} \:\, j = i + 1
\end{array}\right. \label{eq:fluctuation-operator(Box-B)}
\end{equation}}}
and
{\small{\begin{equation} (\Box_\mathbf{\Phi}^{mn})_{ij} =
\left\{
\begin{array}{ll} -2 \, \Big( (\overline{a_{(i-1),i}})_{nn}
\hspace{0.4mm} (a_{(i-1),i})_{mm} + (b_{i,(i-1)})_{nn}
\hspace{0.4mm} (\overline{b_{i,(i-1)}})_{mm}
\Big) & \hspace{0.6cm} \mathrm{for} \:\, j = i - 1 \vspace{0.45cm} \\
-\partial^2 - 2i(\alpha_i^n - \alpha_i^m) \hspace{0.4mm}
\partial_0 + (\alpha_i^n - \alpha_i^m)^2 + R^{-2} & \vspace{0.05cm} \\
\hspace{0.3cm} + \hspace{1.5mm} 2 \hspace{0.4mm} \Big(
(a_{i,(i+1)})_{nn} \hspace{0.4mm} (\overline{a_{i,(i+1)}})_{nn}
+ (a_{(i-1),i})_{mm} \hspace{0.4mm} (\overline{a_{(i-1),i}})_{mm} & \vspace{0.15cm} \\
\hspace{1.2cm} + \, (b_{i,(i-1)})_{nn} \hspace{0.4mm}
(\overline{b_{i,(i-1)}})_{nn}  + (b_{(i+1),i})_{mm} \hspace{0.4mm}
(\overline{b_{(i+1),i}})_{mm} & \vspace{0.15cm} \\
\hspace{1.2cm} + \, \big( (\phi_i)_{nn} - (\phi_i)_{mm} \big)
\hspace{0.5mm} \big( (\overline{\phi_i})_{nn}
- (\overline{\phi_i})_{mm}\big) \Big) & \hspace{0.6cm} \mathrm{for} \:\, j = i \vspace{0.45cm} \\
-2 \, \Big( (a_{i,(i+1)})_{nn} \hspace{0.4mm}
(\overline{a_{i,(i+1)}})_{mm} + (\overline{b_{(i+1),i}})_{nn}
\hspace{0.4mm} (b_{(i+1),i})_{mm} \Big) & \hspace{0.6cm}
\mathrm{for} \:\, j = i + 1
\end{array}\right. \label{eq:fluctuation-operator(Box-Phi)}
\end{equation}}}
In the general vacuum
(\ref{eq:quivervacuum(a,apriori)})-(\ref{eq:quivervacuum(alpha,apriori)})
these operators are tridiagonal, periodically continued matrices
(assuming $M \geq 3$). The determinant of this class of matrices
was considered in Ref. \cite{braun-1998} (Appendix B) who found
the following result, valid for $M \geq 3$:
\begin{eqnarray}
\det \Box^{mn} &=& \mathrm{tr} \prod_{i=M}^1 \left(
\begin{array}{cc} (\Box^{mn})_{ii} & -(\Box^{mn})_{i,(i-1)} (\Box^{mn})_{(i-1),i} \\
1 & 0 \end{array} \right) \nonumber \\
&\phantom{=}& \hspace{0.5cm} + \hspace{0.7mm} (-1)^{M+1}
\hspace{0.4mm} \mathrm{tr} \prod_{i=M}^1 \left( \begin{array}{cc} (\Box^{mn})_{i,(i-1)} & 0 \\
0 & (\Box^{mn})_{(i-1),i} \end{array} \right) \: .
\label{eq:determinant_of_tridiagonal_matrix}
\end{eqnarray}
The inverse order of the initial and final indices on the product
symbol indicates that the matrix with the highest index $i$ is on
the left of the product.

Fortunately, in the vacuum
(\ref{eq:quivervacuum(a)})-(\ref{eq:quivervacuum(alpha)}) the
fluctuation determinants take a much simpler form. Namely, using
(\ref{eq:quivervacuum(a)})-(\ref{eq:quivervacuum(alpha)}), the
operators $\Box_g^{mn}, \Box_\mathbf{A}^{mn},
\Box_\mathbf{B}^{mn}, \Box_{\boldsymbol{\Phi}}^{mn}$ (for fixed
$m,n$) can be written in the particular form below, and there is a
simple closed expression for the determinant.\footnote{To prove
the formula, note first that the powers of $\omega^k$ appearing in
the super- and subdiagonal mutually cancel according to
(\ref{eq:determinant_of_tridiagonal_matrix}), so the determinant
is independent of the value of $k$. Putting $k=0$, the formula
(\ref{eq:determinant_circulant_matrix}) is a special case of Eq.
(A.1) in Ref. \cite{Nakayama:2005mf}.} That is, defining $\omega
\equiv e^{2\pi i/M}$, we have the determinant formula
\begin{equation}
\det \left( \begin{array}{ccccc} \xi & \hspace{-0.7cm} -\eta &
\phantom{\Big(} & & \hspace{-0.5cm} -\omega^{-k(M-1)}\overline{\eta} \\
-\overline{\eta} & \hspace{-0.7cm} \xi & -\omega^k \eta & & \\
& \hspace{-0.7cm} -\omega^{-k} \overline{\eta} & \xi & \ddots & \\
& & \ddots & \ddots & \hspace{-0.2cm} -\omega^{k(M-2)}\eta \\
-\omega^{k(M-1)}\eta & & & \hspace{-1.4cm}
-\omega^{-k(M-2)}\overline{\eta} & \hspace{-0.5cm} \xi
\end{array}\right) \hspace{1.5mm} = \hspace{1.5mm} \prod_{i=1}^M \hspace{0.2mm}
\big( \xi - \omega^i \hspace{0.2mm} \eta - \omega^{-i}
\hspace{0.2mm} \overline{\eta} \big) \: .
\label{eq:determinant_circulant_matrix}
\end{equation}
Note in particular that the phases $\omega^k$ on the left hand
side cancel out. Therefore, for any value of $k \in \mathbb{Z}$ in
(\ref{eq:quivervacuum(a)})-(\ref{eq:quivervacuum(b)}), one obtains
the same result for the fluctuation determinants.

\subsection{Fermionic case}

In order to compute the fluctuation determinant arising from path
integrating over the fermionic fluctuations we must first
introduce some notation:
{\small{\begin{equation}
J_k^+ (\omega) \equiv \left( \begin{array}{ccccc} 0 \hspace{-0.3cm} & 1 & \phantom{\Big(} & & \\
& 0 & \omega^k & & \\
& & 0 & \ddots & \\
& & & \ddots & \omega^{k(M-2)} \\
\omega^{k(M-1)} \hspace{-0.3cm} & & & & 0
\end{array}\right) \: , \hspace{0.5cm}
J_k^- (\omega) \equiv
\left( \begin{array}{ccccc} 0 & & \phantom{\Big(} & & \hspace{-0.3cm}\omega^{k(M-1)} \\
1 & 0 & & & \\
& \omega^{k} & 0 & & \\
& & \ddots & \ddots & \\
& & & \hspace{-0.5cm} \omega^{k(M-2)} & \hspace{-0.3cm} 0
\end{array}\right)
\end{equation}}}
and
\begin{eqnarray}
w_1 &\equiv& a \hspace{0.5mm} = \hspace{0.5mm} \langle A \rangle
\: , \hspace{3cm} \Omega_1 \hspace{0.8mm} \equiv \hspace{0.75mm} J_k^+ (\omega) \phantom{aaaaa} \\
w_2 &\equiv& b \hspace{0.5mm} = \hspace{0.5mm} \langle B \rangle
\: , \hspace{3cm} \Omega_2 \hspace{0.96mm}
\equiv \hspace{0.8mm} J_k^- (\omega^{-1}) \\
w_3 &\equiv& \phi \hspace{0.5mm} = \hspace{0.5mm} \langle \Phi
\rangle \: , \hspace{3cm} \Omega_3 \hspace{0.65mm} \equiv
\hspace{0.77mm} \mathbf{1}_M
\end{eqnarray}
where it is implied that $A, B, \Phi$ take the $\mathbb{Z}_M$
projection invariant forms given in
(\ref{eq:projinvariantAmuA})-(\ref{eq:projinvariantBFi}).

The fluctuation operator $\mathbf{D}^{mn}_{ij}$ in
(\ref{eq:quadratic-part-of-ferm.-action}) is as given below (where
$c,d = 1,2,3$)

{\footnotesize
\begin{equation}
\mathbf{D}^{mn}_{ij} \hspace{-0.7mm} \equiv \hspace{-0.7mm} \left(
\begin{array}{cc} \hspace{-4.5cm} \frac{i}{2} \delta_{pq}
\overline{\tau_\mu} \big(
\partial_\mu + \delta_{\mu 0} (\alpha_i^n - \alpha_i^m) \big)\delta_{ij} &
\hspace{-4.2cm} \begin{array}{l} -\frac{1}{\sqrt{2}} \alpha^c_{pq}
\left[ \big( (w_c)_{nn} + (\overline{w_c})_{nn} \big)
- \big( (w_c)_{mm} \Omega_c + (\overline{w_c})_{mm} \Omega_c^{-1} \big) \right]_{ij} \\
+ \frac{1}{\sqrt{2}} \beta^c_{pq} \left[ \big( (w_c)_{nn} -
(\overline{w_c})_{nn} \big) - \big( (w_c)_{mm} \Omega_c -
(\overline{w_c})_{mm} \Omega_c^{-1} \big) \right]_{ij}
\end{array} \hspace{-0.15cm} \\ \\
\hspace{-0.2cm} \begin{array}{l} -\frac{1}{\sqrt{2}} \alpha^d_{pq}
\left[ \big( (w_d)_{nn} + (\overline{w_d})_{nn} \big) - \big(
(w_d)_{mm} \Omega_d + (\overline{w_d})_{mm}
\Omega_d^{-1} \big) \right]_{ij} \\
- \frac{1}{\sqrt{2}} \beta^d_{pq} \left[ \big( (w_d)_{nn} -
(\overline{w_d})_{nn} \big) - \big( (w_d)_{mm} \Omega_d -
(\overline{w_d})_{mm} \Omega_d^{-1} \big) \right]_{ij}
\end{array} & \frac{i}{2} \delta_{pq}
\tau_\nu \big( \partial_\nu + \delta_{\nu 0} (\alpha_i^n -
\alpha_i^m) \big) \delta_{ij} \hspace{-0.25cm}
\end{array} \right) \nonumber
\end{equation}}
\vspace{-0.3cm}
\begin{equation}\label{eq:ferm.-fluctuation-operator}
\end{equation}
The reason why the $w_c$ entries labelled by the gauge index $m$
have additional factors of $\Omega_c$ compared to the entries
labelled by $n$ comes from the commutator structure of the Yukawa
coupling (see (\ref{eq:quiveraction(ferm,nf)})). Namely, when
taking the trace over the gauge indices, the $w_c$ entries
labelled by $n$ correspond to the terms where a scalar field
appears between two spinor fields, whereas those labelled with $m$
correspond to the terms where the scalar field appears to the
right of both spinor fields. After substituting the orbifold
projection invariant forms given in Eqs.
(\ref{eq:projinvariantAmuA})-(\ref{eq:projinvariantBFi}) and
(\ref{eq:projinvariantpsichiA})-(\ref{eq:projinvariantchiBpsiPhi}),
the bifundamental scalar VEV's will couple different pairs of
spinor fields depending on whether the VEV appears between the
spinor fields or to the right of them in the Yukawa coupling.
Since the scalar VEV's are mutually related through the vacuum
(\ref{eq:quivervacuum(a)})-(\ref{eq:quivervacuum(alpha)}), this
can be compensated for by appropriately multiplying factors of
$\Omega_c.$

To compute the result of the path integrations it is convenient to
define (for a fixed $c$)
\begin{eqnarray}
F_c &\equiv& \big( (w_c)_{nn} + (\overline{w_c})_{nn} \big) -
\big( (w_c)_{mm} \Omega_c + (\overline{w_c})_{mm} \Omega_c^{-1}
\big) \\
G_c &\equiv& \big( (w_c)_{nn} - (\overline{w_c})_{nn} \big) -
\big( (w_c)_{mm} \Omega_c - (\overline{w_c})_{mm} \Omega_c^{-1}
\big) \: .
\end{eqnarray}
Noting that
\begin{equation}
\big[ F_c, F_d \big] = 0 \: , \hspace{1cm} \big[ F_c, G_d \big] =
0 \: , \hspace{1cm} \big[ G_c, G_d \big] = 0
\end{equation}
one finds, by using the (anti)commutation relations
(\ref{eq:(anti)commutation_for_alpha_beta}) for $\alpha^c$ and
$\beta^d$, that the result of the path integrations over the
fermionic fluctuations $(\lambda_p)_i, (\overline{\lambda_p})_i$
is
\begin{eqnarray}
\det \hspace{0.43mm} (\mathbf{D}^{mn}_{ij}) &=& \det \Big(
\hspace{-0.5mm} -\big(
\partial_\mu + i \delta_{\mu 0} (\alpha_i^n - \alpha_i^m) \big)^2
\nonumber \phantom{aaaaaaaaaaaaaaaaaaaaaaaaaaaaaaaaaaa}\\
&\phantom{=}& \phantom{aaaa,} - \textstyle{\frac{1}{2}}\big(
\textstyle{\frac{1}{2}} \{ \alpha^c, \alpha^d\}_{pr} F_c F_d +
[\alpha^c, \beta^d]_{pr} F_c G_d - \textstyle{\frac{1}{2}} \{
\beta^c, \beta^d \}_{pr} G_c G_d \big)\Big) \\
&=& \det \Big( \hspace{0.1mm} \big( i
\partial_\mu - \delta_{\mu 0} (\alpha_i^n - \alpha_i^m) \big)^2 +
\textstyle{\frac{1}{2}} (F_c F_d - G_c G_d) \hspace{0.7mm}
\delta_{cd} \delta_{pr} \Big) \\
&=& \det \Delta_{ij} \: . \label{eq:det_of_fermD_simplified}
\end{eqnarray}
Here we have defined the $M \times M$ matrix (labelled by $i,j =
1,\ldots,M$) {\small{\begin{equation} \Delta_{ij} = \left\{
\begin{array}{ll} -2 \, \Big( (a_{1,2})_{nn} \hspace{0.4mm}
(\overline{a_{1,2}})_{mm} + (\overline{b_{2,1}})_{nn}
\hspace{0.4mm} (b_{2,1})_{mm}
\Big) \hspace{0.4mm} \omega^{-(i-2) \hspace{0.2mm}k} & \hspace{0.6cm} \mathrm{for} \:\, j = i - 1 \vspace{0.45cm} \\
\hspace{0.1mm} -\partial^2 -2i(\alpha^n_1 - \alpha^m_1) \hspace{0.2mm} \partial_0 + (\alpha^n_1 - \alpha^m_1)^2 & \vspace{0.05cm} \\
\hspace{0.3cm} + \hspace{1.5mm} 2 \hspace{0.4mm} \Big(
(a_{1,2})_{nn} \hspace{0.4mm} (\overline{a_{1,2}})_{nn}
+ (a_{1,2})_{mm} \hspace{0.4mm} (\overline{a_{1,2}})_{mm} & \vspace{0.15cm} \\
\hspace{1.2cm} + \, (b_{2,1})_{nn} \hspace{0.4mm}
(\overline{b_{2,1}})_{nn} + (b_{2,1})_{mm} \hspace{0.4mm}
(\overline{b_{2,1}})_{mm} & \vspace{0.15cm} \\
\hspace{1.2cm} + \, \big( (\phi_1)_{nn} - (\phi_1)_{mm} \big)
\hspace{0.5mm} \big( (\overline{\phi_1})_{nn}
- (\overline{\phi_1})_{mm}\big) \Big) & \hspace{0.6cm} \mathrm{for} \:\, j = i \vspace{0.45cm} \\
-2 \, \Big( (\overline{a_{1,2}})_{nn} \hspace{0.4mm}
(a_{1,2})_{mm} + (b_{2,1})_{nn} \hspace{0.4mm}
(\overline{b_{2,1}})_{mm} \Big) \hspace{0.4mm} \omega^{(i-1)
\hspace{0.2mm}k} & \hspace{0.6cm} \mathrm{for} \:\, j = i + 1
\end{array}\right. \label{eq:fluctuation-operator(Delta)}
\end{equation}}}
where we have used
(\ref{eq:quivervacuum(a)})-(\ref{eq:quivervacuum(alpha)}) to
arrive at the equality (\ref{eq:det_of_fermD_simplified}).
Applying the determinant formula
(\ref{eq:determinant_circulant_matrix}) and using
(\ref{eq:quivervacuum(a)})-(\ref{eq:quivervacuum(alpha)}) again
one finds, after taking the traces over the fermionic Matsubara
frequencies $\omega_k \equiv \frac{(2k+1)\pi}{\beta}$ and over the
$S^3$ spherical harmonics, the expression
(\ref{eq:quantum_effective_action(ferm)}).

\addcontentsline{toc}{section}{\ \ \ \ References}


\providecommand{\href}[2]{#2}\begingroup\raggedright\endgroup

\end{document}